\documentclass[traditabstract]{aa}

\usepackage{graphicx}
\usepackage{rotating}
\usepackage{amsmath}
\usepackage{txfonts}
\usepackage{color}
\usepackage{ulem}
\usepackage{hyperref}

\begin{document}

\title{Ground-based astrometry with wide field imagers}

\subtitle{V. Application to near-infrared detectors:
  HAWK-I@VLT/ESO\thanks{Based on observations with the 8\,m VLT ESO
    telescope.}}

\author{ 
  M.\ Libralato\inst{1,2,3,}\thanks{Visiting Ph.D. Student at STScI
  under the 2013 DDRF program.},
  A.\ Bellini\inst{2},
  L.\ R.\ Bedin\inst{3},
  G.\ Piotto\inst{1,3}, 
  I.\ Platais\inst{4}, 
  M.\ Kissler-Patig\inst{5,6},
  A.\ P.\ Milone\inst{7}
}

\institute{
  Dipartimento\ di Fisica e Astronomia, Universit\`a di Padova, 
  Vicolo dell'Osservatorio 3, Padova, I-35122, Italy \\
  \email{giampaolo.piotto@unipd.it;mattia.libralato@studenti.unipd.it}
  \and 
  Space Telescope Science Institute,
  3700 San Martin Drive, Baltimore, MD-21218, USA \\
  \email{bellini@stsci.edu}	     
  \and
  INAF-Osservatorio Astronomico di Padova,
  Vicolo dell'Osservatorio 5, Padova, I-35122, Italy \\
  \email{luigi.bedin@oapd.inaf.it} 
  \and 
  Department of Physics and Astronomy, The Johns Hopkins University, 
  Baltimore, MD-21218, USA \\ 
  \email{imants@pha.jhu.edu}
  \and
  Gemini Observatory,
  N. A’ohoku Place 670, Hilo, Hawaii, 96720, USA \\
  \email{mkissler@gemini.edu}	     
  \and
  European Southern Observatory,
  Karl-Schwarzschild-Str.\ 2, Garching b. M\"unchen, D-85748, Germany
  \and
  Research School of Astronomy and Astrophysics, The Australian National University, Cotter Road, Weston, ACT, 2611, Australia \\
  \email{milone@mso.anu.edu.au}
}

\date{Received 11 June 2013 / Accepted 18 December 2013}
 
\abstract {

High-precision astrometry requires accurate point-spread function
modeling and accurate geometric-distortion corrections.  This paper
demonstrates that it is possible to achieve both requirements with
data collected at the high acuity wide-field $K$-band imager (HAWK-I),
a wide-field imager installed at the Nasmyth focus of UT4/VLT ESO 8\,m
telescope. Our final astrometric precision reaches $\sim$3 mas per
coordinate for a well-exposed star in a single image with a systematic
error less than 0.1 mas. We constructed calibrated astro-photometric
catalogs and atlases of seven fields: the Baade's Window, NGC~6656,
NGC~6121, NGC~6822, NGC~6388, NGC~104, and the James Webb Space
Telescope calibration field in the Large Magellanic Cloud. We make
these catalogs and images electronically available to the
community. Furthermore, as a demonstration of the efficacy of our
approach, we combined archival material taken with the optical
wide-field imager at the MPI/ESO 2.2\,m with HAWK-I observations. We
showed that we are able to achieve an excellent separation between
cluster members and field objects for NGC~6656 and NGC~6121 with a
time base-line of about 8 years. Using both {\it{HST}} and HAWK-I
data, we also study the radial distribution of the SGB populations in
NGC~6656 and conclude that the radial trend is flat within our
uncertainty.  We also provide membership probabilities for most of the
stars in NGC~6656 and NGC~6121 catalogs and estimate membership for
the published variable stars in these two fields.


}

\titlerunning{Ground-based astrometry with wide field imagers. V.}
\authorrunning{Libralato, M. et al.}
\maketitle

%%%%%%%%
\section{Introduction}
%%%%%%%%
\label{intr}

\begin{table*}[!t]
  \caption{List of the major operative wide-field imagers on 3\,m $+$
    telescopes. The WFI@2.2\,m MPI/ESO has been included as
    reference.}  
  \label{tab:wfi}     
  \centering
  \begin{tabular}{ccccc}          
    \hline\hline
    \textbf{Name} & \textbf{Telescope} & \textbf{Detectors} & \textbf{Pixel Scale [$^{\prime\prime}$/pixel]} & \textbf{FoV} \\
    \hline
    & & & & \\
    \multicolumn{5}{c}{\textbf{OPTICAL REGIME}} \\
    & & & & \\
    WFI & 2.2\,m MPI/ESO & 8$\times$(2048$\times$4096) & 0.238 & $34^{\prime}$$\times$$33^{\prime}$ \\
    Prime Focus Camera & William Herschel Telescope & 2$\times$(2048$\times$4100) & 0.24 & $16^{\prime}\!\!.2$$\times$$16^{\prime}\!\!.2$ \\
    LBC (blue and red) & LBT & 4$\times$(2048$\times$4608) & 0.23 & $23^{\prime}$$\times$$23^{\prime}$ \\
    Suprime-Cam & Subaru Telescope & 10$\times$(2048$\times$4096) & 0.202 & $34^{\prime}$$\times$$27^{\prime}$ \\
    MOSA & KPNO Mayall 4\,m & 8$\times$(2048$\times$4096) & 0.26 & $36^{\prime}$$\times$$36^{\prime}$ \\
    LAICA & Calar Alto 3.5\,m Telescope & 4$\times$(4096$\times$4096) & 0.225 & $44^{\prime}\!\!.36$$\times$$44^{\prime}\!\!.36$ \\
    MegaCam & CFHT & 36$\times$(2048$\times$4612) & 0.187 & $57^{\prime}\!\!.6$$\times$$56^{\prime}\!\!.4$ \\
    OmegaCam & VST & 32$\times$(2048$\times$4102) & 0.21 & $60^{\prime}$$\times$$60^{\prime}$ \\
    DECam & CTIO Blanco 4\,m & 62$\times$(2048$\times$4096)$+$12$\times$(2048$\times$2048) & 0.27 & $132^{\prime}$$\times$$132^{\prime}$ \\
    & & & & \\
    \hline
    & & & & \\
    \multicolumn{5}{c}{\textbf{NIR REGIME}} \\
    & & & & \\
    GSAOI & Gemini & 4$\times$(2048$\times$2048) & 0.02 & $1^{\prime}\!\!.42$$\times$$1^{\prime}\!\!.42$ \\
    HAWK-I & VLT & 4$\times$(2048$\times$2048) & 0.106 & $7^{\prime}\!\!.5$$\times$$7^{\prime}\!\!.5$ \\
    ISPI & CTIO Blanco 4\,m & 1$\times$(2048$\times$2048) & 0.3 & $10^{\prime}\!\!.25$$\times$$10^{\prime}\!\!.25$ \\
    FourStar & Magellan & 4$\times$(2048$\times$2048) & 0.159 & $10^{\prime}\!\!.8$$\times$$10^{\prime}\!\!.8$ \\
    WFCAM & UKIRT & 4$\times$(2048$\times$2048) & 0.4 & $12^{\prime}\!\!.6$$\times$$12^{\prime}\!\!.6$ \\
    Omega2000 & Calar Alto 3.5\,m Telescope & 4$\times$(2048$\times$2048) & 0.45 & $15^{\prime}\!\!.4$$\times$$15^{\prime}\!\!.4$ \\
    WIRCAM & CFHT & 4$\times$(2048$\times$2048) & 0.3 & $20^{\prime}\!\!.5$$\times$$20^{\prime}\!\!.5$ \\
    NEWFIRM & CTIO Blanco 4\,m & 4$\times$(2048$\times$2048) & 0.4 & $27^{\prime}\!\!.6$$\times$$27^{\prime}\!\!.6$ \\
    VIRCAM & VISTA & 16$\times$(2048$\times$2048) & 0.339 & $35^{\prime}\!\!.4$$\times$$35^{\prime}\!\!.4$ \\
    & & & & \\
    \hline
  \end{tabular}
\end{table*}

Multiple fields within astronomy are driving the execution of larger
and yet larger surveys of the sky. Over the last two decades, this
scientific need has stimulated the construction of instruments
equipped with mosaics of large-format digital detectors for wide-field
imaging at both the optical and near-infrared (NIR) wavelengths. The
most recent generation of these wide-field imagers now competes with
the older technology of Schmidt telescope and photographic plates in
terms of number of resolution elements on sky but does so with
order-of-magnitude greater sensitivity and efficiency. \looseness=-2

A list of some widely-used wide-field imagers was given by Anderson et
al. (2006, hereafter Paper~I). Since then, however, many wide-field
imagers have been upgraded or decommissioned, and additional new
wide-field imagers have begun their operations. In the top-half of
Table~\ref{tab:wfi}, we provide a brief list of the major operative
wide-field imagers on 3\,m$+$ telescopes (we also included the
WFI@2.2\,m MPI/ESO as reference). \looseness=-2

In addition, two wide-field imagers mounted on 1\,m telescopes should
be mentioned. The LaSilla-QUEST Variability survey is a project that
uses the ESO 1.0-m Schmidt Telescope at the La Silla Observatory of
the European Southern Observatory in Chile with the new large area
QUEST camera. It is a mosaic of 112 600$\times$2400 pixels CCDs
covering a field of view of about
$4^{\circ}\!\!.6$$\times$$3^{\circ}\!\!.6$.  The camera, commissioned
in early 2009 has been built at the Yale and Indiana University.  La
Silla-QUEST survey is expected to cover about 1000 square degrees per
night repeated with a 2-day cadence (Hadjiyska et
al. 2012). \looseness=-2

The Panoramic Survey Telescope and Rapid Response System (Pan-STARRS)
also is of great interest for the astronomical community. The
Pan-STARRS survey will cover the sky using wide-field facilities and
provide astrometric and photometric data for all observed objects.
The first Pan-STARRS telescope, PS1, is located at the summit of
Haleakala on Maui, Hawaii and began full time science observations on
May 13, 2010 (Kaiser et al., 2010). With its 1.8\,m primary mirror, it
covers a FoV of $\sim$7 square degrees. \looseness=-2

Among wide-field imagers planned for the future, the
LSST\footnote{{\url{http://www.lsst.org/lsst/}}.} (Large Synoptic
Survey Telescope) represents the most significant step forward for
wide-field imagers in modern astrophysics. It will be a 8.4\,m
wide-field ground-based telescope with a FoV of about 9.6 square
degrees. With its 189 4k$\times$4k CCDs, it will observe over 20\,000
square degrees of the southern sky in six optical bands. Construction
operations should begin in 2014; the survey will be taken in
2021. \looseness=-2

While a great number of papers have presented photometry obtained with
these facilities over the last decade, their astrometric potential has
remained largely unexploited. Our team is committed in pushing the
astrometric capabilities of wide-field imagers to their limits.
Therefore, we have begun to publish in this Journal a series of papers
on \textit{Ground-based astrometry with wide field imagers}. In
\textit{Paper~I}, we developed and applied our tools to data collected
with the WFI@2.2\,m MPI/ESO telescope. The techniques used in
\textit{Paper~II} (Yadav et al. 2008) and \textit{Paper~III} (Bellini
et al. 2009) produced astro-photometric catalogs and proper motions of
the open cluster M\,67 and of the globular cluster NGC~5139,
respectively. In \textit{Paper~IV} (Bellini \& Bedin 2010), we applied
the technique to the wide-field camera on the blue focus of the
LBC@LBT 2$\times$8.4\,m. \looseness=-2

In this paper, we turn our attention to wide-field imagers equipped
with NIR detectors. Indeed, the pioneering work of 2MASS has shown the
great potential of these instruments (Skrutskie et al. 2006). The
number and the quality of NIR detectors has improved considerably
since then. New generations of 2k$\times$2k arrays are now mounted at
the foci of various telescopes (bottom-half of
Table~\ref{tab:wfi}). \looseness=-2

These wide-field imagers enable wide surveys, such as the VISTA
variables in the Via Lactea (VVV, Minniti et al 2010). The VVV is
monitoring the Bulge and the Disk of the Galaxy. The survey will map
562 square degrees over 5 years (2010-2014) and give NIR photometry in
$Z$, $Y$, $J$, $H$, and $K_{\rm S}$ bands. The first data set of the
VVV project has already been released to the community (Saito et
al. 2012). It contains 348 individual pointings of the Bulge and the
Disk, taken in 2010 with $\sim$$10^8$ stars observed in all
filters. Typically, the declared astrometric precisions vary from
$\sim$25 mas for a star with $K_{\rm S} = 15$, to $\sim$175 mas for a
star with $K_{\rm S} = 18$ mag. \looseness=-2

Ground-based telescopes are not alone in focusing their attention on
this kind of detector. The James Webb Space Telescope (JWST) will be a
6.5-meter space telescope optimized for the infrared regime. It will
orbit around the Earth's second Lagrange point (L2), and it will
provide imaging and spectroscopic data. The wide-field imager, NIRCam,
will be made up by a short- (0.6 -- 2.3 $\rm \mu m$) and a
long-wavelength (2.4 -- 5.0 $\rm \mu m$) channel with a FoV of
$2^{\prime}\!\!.2$$\times$$4^{\prime}\!\!.4$ each. \looseness=-2

In this paper, we explore the astrometric performance of the
HAWK-I@VLT facility and provide astrometric catalogs (together with
stacked images) of seven dense stellar fields along with the tools
required to correct any observed field for geometric distortion. We
adopt the UCAC~4 catalog (Zacharias et al. 2013) as a reference frame
to determine the linear terms of the distortion and to put all the
objects in on the International Celestial Reference System but at the
epoch of our observations. \looseness=-2

The paper is organized as follows: In Sect~\ref{telescope}, we briefly
describe the instrument. Section~\ref{OBS} presents the
observations. Section~\ref{psf} describes the tools developed to
extract the position and flux of the sources. In Sect.~\ref{GDC}, we
derive our geometric-distortion solution. In Sect.~\ref{expl}, we
discuss the periodic feature of the residual highlighted during the
geometric-distortion correction and give a possible
explanation. Sections~\ref{weed} and \ref{calib} describe how we
removed the point-spread function (PSF) artifacts from the catalogs
and the photometric calibration, respectively. Section~\ref{app} shows
some possible applications of our calibrations. Finally, we describe
the catalogs that we release with this paper in
Sect.~\ref{catalog}. \looseness=-2

\begin{figure}
  \centering
  \includegraphics[width=9.cm]{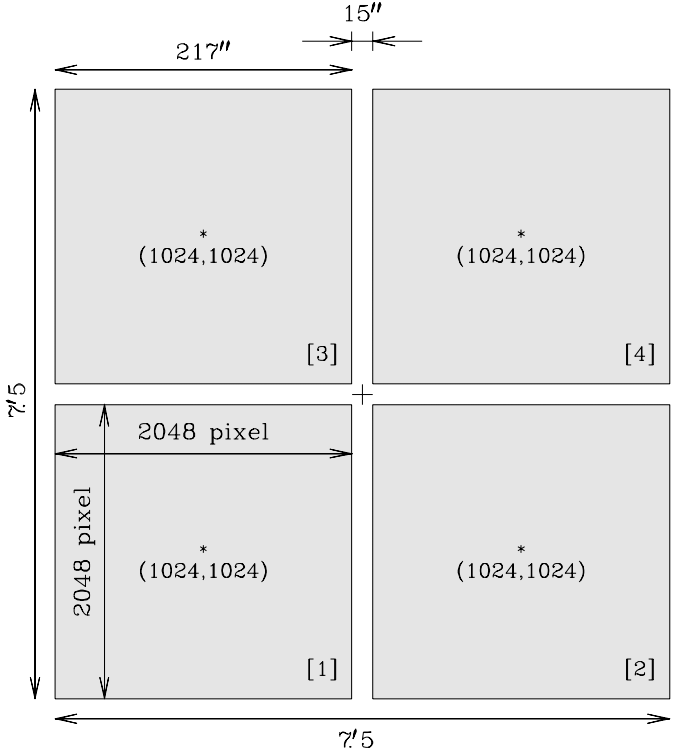}
  \caption{HAWK-I layout. The labels give the dimensions in arcsec and
    arcmin (and in pixels) of the four detectors and of the gaps.
    Numbers in square brackets label the chip denomination used in
    this work (note that the choice is different from that of Fig.~9
    of Kissler-Patig et al. 2008). In each chip, we indicate the
    coordinate of the chip center. This is also the reference position
    that we used while computing the polynomial correction described
    in Sect.~\ref{poly}. The black cross in the middle shows the
    center of the field of view in a single exposure that we used in
    Fig.~\ref{fig:dit}.}
  \label{fig:dev}
\end{figure}

%%%%%%%%
\section{HAWK-I@VLT}
%%%%%%%%
\label{telescope}

An exhaustive description of HAWK-I is given in Kissler-Patig et
al.\ (2008). Here, we only provide a brief summary. \looseness=-2

The HAWK-I focal plane is equipped with a mosaic of four
2048$\times$2048 pixels Rockwell HgCdTe Molecular Beam Epitaxy
HAWAII-2RG arrays.  The pixel-scale (Kissler-Patig et al. 2008) is
about 106 mas pixel$^{-1}$, resulting in a total FoV of about
$7^{\prime}\!\!.5$$\times$$7^{\prime}\!\!.5$ (with gaps of
$\sim$$15^{\prime \prime}$ between the detectors). A sketched outline
of the HAWK-I FoV layout is shown in Fig.~\ref{fig:dev}. The detectors
and the filter wheel unit are connected to the second stage of the
Closed Cycle Cooler and operated at a temperature close to 75--80
K. The remaining parts of the instrument are cooled to a temperature
below 140 K. The acquisition system is based on the IRACE system
(Infrared Array Control Electronics) developed at ESO. HAWK-I also is
designed to work with a ground-layer adaptive optics module (GRAAL) as
part of the Adaptive Optics Facility (Arsenault et al. 2006) for the
VLT (scheduled to be installed in the second half of 2014). HAWK-I
broad band filters follow the Mauna Kea Observatory
specification. \looseness=-2

\begin{table*}[!t]
  \caption{List of the HAWK-I@VLT data set used for this work. N$_{\rm
      dither}$ is the number of dithered images per observing
    block. ``Step'' is the dither spacing (shift in arcsec from one
    exposure to the next one). The integration time (DIT) times the
    number of individual integrations (NDIT) gives the exposure
    time. $\sigma$(Radial residual) gives an assessment of the
    astrometric accuracy reached (see Sect.~\ref{acc} for the full
    description).}
  \label{tab:obs}     
  \centering
  \begin{tabular}{ccccccc}          
    \hline\hline
    \textbf{Filter} & \textbf{N$_{\rm dither}$} & \textbf{Step} & \textbf{Exposure Time} & \textbf{Image-quality} & \textbf{Airmass} & \textbf{$\sigma$(Radial residual)} \\ 
    & & (arcsec) & (NDIT$\times$DIT) & (arcsec) & ($\sec z$) & (mas) \\
    \hline  
    & & & \\        
    \multicolumn{7}{c}{\it \textbf{Commissioning 1}, August 3-6, 2007} \\
    & & & \\
    \multicolumn{7}{c}{ {\bf Bulge --- Baade's Window (\#1)}} \\
    $J$ & 25 & 95 & (6$\times$10\,s) & $0.56$-$1.07$ & 1.038-1.085 & 4.5 \\
    $H$ & 25 & 95 & (6$\times$10\,s) & $0.40$-$0.76$ & 1.081-1.149 & 4.3 \\
    $K_{\rm S}$ & 25 & 95 & (6$\times$10\,s) & $0.25$-$0.45$ & 1.042-1.091 & 2.8 \\
    & & & \\
    \multicolumn{7}{c}{ {\bf Bulge --- Baade's Window (Rotated by 135$^{\circ}$) (\#2)}} \\
    $K_{\rm S}$ & 25 & 95 & (6$\times$10\,s) & $0.49$-$0.85$ & 1.015-1.044 & 5.6 \\
    & & & \\
    \multicolumn{7}{c}{ {\bf NGC~6121 (M~4)} } \\
    $J$  & 4$\times$25 & 95 & (6$\times$10\,s) &  $0.36$-$1.04$ & 1.010-1.540 & 6.5 \\
    $K_{\rm S}$  & 5 & 140 & (6$\times$10\,s) &  $0.40$-$0.51$ & 1.050-1.056 & 3.8 \\
    & & & \\
    \multicolumn{7}{c}{ {\bf NGC~6822} } \\
    $J$ & 9 & 190 & (12$\times$10\,s) &  $0.61$-$0.83$ & 1.028-1.049 & 5.3 \\
    $K_{\rm S}$ & 9 & 190 & (12$\times$10\,s) & $0.43$-$0.75$ & 1.050-1.082 & 4.8 \\
   & & & \\
   \hline
   & & & \\
   \multicolumn{7}{c}{\it \textbf{Commissioning 2}, October 14-19, 2007} \\
   & & & \\
   \multicolumn{7}{c}{ {\bf NGC~6656 (M~22)}} \\
   $K_{\rm S}$ & 25 & 47.5 & (6$\times$10\,s) & $0.28$-$0.41$ & 1.252-1.420 & 3.1 \\
   & & & \\
   \multicolumn{7}{c}{ {\bf NGC~6388} } \\
   $J$ & 25 & 95 & (6$\times$10\,s) &  $0.64$-$0.94$ & 1.287-1.428 & 9.7 \\
   $K_{\rm S}$ & 25 & 95 & (6$\times$10\,s) & $0.50$-$0.75$ & 1.436-1.637 & 12.2 \\
   & & & \\
   \multicolumn{7}{c}{ {\bf JWST calibration field (LMC)} } \\
   $J$ & 25 & 95 & (6$\times$10\,s) &  $0.51$-$0.65$ & 1.408-1.412 & 5.3 \\
   $K_{\rm S}$ & 24 & 95 & (6$\times$10\,s) &  $0.45$-$0.60$ & 1.411-1.429 & 4.8 \\
   & & & \\
   \hline
   & & & \\
   \multicolumn{7}{c}{\it \textbf{Commissioning 3}, November 28-30, 2007} \\
   & & & \\
   \multicolumn{6}{c}{ {\bf NGC~104 (47\,Tuc)} }\\
   $J$ & 25 & 47.5 & (6$\times$10\,s) & $0.51$-$0.82$ & 1.475-1.479 & 7.1 \\
   $K_{\rm S}$ & 23 & 47.5 & (6$\times$10\,s) & $0.54$-$1.01$ & 1.475-1.483 & 15.0 \\
    & & & \\
    \hline
  \end{tabular} 
\end{table*} 

\begin{figure}
  \centering
  \includegraphics[width=9.cm]{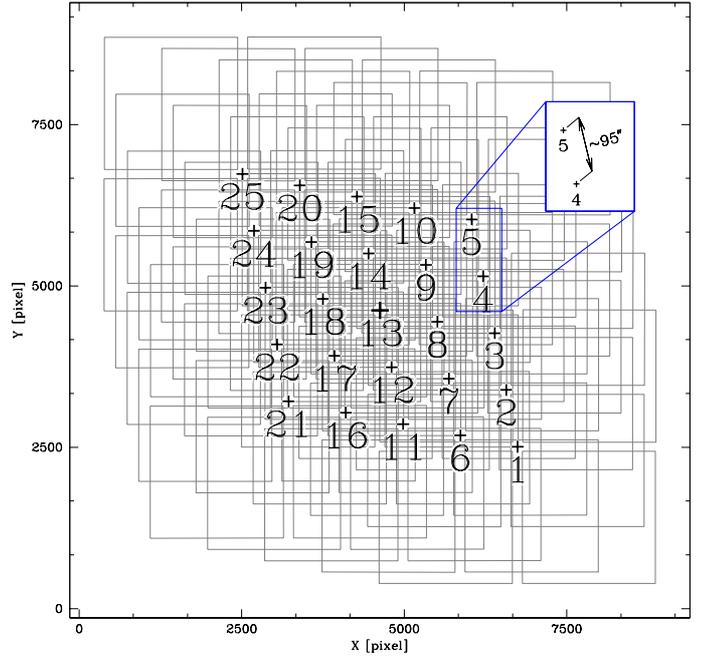}
  \caption{Outline of the relative positions of pointings in our
    adopted dither-pattern strategy.  The 25 images are organized in a
    5$\times$5 array, where the center of the field falls in the
    central position 13.  The other pointings are taken in a way that
    the gaps between the four detectors never cover the same point of
    sky more than once. The 25 images were designed for astrometric
    purposes allowing stars in frame 13 to be imaged in as many
    different locations of the detectors as possible. This enables us
    to self-calibrate the geometric distortion. The zoom-in in the
    blue panel shows an example of the adopted dither between two
    pointings. As described in Table~\ref{tab:obs}, the shift step can
    change from field to field.}
  \label{fig:dit}
\end{figure}

%%%%%%%%
\section{Observations}
%%%%%%%%
\label{OBS}

In Table~\ref{tab:obs} we provide a detailed list of the
observations. \looseness=-2

All of the HAWK-I images used here were collected during the
instrument commissioning, when several fields were observed with the
aim of determining an average optical geometric-distortion solution
for HAWK-I and for monitoring its stability in the short- and
mid-term. \looseness=-2

To this end, the fields were observed with an observing strategy that
would enable a self-calibration of the distortion. Briefly, the
strategy consists of observing a given patch of sky in as many
different parts of the detectors as possible. Each observing block
(OB) is organized in a run of 25 consecutive images. The exposure time
for each image was the integration time (DIT in s) times the number of
individual integrations (NDIT). Figure~\ref{fig:dit} shows the outline
of the adopted dither-pattern strategy\footnote{Note that this
  strategy had been modified for some fields. We specify these
  changes, when necessary, in the following
  subsections.}. \looseness=-2

Important by-products of this effort are \textit{astrometric standard
  fields} (i.e., catalogs of distortion-free positions of stars),
which in principle could be pointed by HAWK-I anytime in the future to
efficiently assess whether the distortion has varied and by how much.
Furthermore, these astrometric standard fields might serve to
calibrate the geometric distortions of many other cameras on other
telescopes (including those equipped with AO, MCAO, or those
space-based). However, the utility of our fields deteriorate over time
since a proper motion estimate for stars in our catalogs is only
provided for those stars that are in common with UCAC~4 catalog. In
this paper, we make these astrometric standard fields available to the
community. \looseness=-2

\begin{figure*}[t]
  \centering
  \includegraphics[width=\textwidth]{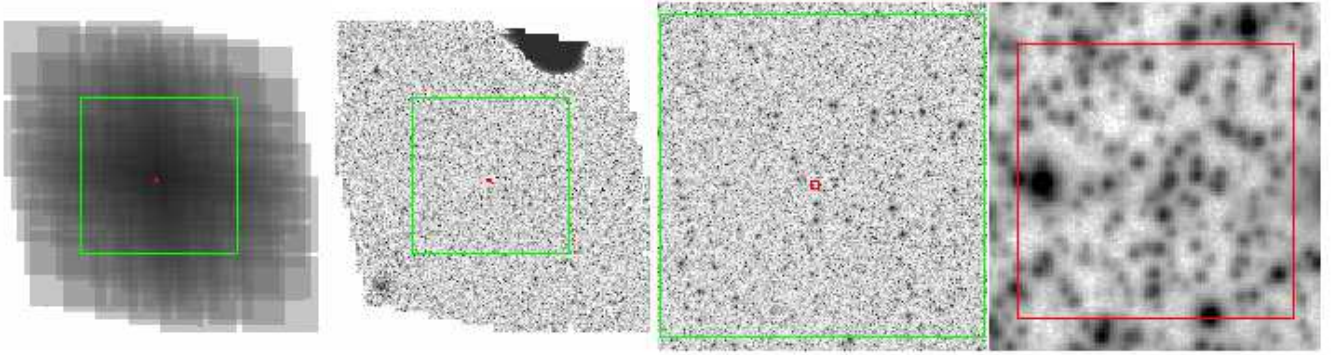}
  \caption{(\textit{From Left to Right}): The \textit{First Panel} is
    a depth-of-coverage map of 25 $K_{\rm S}$ HAWK-I's images
    collected in Baade's Window during the run of August 3-6,
    2007. The gray-scale goes linearly from 1 to 25. The green box is
    the $7^{\prime}\!\!.5\times7^{\prime}\!\!.5$ patch of sky within
    which there are always at least five images. The \textit{Second
      Panel} shows the resulting stack of the 25 images.  The dark
    spot on the top-right is the signature left by the ``shadow'' of
    the probe, which pick-up the star used for the simultaneous Active
    Optic correction of the VLT/UT4's primary mirror.  On the
    bottom-left, there is the globular cluster NGC~6522. Note that
    neither the dark spot nor NGC~6522 are inside the region enclosed
    by the green box. The \textit{Third Panel} focuses on the green
    region and shows that the distribution of stars in this field is
    remarkably homogeneous. The \textit{Fourth Panel} is a zoom-in of
    a representative sub-set of the field (indicated by the
    $10^{\prime\prime}\times10^{\prime\prime}$ red box in all panels),
    which is able to show a better resolved image.}
  \label{fig:bulgedit}
\end{figure*}

%%%%%%%%%%%
\subsection{The Baade's Window astrometric field} 
%%%%%%%%%%% 

The first selected astrometric field is located in Baade's Window. Our
field is centered on coordinates $(\alpha,\delta)_{\rm
  J2000.0}\sim(\rm
18^h03^m10^s\!\!.4,-29^{\circ}56^{\prime}48^{\prime\prime}\!\!.4)$.
Most of this field has a smooth, uniform distribution of Galactic
Bulge stars. Stars just below saturation in a 60\,s $K_{\rm S}$
exposure are typically separated by a few arcseconds, so that there
are many of them in each field. In general, however, they are
sufficiently isolated to allow us to compute accurate
positions. \looseness=-2
 
The choice of a cumulative integration time of 60\,s was driven by two
considerations. First, we wanted to have the upper main sequence in a
CMD of all chosen targets to be optimally exposed with low-luminosity
RGB stars still below the saturation threshold. Second, the
large-scale semi-periodic and correlated atmospheric noise (with
estimated scale length at $\sim$3-5$\arcmin$) noted by Platais et
al. (2002, 2006) essentially disappears at exposures exceeding
30\,s. Thus, 60\,s was a good compromise of integration
time. According to the formula developed by Lindegren (1980) and Han
(1989), a standard deviation due to atmospheric noise on the order of
15 mas is expected over the angular extent of HAWK-I FoV. This
certainly is an upper limit of the actual standard deviation because
the seeing conditions of our NIR observations were 2-3 times better
than those considered by the aforementioned authors for visual
wavelengths. \looseness=-2

The images were taken close to the zenith in an effort to minimize
differential refraction effects, which plague ground-based images (and
consequently affect the estimate of the low-order terms of the
distortion solution). \looseness=-2

The Baade's Window field is the main field we use to derive the
geometric-distortion solution that is tested for stability --or
refined-- with the other fields.  In Sect.~\ref{GDC}, we derive the
geometric-distortion solution in this field for each of the three
available filters, $J$, $H$ and $K_{\rm S}$, using 25 images dithered
with a step of about 95$^{\prime\prime}$. In addition to this, we also
collected 25 $K_{\rm S}$ images of the same field but with the
de-rotator at a position of 135$^\circ$ clockwise. We used this field
to perform a check of the distortion with different angles (see
Sect.~\ref{extcheck}). \looseness=-2

In Fig.~\ref{fig:bulgedit}, we show a summary of one of these
observing runs in filter $K_{\rm S}$ from left to right: the overlap
of the different pointings, the stacked image, a zoom-in of the region
actually used to calibrate the geometric distortion (the region
highlighted in green), and a further zoom-in at a resolution able to
reveal individual pixels (region indicated in red in the other
panels). \looseness=-2

%%%%%%%%%%%
\subsection{The star-cluster astrometric fields} 
%%%%%%%%%%% 

The tangential internal motions of Bulge stars is on average 100 km
s$^{-1}$, and assuming an average distance of 8 kpc, this yields a
proper motion dispersion of $\sim$3 mas yr$^{-1}$ (see, for example,
Bedin et al. 2003). In just a few years, proper motions this large can
mask out systematic distortion trends that have amplitudes below the
3-mas-yr$^{-1}$ level (such as those discussed in
Sect.~\ref{step}). It is therefore important in some applications to
have more stable astrometric fields. \looseness=-2

For this reason, we also observed four globular clusters. Stars
gravitationally bound in globular clusters have an internal velocity
dispersion $\lesssim$20 km s$^{-1}$ in their cores and are even
smaller in their outskirts. Although the systemic motion of star
clusters is usually different to (and larger than) the Galactic field
dispersion, their common rest-frame motions are generally more than 10
times smaller than the internal motions of Bulge stars, so clusters
members can be expected to serve as astrometric standards with much
smaller internal proper motions. \looseness=-2

%%%%%%%%%%%%%%
\subsubsection{NGC~6656 (M~22)} 
%%%%%%%%%%%%%%

The second field was centered on the globular cluster NGC~6656 (M~22).
At a distance of about 3.2 kpc, M~22 $(\alpha,\delta)_{\rm
  J2000.0}=(\rm
18^h36^m23^s\!\!.94,-23^{\circ}54^{\prime}17^{\prime\prime}\!\!.1$,
Harris 1996, 2010 edition) is one of the closest globular clusters to
the Sun. \looseness=-2

These data were impacted by an internal reflection of the Moon in the
optics, causing an abnormally-high sky value on the rightmost 300
pixels of the detector. In spite of this, the exquisite image quality
of these data makes them among the best in our database. We used this
field to test the solution of the geometric distortion (see
Sect.~\ref{stab} for detail). \looseness=-2

%%%%%%%%%%%%%%
\subsubsection{NGC~6121 (M~4)} 
%%%%%%%%%%%%%%

The third field is centered on globular cluster NGC~6121 (M~4),
$(\alpha,\delta)_{\rm
  J2000.0}=(\rm16^h23^m35^s\!\!.22,-26^{\circ}31^{\prime}32^{\prime\prime}\!\!.7$,
Harris 1996, 2010 edition). It is the closest globular cluster to the
Sun, and its rich star field has a small angular distance from the
Galactic Bulge. \looseness=-2

The observing strategy for the $J$-filter is similar to that described
before. Each OB is organized in a run of 25 consecutive exposures and
the the same block was repeated four times in four different nights,
shifting the grid by few arcsec each time. \looseness=-2

This field was also observed in the $K_{\rm S}$-filter but with a
dither pattern completely different from the others. There are only
five exposures dithered with steps of $100^{\prime\prime}$, which are
taken with the purpose of estimating stars' color. \looseness=-2

%%%%%%%%%%%%%% 
\subsubsection{NGC~6388} 
%%%%%%%%%%%%%%

NGC~6388 is a globular cluster located in the Galactic Bulge at
$(\alpha,\delta)_{\rm J2000.0}=(\rm
17^h36^m17^s\!.23,-44^o44^{\prime}07^{\prime\prime}\!.8$) (Harris
1996, 2010 edition). Some exposures of this field show the same dark
spot due to the probe as in the Baade's window (see
Fig.~\ref{fig:bulgedit}).

%%%%%%%%%%%%%%
\subsubsection{NGC~104 (47\,Tuc)} 
%%%%%%%%%%%%%%

The last globular cluster observed during the \mbox{HAWK-I}
commissioning is NGC~104 (47\,Tuc), $(\alpha,\delta)_{\rm
  J2000.0}=(\rm00^h22^m05^s\!\!.67,-72^{\circ}04^{\prime}52^{\prime\prime}\!\!.6$)
(Harris 1996, 2010 edition). Two of the 25 pointings of the $K_{\rm
  S}$-filter data were not usable.

%%%%%%%%%%%
\subsection{The Extra-Galactic astrometric fields} 
%%%%%%%%%%%

Extra-galactic fields are more stable than Galactic fields, since
their internal proper motions are negligible compared to foreground
stars, even with a 10-yr time baseline. The downside of such
extra-galactic fields is the need to increase the integration time to
compensate for the faintness of the targets. \looseness=-2

%%%%%%%%%%%%%%
\subsubsection{NGC~6822} 
%%%%%%%%%%%%%%

The first extra-galactic field is centered on the Local Group dwarf
irregular galaxy NGC~6822 at a distance of $\sim$500 kpc (Madore et
al. 2009). \looseness=-2

For this galaxy, we took fewer pointings (9 in a 3$\times$3 array) but
with a longer integration time. Adopting the same numbers as in
Fig.~\ref{fig:dit}, we only used dithers labeled 1, 3, 5, 11, 13, 15,
21, 23, and 25. The integration time was 120\,s with NDIT=12 and
DIT=10\,s. \looseness=-2

%%%%%%%%%%%%%%
\subsubsection{An astrometric field for JWST in LMC} 
%%%%%%%%%%%%%%

In 2005, a field near the center of the Large Magellanic Cloud (LMC)
was selected as reference field to solve for the geometric distortion
and to eventually help calibrate the relative positions of JWST's
instruments in the focal plane. This field is in the JWST continuous
viewing zone and it can be observed whenever necessary. In 2006, it
was observed with the Advanced Camera for Surveys (ACS) Wide Field
Channel (WFC) to create a reference catalog in F606W. \looseness=-2

The JWST calibration field is centered at $(\alpha,\delta)_{\rm
  J2000.0}=(\rm
5^h21^m55^s\!\!.87,-69^o29^{\prime}47^{\prime\prime}\!\!.05)$.  The
distortion-corrected star catalog we provide in this paper can be used
as distortion-free frame to compute the geometric distortion of the
JWST's detectors when the time comes. We adopted the same observing
strategy as above with 25 images organized in a 5$\times$5
array. Unfortunately, one of the pointings in the $K_{\rm S}$-filter
data set was not usable. \looseness=-2

%%%%%%%%
\section{PSF-modeling, fluxes and positioning}
%%%%%%%%
\label{psf}

In our reductions, we used the custom-made software tools. It is
essentially the same software used in the previous papers of this
series. We started from a raw multi-extension FITS image. Each
multi-extension FITS image stores all four chips in a datacube. We
kept this FITS format up to the sky-subtraction phase. \looseness=-2

First, we performed a standard flat-field correction\footnote{The
  correction was performed using a single master flat field for each
  of the three filters. We did not use a flat field tailored to each
  epoch because some of them were not collected.} on all the
images. In the master flat fields, we built a bad-pixel mask by
flagging all the outliers respect to the average counts. We used the
bad-pixel-mask table to flag warm/cold/dead pixels in each
exposure. Cosmic rays were corrected by taking the average value of
the surrounding pixels if they were not inside a star's
region\footnote{Cosmic rays close to the star's center increase the
  apparent flux and shift the center of the star, resulting in a large
  {\sf QFIT} value (see Appendix~\ref{appendixA} for detail).}; bad
columns were replaced by the average between the previous and
following columns. \looseness=-2

Digital saturation in our images starts at $32\,768$ counts. To be
safe, we adopted a saturation limit of $30\,000$ counts to minimize
deviations from linearity close to the saturation regime (accordingly
to Kissler-Patig et al. 2008) and flat-field effects. Each pixel for
which the counts exceed the saturation limit was flagged and not
used. \looseness=-2

Finally, we subtracted the sky from the images, computing the median
sky value in a 10$\times$10 grid and then subtracting the sky
according to the table (bi-linear interpolation was used to compute
the sky value in a given location). After the sky subtraction, we
split each multi-extension FITS file in four different FITS files, one
per chip. The next step was to compute the PSF models. \looseness=-2

HAWK-I's PSF is always well sampled, even in the best-seeing
condition. To compute PSF models, we developed the software {\sf
  img2psf\_HAWKI} in which our PSF models are completely
empirical. This is derived from the WFI@2.2m reduction package
(Paper~I). They are represented by an array of 201$\times$201 grid
points, which super-sample PSF pixels by a factor of 4 with respect to
the image pixels. The fraction of flux contained in the central pixel
of a star is given by the central PSF pixel. A bi-cubic spline is used
to interpolate the value of the PSF in between the grid points. The
value of a given pixel $P_{i,j}$ in the vicinity of a star of total
flux $z_{\ast}$ that is located at position $(x_{\ast},y_{\ast})$ is:
\looseness=-2

\begin{displaymath}
  P_{i,j}=z_{\ast}\cdot \psi(i-x_{\ast},i-y_{\ast})+s_{\ast}
  \phantom{1} ,
\end{displaymath}

where $\psi(\Delta x,\Delta y)$ is the instrumental PSF, or
specifically, the fraction of light (per unit pixel area) that falls
on the detector at a point offset $(\Delta x,\Delta
y)=(i-x_{\ast},j-y_{\ast})$ from the star's center, and $s_{\ast}$ is
the local sky background value. For each star, we have an array of
pixels that we can fit to solve for the triplet of parameters:
$x_{\ast}$, $y_{\ast}$, and $z_{\ast}$. The local sky $s_{\ast}$ is
calculated as the 2.5$\sigma$-clipped median of the counts in the
annulus between 16 and 20 pixels from the location where the star's
center falls. The previous equation can be inverted (with an estimate
of the position and flux for a star) to solve for the PSF:
\looseness=-2

\begin{displaymath}
  \psi(\Delta x,\Delta y)=\frac{P_{i,j}-s_{\ast}}{z_{\ast}} \phantom{1} .
\end{displaymath}

This equation uses each pixel in a star's image to provide an estimate
of the 2-dimensional PSF at the location of that pixel, $(\Delta
x,\Delta y)$. By combining the array of sampling from many stars, we
can construct a reliable PSF model. As opposed to the pioneering work
of Stetson and his DAOPHOT code (Stetson 1987) that combines an
empiric and semi-analytic PSF model, we created a fully-empirical PSF
model, as described in Paper~I. \looseness=-2

The software {\sf img2psf\_HAWKI} iterates to improve both the PSF
model and stellar parameters. The starting point is given by simple
centroid positions and aperture-based fluxes. A description of the
software is given in detail in Paper~I. \looseness=-2

To model the PSFs in both the core and the wings, we use only stars
with a high S/N (signal-to-noise ratio). This is done by creating a list
of stars that have a flux of at least 5000 counts above the local sky
(i.e., S/N$>$60-70 in the central pixel) and also have no brighter
neighbors within 15 pixels.  We need at least 50 such stars for each
PSF model, so that we can iteratively reject stars that may be
compromised by nearby neighbors, cosmic rays, or detector defects
(e.g., bleeding columns). \looseness=-2

Determining both a good model for the PSF and determining stellar
positions and fluxes requires an iterative solution, so the software
iterates this process several times until convergence is reached with
both good PSF models and stellar parameters that fit well. The result
is a 5$\times$5 grid of PSF models for each chip, which are
bi-linearly interpolated to provide a model PSF at any pixel
position. \looseness=-2

With an array of PSF models, we are able to measure stars' positions
and fluxes for all the stars in the image by using a software analog
to that described in Paper~I. As an input, we need to give the
faintest level above the sky for a star to be found and determine how
close this star can be to brighter neighbors. The program finds and
measures all stars that fit these criteria. The final catalogs (one
for each chip) contain positions, instrumental magnitudes, and another
quantity called quality-of-PSF-fit ({\sf QFIT}, which represents the
fractional error in the PSF-model fit to the star). For each pixel of
a star within the fitting radius (2.5 pixels), the {\sf QFIT} is
defined as the sum of the absolute value of the difference between the
pixel values $P_{i,j}$ (sky subtracted) and what the local PSF model
predicts at that location $\psi(i-x_{\ast},j-y_{\ast})$, normalized
with respect to the sky-subtracted $P_{i,j}$: \looseness=-2

\begin{displaymath} 
{\sf QFIT}=\sum_{i,j}\left|\frac{({\rm P}_{i,j}-{\rm
    sky})-z_{\ast}\cdot\psi(i-x_{\ast},j-y_{\ast})}{{\rm P}_{i,j}-{\rm
    sky}}\right|,
\end{displaymath}

\noindent where $(x_{\ast},y_{\ast})$ is the star's center. The {\sf
  QFIT} is close to zero for well-measured stars and close to unity
for ones that are badly-measured (or not star-like). Typically we
found ${\sf QFIT}\lesssim 0.05$ for well-measured stars in our
images. Saturated stars are also measured in our pipeline. For these,
stars we only fitted the PSF on the wings of the stars using
unsaturated pixels. In this way, we are able to measure a flux and a
position for saturated stars, even if they are less accurate (high
{\sf QFIT}) than for unsaturated stars. \looseness=-2

%%%%%%%%
\section{Geometric distortion correction}
%%%%%%%%
\label{GDC}

\begin{figure*}[t]
  \centering
  \includegraphics[width=\columnwidth]{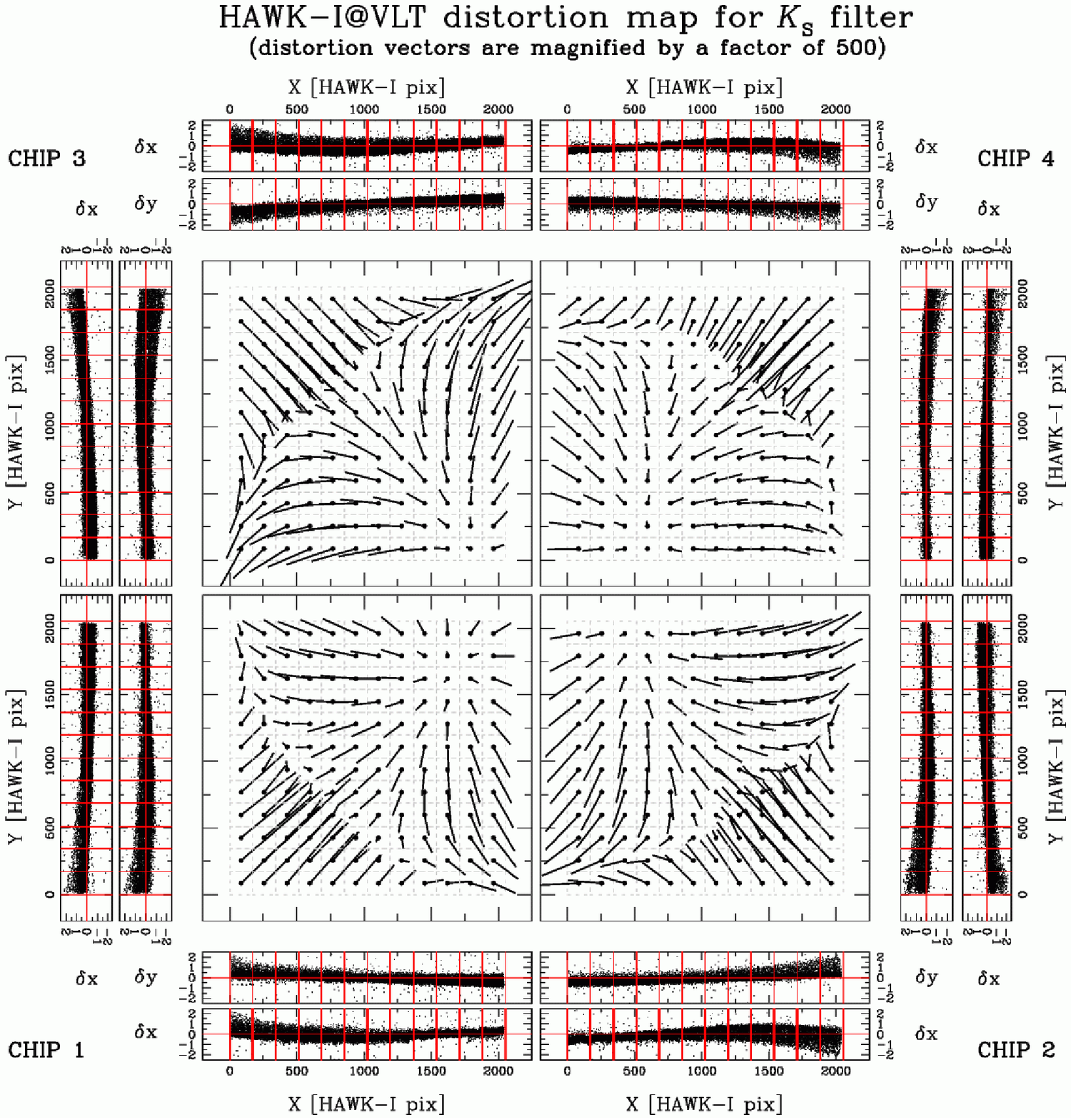}
  \includegraphics[width=\columnwidth]{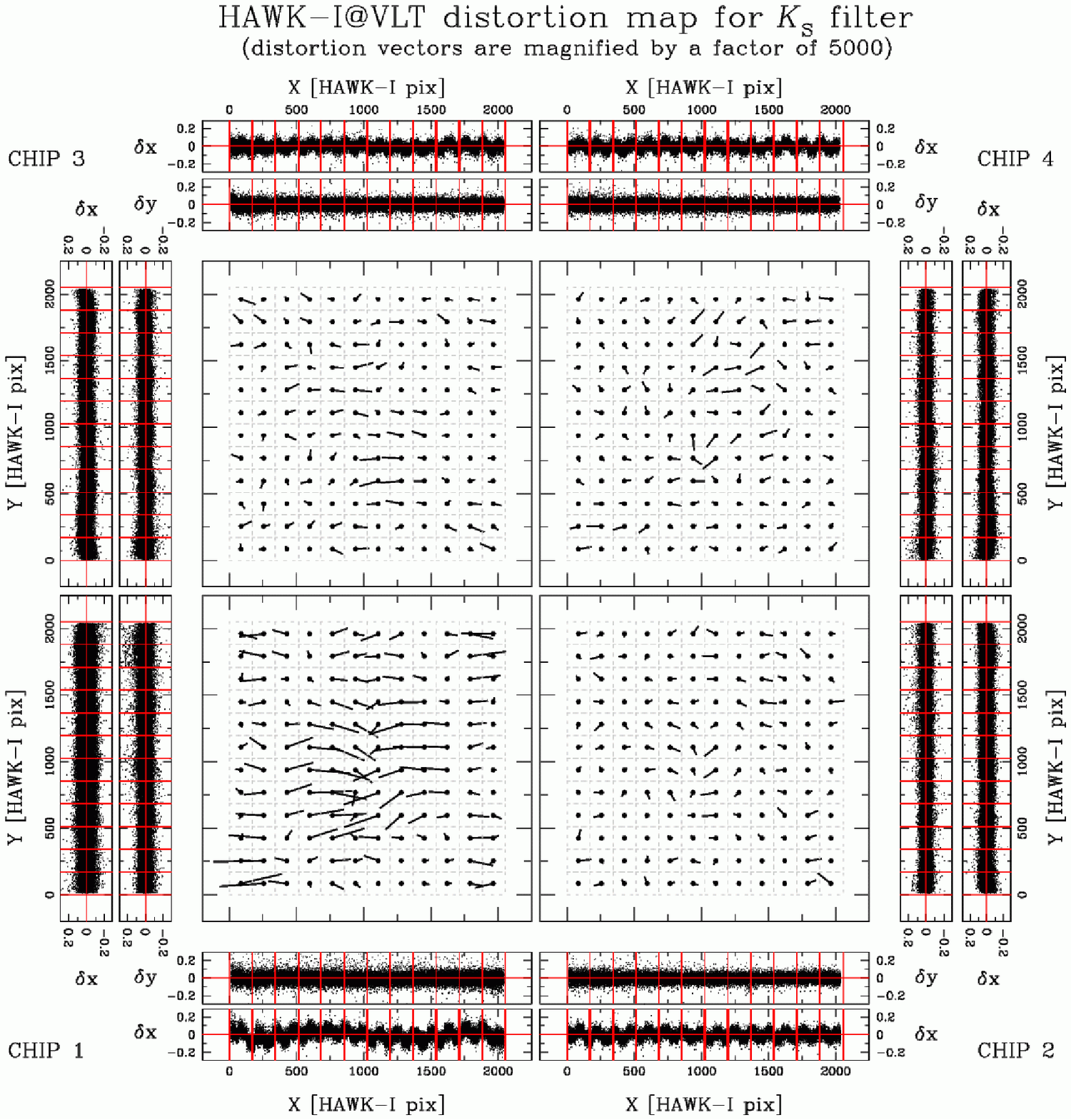}
  \caption{(\textit{Left}): Residual trends for the four chips when we
    use uncorrected stars' positions. The size of the residual vectors
    is magnified by a factor of 500. For each chip, we also plot the
    single residual trends along X and Y axes. Units are expressed in
    HAWK-I raw pixel. (\textit{Right}): Residuals after our polynomial
    correction is applied. The size of the residual vectors is now
    magnified by a factor 5000.}
  \label{fig:gdcor}
\end{figure*}

In this section, we present a geometric-distortion solution for the
HAWK-I in three broad band filters ($J$, $H$, and $K_{\rm S}$) derived
using exposures of the Baade's Window field. No astrometric reference
data is available for the Baade's Window field, so we iteratively
constructed our own. \looseness=-2

Adopting the observing strategy described in Sect.~\ref{OBS}, the
systematic errors in the measure of stars' positions from one exposure
to the other have a random amplitude and the stars' averaged positions
provide a better approximation of their true positions in the
distortion-free master frame. \looseness=-2

To build the master frame, we cross-identified the star catalogs from
each individual HAWK-I chip. Conformal transformations (four-parameter
linear transformations, which include rigid shifts in the two
coordinates, one rotation, and one change of scale, so the shape is
preserved) were used to bring the stars' positions, as measured in
each image, into the reference system of the master frame. We
considered only well-measured, unsaturated objects with a stellar
profile and measured in at least three different images. \looseness=-2

Our geometric-distortion solution for HAWK-I is made up of five parts:
(1) a linear transformation to put the four chips into a convenient
master frame (Hereafter, we refer to the transformation from chip $k$
of the coordinate system of image $j$ to the master system $T_{\it
  j,k}$.), (2) two fifth-order polynomials to deal with the general
optical distortion (hereafter, the ``P'' correction), (3) an analytic
correction for a periodic feature along the x-axis, as related to the
detector read-out amplifiers (the ``S'' correction), (4) a fine-tuning
to correct second-order effects on the x-residuals of the S correction
(the ``FS'' correction), and (5) a table of residuals that accounts
for both chip-related anomalies and a fine-structure introduced by the
filter (the ``TP'' correction). \looseness=-2

The final correction is better than $\sim$0.027 pixel ($\sim$2.8 mas)
in each coordinate.  We provide the solution in two different forms: a
FORTRAN subroutine and a set of FITS files for each
filter/chip/coordinate. Since focus, flexures, and general conditions
of the optics and telescope instrumentation change during the
observations (within the same night and even between consecutive
exposures), we derive an {\it{average}} distortion
correction. \looseness=-2

Here, we describe our correction procedure for filter $K_{\rm S}$. The
procedure for filters $J$ and $H$ is identical, and the results are
presented in the Sect.~\ref{filt}. \looseness=-2

\begin{figure*}[!t]
  \centering
  \includegraphics[scale=.78]{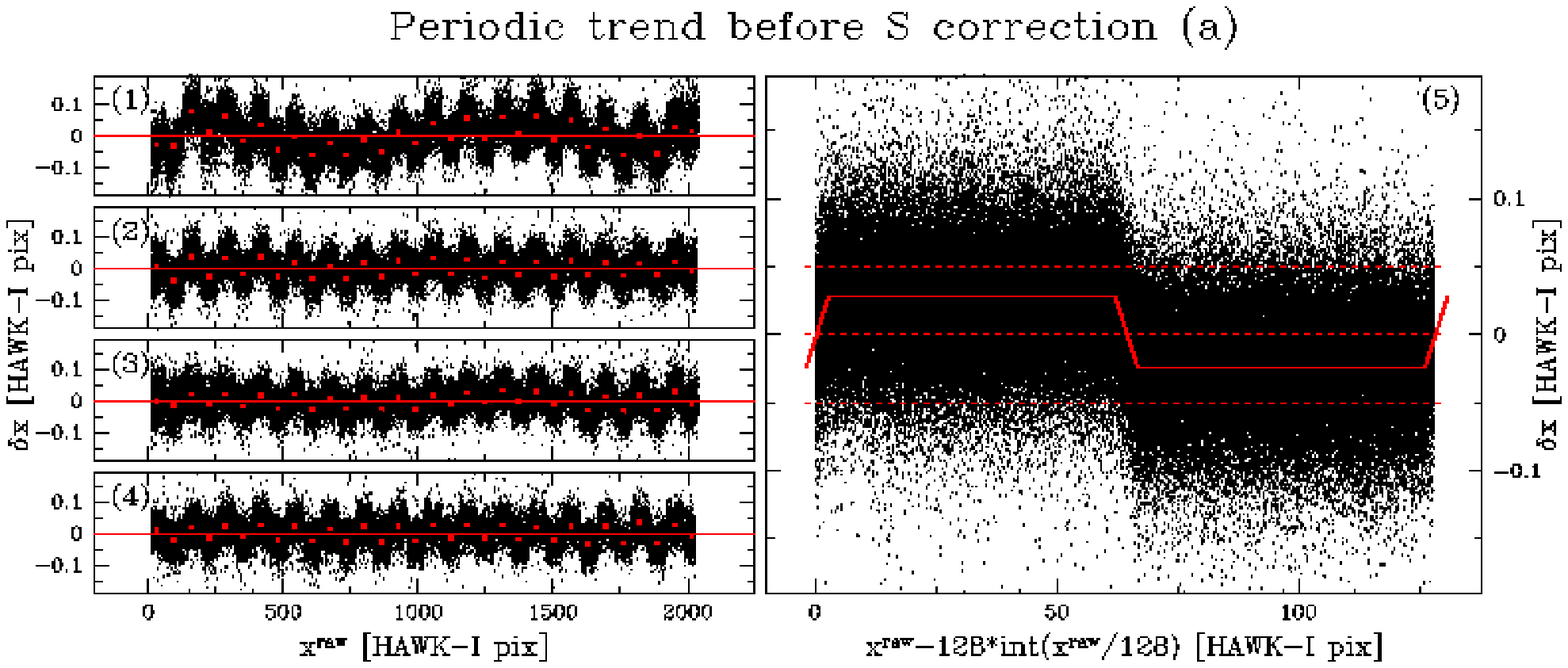}
  \includegraphics[scale=.78]{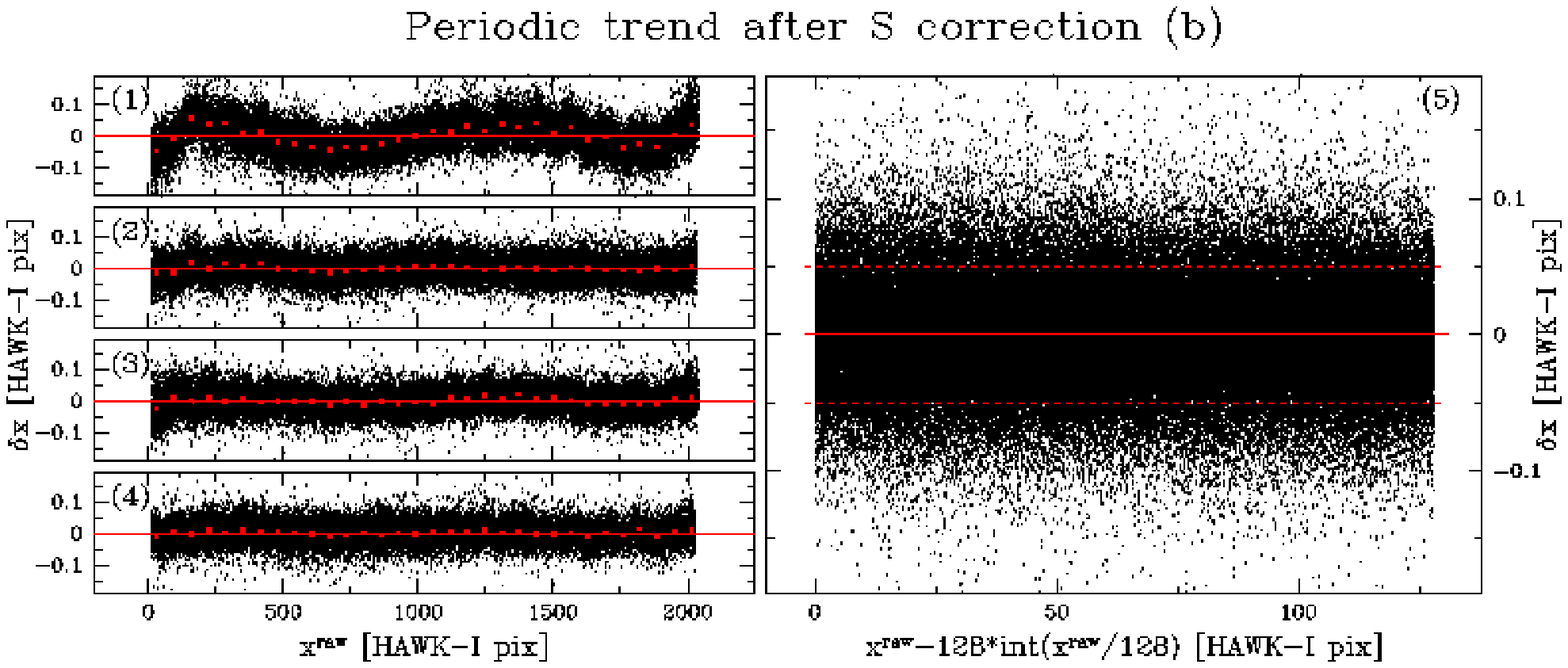}
  \includegraphics[scale=.78]{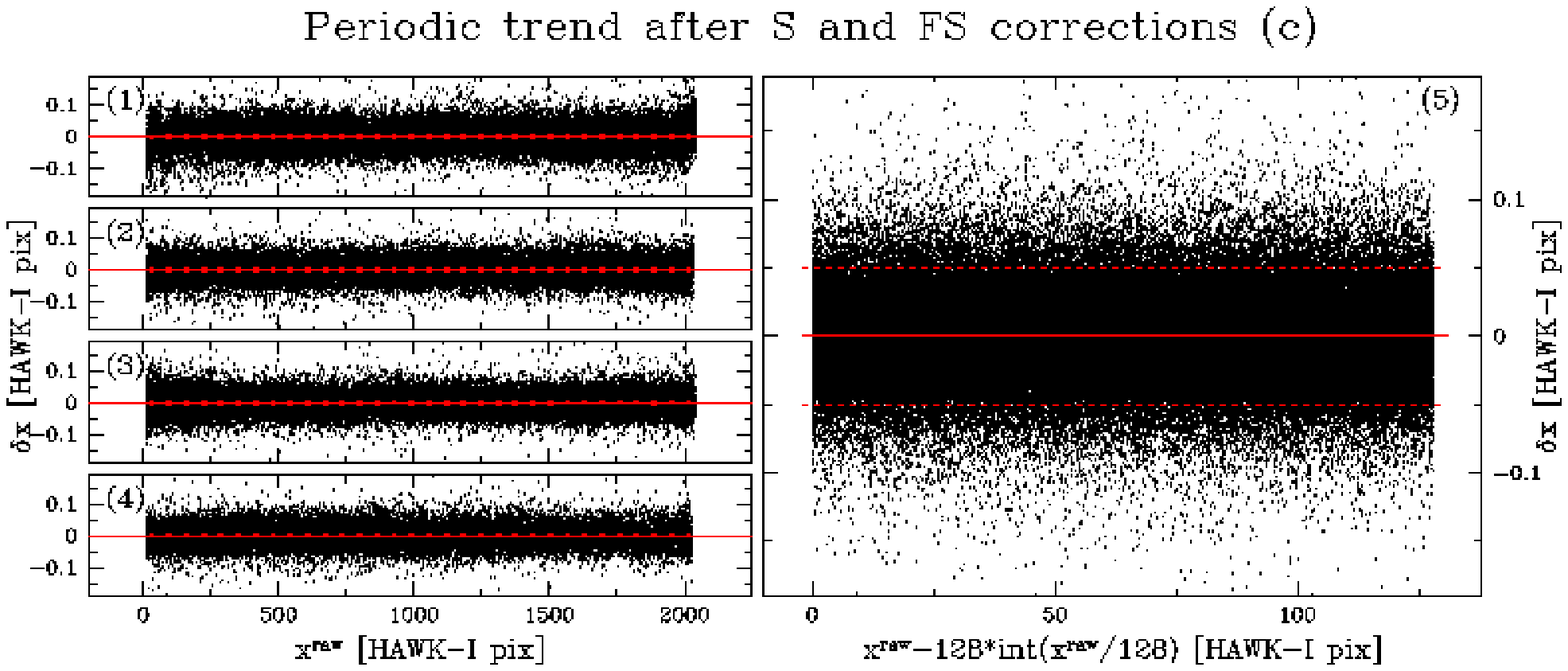}
  \caption{(\textit{Top}): $\delta x$ as a function of X in units of
    HAWK-I pixels before S correction. (\textit{Middle}): As above but
    after S correction. (\textit{Bottom}): Same as above but after S
    and FS corrections.  In the left panels (from 1 to 4), we took 32
    bins of 64 pixels each and computed the median of residuals in
    each bin (red squares). In the right panels (5), we show the
    periodogram with a period of 128 columns containing all the points
    plotted in the left panels. The red dashed lines show $+0.05$,
    $0$, and $-0.05$ HAWK-I pixel.}
  \label{fig:stepcor}
\end{figure*}

%%%%%%%%%%%
\subsection{Polynomial correction (P)}
%%%%%%%%%%%
\label{poly}

We followed the method given in Anderson \& King (2003) for the
Wide-Field Planetary Camera 2 (WFPC2). This method was subsequently
used to derive the distortion correction for the ACS High Resolution
Channel (Anderson \& King 2004) and for the Wide Field Camera 3 (WFC3)
Ultraviolet-Visual (UVIS) channel (Bellini \& Bedin 2009; Bellini,
Anderson \& Bedin 2011). The same strategy was also used by two of us
to calibrate the blue prime-focus camera at the LBT (Paper~IV). We
treated each chip independently, and we solved the $5^{\rm th}$-order
polynomial that provided the most correction. We chose pixel
(1024,1024) in each chip as a reference position and solved for the
distortion with respect to it. \looseness=-2

The polynomial correction is performed as follows:
\begin{itemize}
\item In each of the list of unsaturated stars found in each chip of
  each exposure (4 $\times$ 25 lists), we first selected stars with an
  instrumental magnitude brighter than $K_{\rm S} \simeq -11$ and with
  a {\sf QFIT} lower than 0.05, to ensure that the master list would
  be free from poorly measured stars, which would harm the distortion
  solution.
\item We computed the linear transformation ($T_{\it j,k}$) between
  stars in each chip of each exposure and the current master frame.
\item Each star in the master frame was conformally transformed in the
  raw-coordinate system of each chip/image ($T^{-1}_{\it j,k}$) and
  cross-identified with the closest source. Each such
  cross-identification generates a pair of positional residuals
  ($\delta x$,$\delta y$), which correspond to the difference between
  the observed position and the transformed reference-frame position.
\item These positional residuals were distilled into a look-up table
  made up of 12$\times$12 elements of 170.7$\times$170.7 pixels
  each. This setup proved to be the best compromise between the need
  of an adequate number of grid points to model the polynomial part of
  the distortion solution and an adequate sampling of each grid
  element. We found about $19\,000$ pairs of residuals in each chip
  with a median number of 135 pairs per cell (the number varied
  between 30, which occurred in a corner grid element, and 170, which
  was near the chips' center).
\item We performed a linear least-square fit of the average positional
  residual of each of the 144 cells to obtain the coefficients for the
  two fifth-order polynomials in each chip (see Paper IV for a
  detailed description).
\item We applied this P correction to all stars' positions.
\item Finally, we iterated the entire process, deriving a new and
  improved combination of a master frame and distortion solution. The
  residuals improved with each iteration.
\end{itemize}
The iterative process was halted when the polynomial coefficients from
one iteration to the next differed by less than 0.01\%. \looseness=-2

The final P correction reduced the average distortion residuals (from
the center of the detector to the corner) from $\sim$2.1 pixels down
to $\sim$0.2 pixel. By applying the P correction, the accuracy of our
distortion solution (defined as $68.27^{\rm th}$ percentile of the
$\sigma$(Radial residual), see Sect.~\ref{acc} for detail) improves
from $\sim$0.336 to $\sim$0.043 pixel per coordinate for a
well-exposed star, which translates from $\sim$35.6 mas to $\sim$4.5
mas. In Fig.~\ref{fig:gdcor}, we show the HAWK-I distortion map before
and after our P correction. Although the aim of our work is not to
analyze what makes the distortion happen, the distortion pattern
before the correction appears to be primarily a radial distortion in
the focal plane with some vignetting at the edges plus some
shift-rotation-shear in the detector positioning.  \looseness=-2

%%%%%%%%%%%
\subsection{Average periodic ``step'' correction (S)}
%%%%%%%%%%%
\label{step}

Our P correction reveals a high-frequency, smaller-amplitude effect,
which is a periodic pattern in the x-positional residuals as a
function of the x positions in all HAWK-I chips. This effect is
clearly shown in the distortion map after the P correction is applied
($\delta x$ vs. $X$ panels in Fig.~\ref{fig:gdcor}). For every 128
columns, stars' positions have positive residuals (about 0.075 HAWK-I
pixel) in the first 64 pixels and negative residuals (about 0.045
HAWK-I pixel) in the second 64 pixels (see panels (a) of
Fig.~\ref{fig:stepcor}). At first glance, this residual pattern has
the appearance of being caused by irregularities in the pixel grid of
the detectors. However, a detailed analysis (see Sect.~\ref{expl})
leads us to conclude that it is instead a pattern caused by a
``periodic lag'' in the readout process, which is offset in opposite
directions in alternating 64 pixel sections of the detector addressed
by each of the 32 read-out amplifiers.  \looseness=-2

We adopted an iterative procedure to empirically correct for this
periodic pattern that shows up only along the x axis. We started with
the master frame made by using catalogs corrected with our P
correction. We then transformed the position of each star ($i$) from
the master frame back into the raw coordinate system of each chip
($k$) of each image ($j$). We determined the quantity: \looseness=-2

\begin{displaymath}
  \centering
  \delta x_{i} = x_{i}^{\rm raw} - x_{i,j}^{\rm P^{-1}(T^{-1}_{\it j,k})},
\end{displaymath}

where $x_{i}^{\rm raw}$ are the raw x-coordinates, and $x_{i,j}^{\rm
  P^{-1}(T^{-1}_{\it j,k})}$ are the x-coordinates on the master frame
transformed to the raw coordinate system and corrected with the
inverse P correction. We assumed that the periodic trend had a
constant amplitude across the detector. Panels (a) in
Fig~\ref{fig:stepcor} show $\delta x$ vs $x^{\rm raw}$ for each chip
(from 1 to 4) and $\delta x$ vs. $x^{\rm raw}$ modulus 128, in which
we collect together all the residuals (panel 5) before applying the S
correction. \looseness=-2

To model the trend in the residuals, we used a square-wave function
(panel (5a) of Fig.~\ref{fig:stepcor}). The amplitude of this function
is defined as the 3$\sigma$-clipped median value of the residuals
between pixels 2.8--62.2 and 66.8--126.2. To model the average
periodicity between 62.2$\le x^{\rm raw} \le$ 66.8 pixels, we fitted
the data points with a straight line using by linear least
squares. \looseness=-2

We corrected the stars' positions by applying 75\% of the S correction
(to encourage smooth convergence) and the P correction. We computed an
improved master frame and calculated new, generally smaller,
residuals. New square-wave-function amplitudes were derived and added
to the previous corrections to improve the S correction. The procedure
was iterated until the observed average periodicity residuals had an
amplitude smaller than $10^{-4}$ pixel. \looseness=-2

Combining the two corrections (S$+$P, applied in this order to the raw
coordinates), we are able to reduce the $68.27^{\rm th}$ percentile of
the $\sigma$(Radial residual) down to $\sim$$4.0$ mas (0.038
pixel). \looseness=-2

%%%%%%%%%%%
\subsection{Fine-tuned correction of the residual periodicity (FS)}
%%%%%%%%%%%
\label{fine}

In panels (b) of Fig.~\ref{fig:stepcor}, we show the residual trend
after the S$+$P correction is applied. Looking at panels (1--4b), it
is obvious that the amplitude of the $\delta$x periodicity pattern is
not constant from chip to chip. In addition to this, there is still a
polynomial residual that needs to be removed. For this reason, we
applied a fine-tuned residual correction as follows. \looseness=-2

\begin{figure}[t]
  \centering
  \includegraphics[width=9.cm]{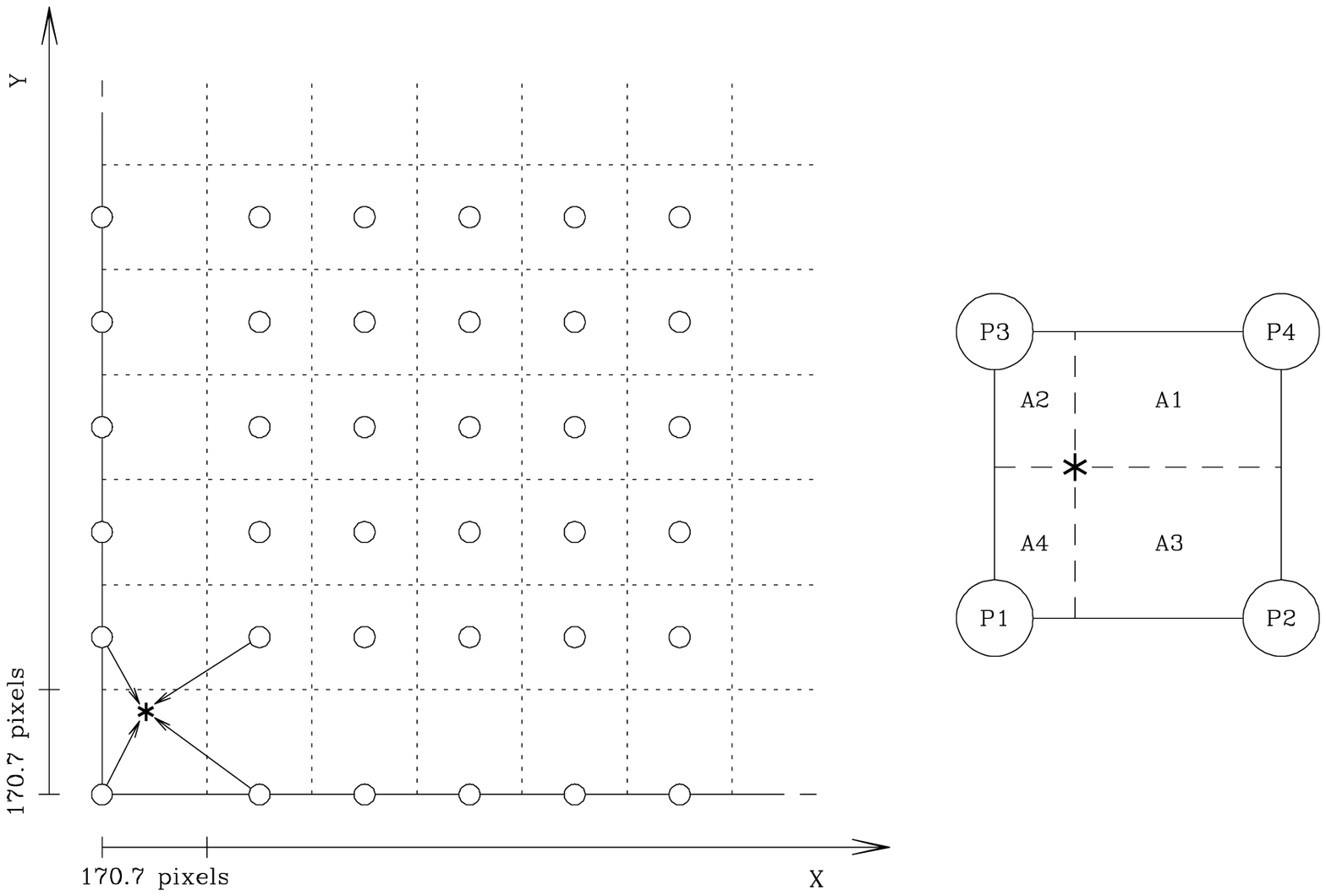}
  \caption{(\textit{Left}): Example of cell and grid-point locations
    on the bottom-left area of chip[1]. Dotted lines mark the
    170.7$\times$170.7 pixels square elements inside which we computed
    the median value of the distortion residual to use as grid point
    (empty circles) in the look-up table. For a given star (marked
    with an $*$), we used the four surrounding closest grid point to
    perform the bi-linear interpolation (sketched with the arrows) and
    evaluate the residual geometric distortion in that location of the
    detector. (\textit{Right}): Bi-linear interpolation outline. Each
    grid point $\rm P_1,\dots,P_4$ is weighted by the corresponding
    area $\rm A_1,\dots,A_4$ to associate the correction in $*$.}
  \label{fig:griglia}
\end{figure}

We first computed a master frame by applying the S$+$P correction to
the raw positions of each chip/exposure. We then determined the
residuals as the difference between the raw x-coordinates corrected
with the S correction and $x_{i,j}^{\rm P^{-1}(T^{-1}_{\it
    j,k})}$. Next, we divided each chip into 32 bins of 64 pixels each
along the x axis, computed the 3$\sigma$-clipped average value of the
residuals, and subtracted the 75\% of it from the $\delta x$ residuals
in each bin. Then, we iterated the procedure until the difference
between the 3$\sigma$-clipped average value of the residuals in all
bins of all chips from one iteration to the next one was smaller than
$10^{-3}$ pixel. In panels (c-1) to (c-4) of Fig.~\ref{fig:stepcor},
we show the residual trends for all chips after our S$+$FS$+$P
corrections are applied. \looseness=-2

This approach was able to provide accuracies ($68.27^{\rm th}$
percentile of the $\sigma$(Radial residual)) down to 0.035 pixel
($\sim$3.7 mas) level. \looseness=-2

%%%%%%%%%%%
\subsection{Table of residual geometric-distortion correction (TP)}
%%%%%%%%%%%
\label{interp}

The final step of our distortion-solution model consists of four
look-up tables (one for each chip) to minimize all the remaining
detectable systematic residuals that were left. We constrained the
look-up tables using the same procedure that Bellini, Anderson \&
Bedin (2011) used to derive the distortion correction for the
WFC3/UVIS camera. \looseness=-2

First, we corrected all stars' positions by applying the S, FS, and P
corrections (in this order).  We then built a new master frame and
computed the residuals, as described in Sect.~\ref{poly}. We
subdivided again each chip into 12$\times$12 square elements.  We used
the stars' residuals within each cell to compute a 3$\sigma$-clipped
median positional residuals and assigned these values to the
corresponding grid points (open circles in
Fig~\ref{fig:griglia}). When a cell adjoins detector edges, the grid
point is displaced to the edge of the cell, as shown. For the grid
point on the edges, the value of the median only at the first
iteration is computed at the center and shifted to the edge. Then, we
iteratively found the value that the grid-point element on the edge
should have to remove the systematic errors. We built a look-up table
correction for any given location of the chip, using a bi-linear
interpolation among the surrounding four grid
points. Figure~\ref{fig:griglia} shows an example of the geometry
adopted for the look-up table and of the bi-linear
interpolation. \looseness=-2

We corrected stars' positions using only 75\% of the recommended
grid-point values, computed an improved master frame, and calculated
new (generally smaller) residuals. We calculated new grid-point values
and added them to the previous values. The procedure was iterated
until the bi-linear interpolation offered negligible improvement of
the positional residuals r.m.s. from one iteration to the
next. \looseness=-2

\begin{figure*}[!t]
  \centering
  \includegraphics[width=\textwidth]{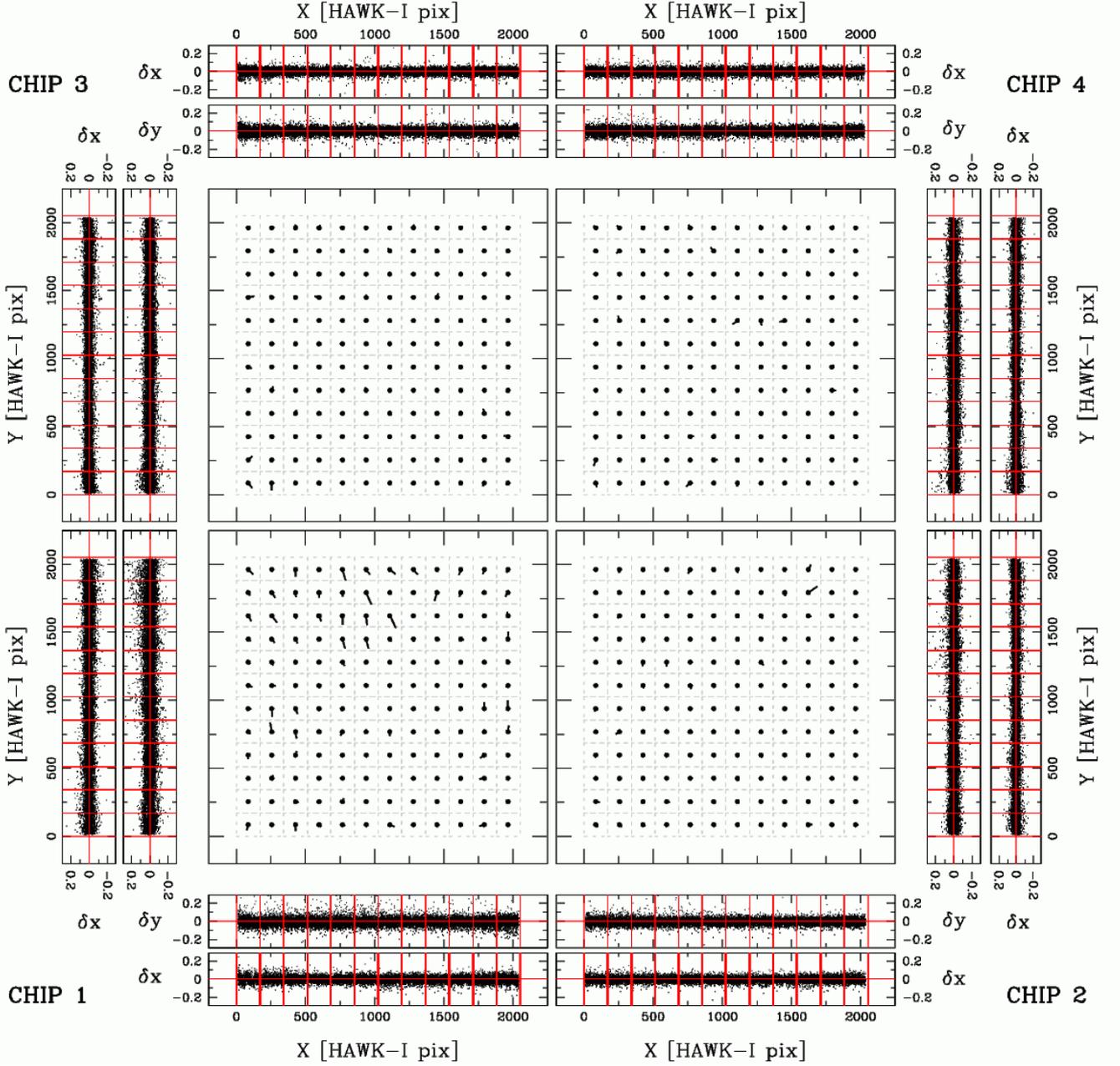}
  \caption{As Fig.~\ref{fig:gdcor} but after we have applied all the
    distortion corrections. The size of the residual vectors are now
    magnified by a factor $10\,000$.}
  \label{fig:allcor}
\end{figure*}

%%%%%%%%%%%
\subsection{Accuracy of the geometric-distortion correction}
%%%%%%%%%%%
\label{acc}

In Fig.~\ref{fig:allcor}, we show the final HAWK-I distortion map
after we applied our full distortion solution (S$+$FS$+$TP$+$P). To
have a reliable assessment of the errors in the distortion correction,
we computed the r.m.s. of the position residuals of each star
(\textit{i}) observed in each chip (\textit{k}) of the image
(\textit{j}), which have been distortion corrected and conformally
transformed into the master-frame reference system
($x_{i,j,k}^{T_{j,k}},y_{i,j,k}^{T_{j,k}}$). The difference between
these positions and the distortion-free positions ($X_{i}^{\rm
  master},Y_{i}^{\rm master}$) directly quantifies how close we are to
reach the ideal distortion-free system. We defined the $\sigma$(Radial
residual) as: \looseness=-2

\begin{displaymath}
  \sigma(\textrm{Radial residual})_{i,j} = \sqrt{\frac{ (x_{i,j,k}^{T_{j,k}}
    - X_{i}^{\rm master})^2 + (y_{i,j,k}^{T_{j,k}} - Y_{i}^{\rm
      master})^2 }{ 2 }}\phantom{1} .
\end{displaymath}

\begin{figure}[t]
  \centering \includegraphics[width=\columnwidth]{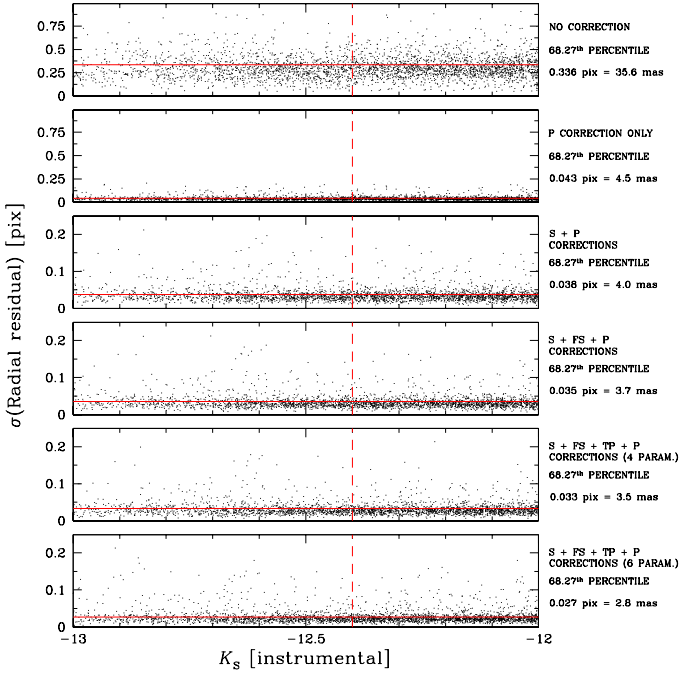}
  \caption{$\sigma$(Radial residual) versus instrumental $K_{\rm S}$
    magnitude after each step of our solution. The red solid
    horizontal line shows the $\sigma_{\rm perc}$; the red dashed
    vertical line indicates the magnitude cut-off $K_{\rm S} =
    -12.4$.}
  \label{fig:mdr}
\end{figure}

In Fig.~\ref{fig:mdr}, we show the size of these $\sigma$(Radial
residual) versus instrumental $K_{\rm S}$ magnitude after each step of
our solution. To test the accuracy of the geometric-distortion
solution, we only used unsaturated stars with an instrumental
magnitude $K_{\rm S}\le-12.4$ (red dashed line) in the master list,
which is observed in at least three images and with a $\sf QFIT \le$
0.05. Faint stars have larger residuals due to an increasing
contribution of random errors. The $3\sigma$-clipped $68.27^{\rm
  th}$-percentile value of these residuals is shown on the right of
each panel. The 3$\sigma$ clipping rule excludes outliers, which can
bias the percentile value. These outliers can have different
explanations. For example, most of the outliers for the Bulge field
are close to the edge of the FoV, where the distortion solution is
less constrained. In the case of NGC~6656, most of these outliers are
close to the center of the cluster (crowding effects) or are located
in the region affected by the internal reflection of the Moon in the
optics. Hereafter, we refer in the text with $\sigma_{\rm perc}$ to
the $3\sigma$-clipped $68.27^{\rm th}$-percentile value of the
$\sigma$(Radial residual). The distributions of the r.m.s. is very
non-Gaussian and the $68.27^{\rm th}$-percentile is an arbitrary
choice to represent the errors. Although it is not absolutely correct
mathematically, it gives a good indications of where an outlier will
lie.  \looseness=-2

In the bottom panel, we plot the $\sigma$(Radial residual) obtained
using more-general 6-parameter linear transformations to compute the
master-frame average positions. These transformations also include
other two terms that represent the deviation from the orthogonality
between the two axes and the change of relative scale between the two
axes (the shape is not preserved anymore). When general linear
transformations are applied, most of the residuals introduced by
variations in the telescope$+$optics system and differential
atmospheric refraction are removed, and $\sigma_{\rm perc}$ further
reduces to 0.027 pixel ($\sim$$2.8$ mas). \looseness=-2

%%%%%%%%%%%
\subsection{Geometric-distortion correction for $J$ and $H$ filters}
%%%%%%%%%%%
\label{filt}

Each HAWK-I filter constitutes a different optical element, which
could slightly change the optical path and introduce changes in the
distortion. To test the filter-dependency of our $K_{\rm S}$-based
distortion solution, we corrected the positions measured on each $J$-
and $H$-filter images of Baade's Window field with our $K_{\rm
  s}$-filter-derived distortion solution and studied the residuals. We
found $\sigma$(Radial residual) significantly larger than those
obtained for the $K_{\rm s}$-band images. We also tried to apply the
$K_{\rm s}$-filter distortion solution plus an ad-hoc table of
residual (TP correction) for each filter without significant
improvements. For these reasons, we decided to independently solve for
the distortion for the $J$ and $H$ images. \looseness=-2

We built the $J$-filter master frame using only stars with an
instrumental magnitude brighter than $J \simeq -11.5$ and with a $\sf
QFIT \le$ 0.05. For the $H$-filter, we built the master frame using
only stars with $H \simeq -12.5$ and, again, $\sf QFIT \le$ 0.05. We
adopted these selection criteria to sample each chip with an adequate
number of stars (at least 12\,000 and 19\,000 stars for the $J$- and
$H$-filter, respectively). The distortion corrections were performed
as described in the previous sections. \looseness=-2

As shown in Table~\ref{tab:obs}, the image quality of the $J$-filter
Bulge images changed dramatically during the night of the observation,
reaching 1.07 arcsec.  We initially used all $J$-filter images to
compute the distortion correction and obtained a $\sigma_{\rm perc}$
(using general transformations) of $\sim$6.8 mas. We then only
considered those exposures with an image quality better than
FWHM$=$0.80 arcsec and re-derived the distortion correction (57 out of
100 catalogs were excluded this way). \looseness=-2

\begin{figure}[t]
  \centering 
  \includegraphics[width=\columnwidth]{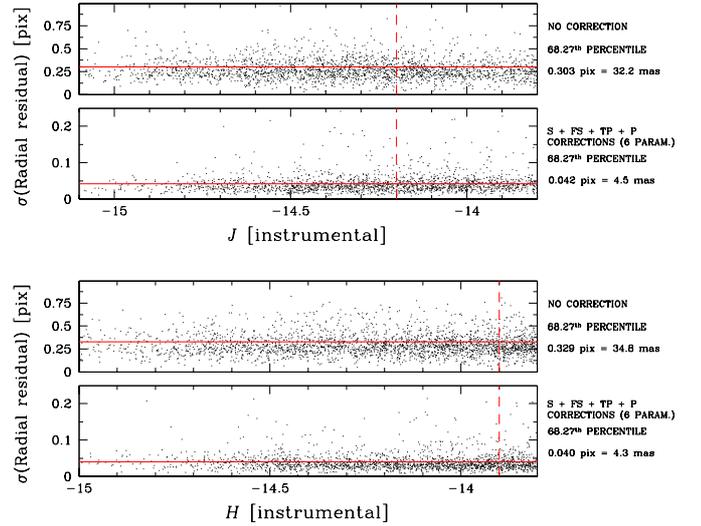}
  \caption{In each half of the figure, we show the $\sigma$(Radial
    residual) vs. instrumental magnitude before and after we applied
    our distortion correction for the $J$ (\textit{Top}) and $H$
    (\textit{Bottom}). The red lines have the same meaning as in
    Fig.~\ref{fig:mdr} but the red dashed vertical lines are set at
    $J=-14.2$ (\textit{Top}) and $H=-13.9$ (\textit{Bottom}),
    respectively.}
  \label{fig:mdr2}
\end{figure}

In Fig.~\ref{fig:mdr2}, we show that the $\sigma$(Radial residual)
before and after applying the distortion correction for the $J$ and
$H$ images. The $\sigma_{\rm perc}$ for well-measured unsaturated
stars is shown on the right of each panel. Using general linear
transformations, we obtained $\sigma$(Radial residual) of $\sim$4.5
mas and $\sim$4.3 mas for $J$ and $H$ filter,
respectively. \looseness=-2

\begin{figure*}[t]
  \centering
  \includegraphics[width=0.95\columnwidth]{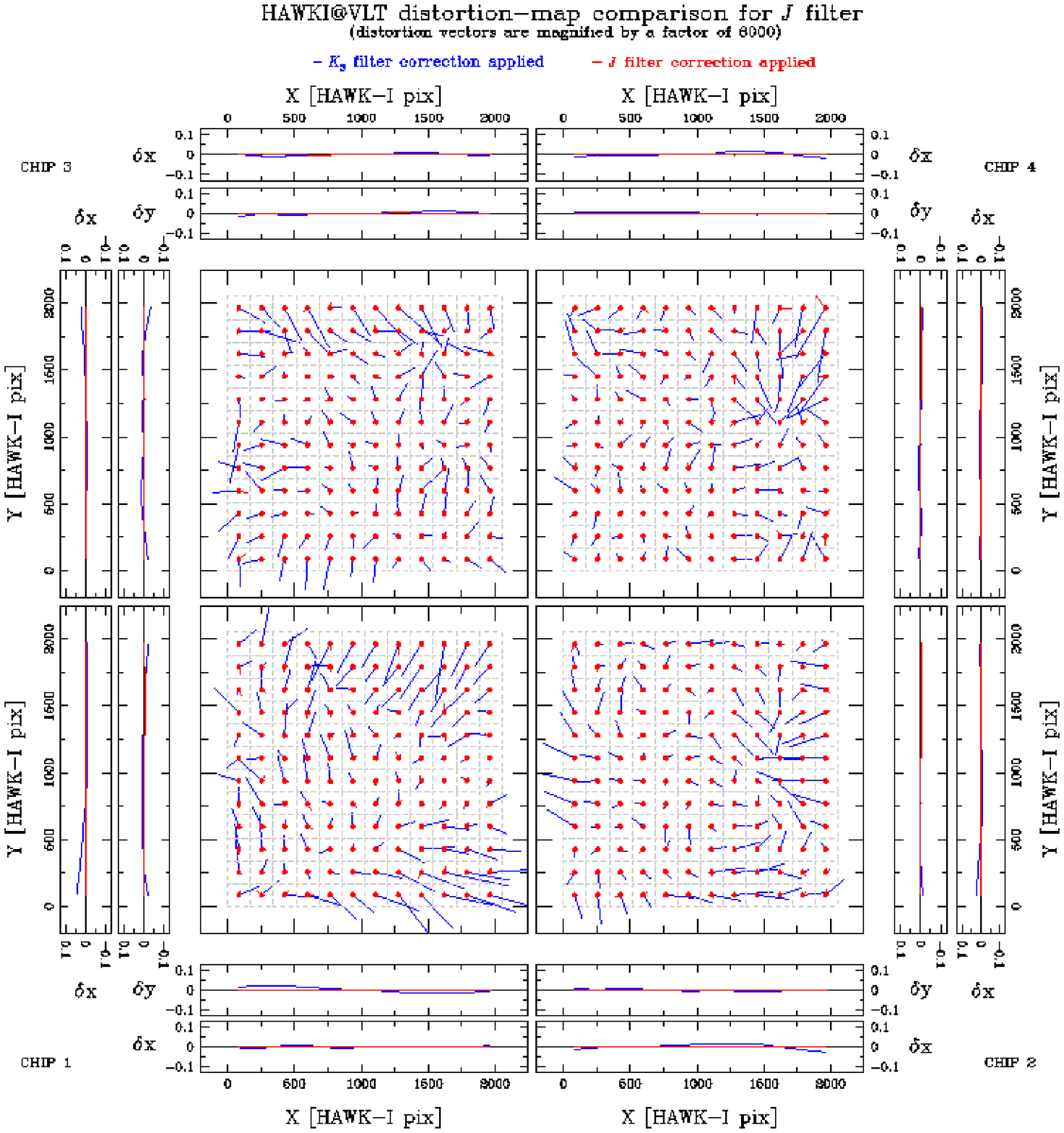}
  \includegraphics[width=0.95\columnwidth]{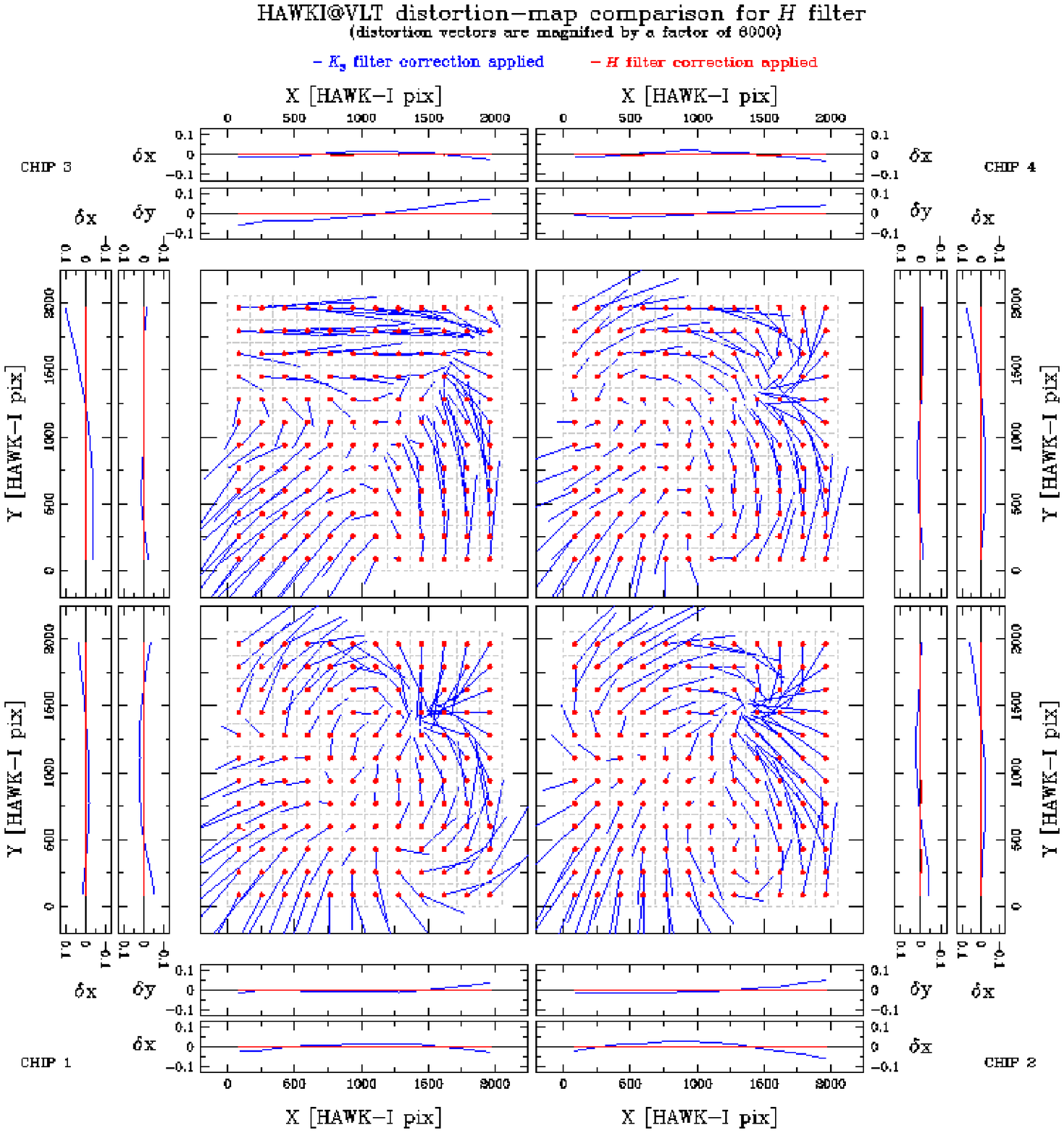}
  \caption{(\textit{Left}): $J$-filter distortion map comparison. In
    blue, we plot the vectors when the $K_{\rm S}$-filter correction
    is applied; in red, we show when the $J$-filter correction is
    used. We plot the single residual trends along X and Y axes with
    the same colors, but we do not plot the single stars to not create
    confusion in the plot.  (\textit{Right}): The same but for the $H$
    filter. The size of the residual vectors is magnified by a factor
    6000.}
  \label{fig:filtcomp}
\end{figure*}

On the left panel in Fig.~\ref{fig:filtcomp}, we compare the residual
trends obtained by applying the $K_{\rm S}$-filter correction to
$J$-filter Bulge images (blue vectors) to the residual trends obtained
by applying the $J$-filter correction instead (red vectors). In the
right panel, we show the same comparison for the $H$-filter case. A
clear residual trend (up to 0.1 pixel) of is present when a correction
made for a different filter is applied to a given data set of
images. The optical system performances are different at different
wavelengths, so it is not surprising that the $K_{\rm S}$ solution is
not completely suitable for the $J$- and $H$-filter data. The filter
also introduces an additional optical element that leads to a
different distortion on the focal plane. Both $\sigma$(Radial
residual) and distortion maps tell us that an auto-calibration for the
distortion correction in each filter is required for high-precision
astrometry.   \looseness=-2

%%%%%%%%%
\subsection{Stability of the correction}
%%%%%%%%%
\label{stab}

Different factors (e.g., the contribution given by light-path
deviations caused by filters, alignment errors of the detector on the
focal plane) change the HAWK-I distortion over time. To explore the
stability of our derived distortion solution over time, we observed
the astrometric precision obtained by applying our distortion
correction to images taken several months apart. \looseness=-2

\begin{figure}
  \centering
  \includegraphics[width=9.cm]{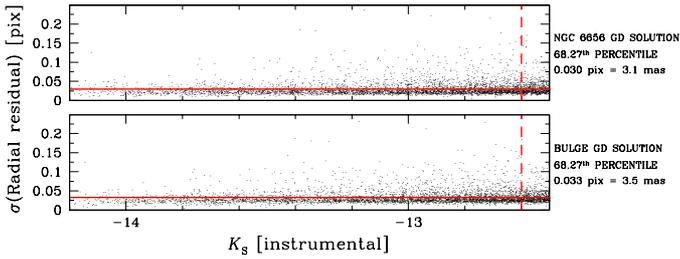}
  \caption{Comparison of NGC~6656 $\sigma$(Radial residual) after
    application of the Bulge-based (\textit{Bottom}) and the
    NGC~6656-based (\textit{Top}) correction. The red solid horizontal
    line shows the 3$\sigma$-clipped value of the $\sigma$(Radial
    residual); the red dashed vertical line indicates the magnitude
    limit $K_{\rm S} = -12.6$.}
  \label{fig:m22mdr}
\end{figure}

\begin{figure}[!t]
  \centering \includegraphics[width=0.95\columnwidth]{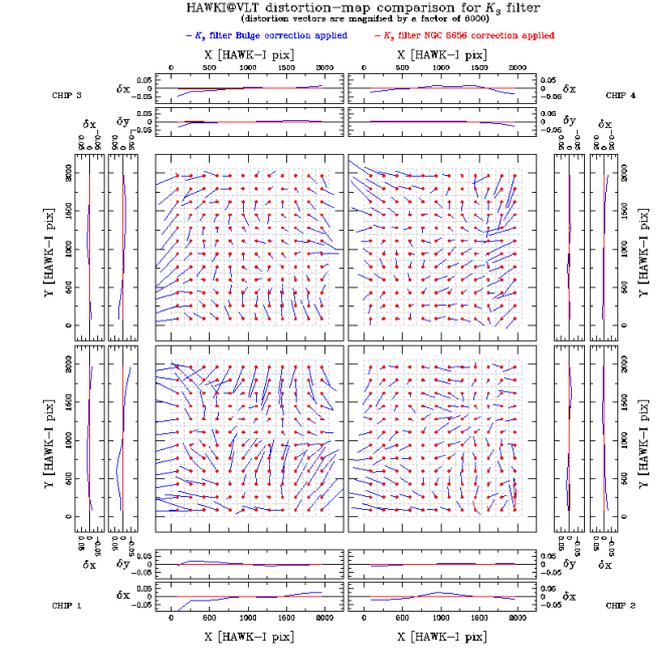}
  \caption{As in Fig.~\ref{fig:filtcomp} but for the NGC~6656 case. In
    blue, we plot the vectors when the Bulge correction is applied,
    and in red, we show when the NGC~6656-made solution is
    applied. The size of the residual vectors is magnified by a factor
    6000.}
  \label{fig:m22bulgecomp}
\end{figure}

We applied our distortion solution to images of NGC~6656 (M~22) taken
during the second commissioning. The $\sigma_{\rm perc}$ (computed as
described in Sect.~\ref{acc}) was found to be $\sim$3.5 mas for
well-measured unsaturated stars. To estimate the stability of the
distortion correction over the 3-month time baseline between the first
and second commissioning, we derived an independent distortion
solution from the second commissioning run images and compared the
results. We adopted the same auto-calibration method described
above. In this way, we were able to reduce the 1-D r.m.s down to
$\sim$3.1 mas. In Fig.~\ref{fig:m22mdr}, we show the comparison
between the $\sigma$(Radial residual) after we applied the Bulge-based
distortion correction (bottom) and the newly made NGC~6656-based
correction (top). The difference between these distortion solutions is
only 0.003 pixel. Therefore, our Bulge distortion correction should be
stable at a 3-mas level on a 3-month scale for general uses. In
Fig.~\ref{fig:m22bulgecomp}, we show the distortion-map
comparison. There are systematic trends when the $K_{\rm S}$-filter
Bulge solution is applied to this data set. Nevertheless, for
high-precision astrometry, we suggest auto-calibrating the distortion
correction for each data set, as it is continuously
evolving. \looseness=-2

We note that the positions of the nodes of the periodic trend did not
change over this trim baseline, adding support to our conclusion that
this periodic residual is linked to the {\it{detector's}} properties
and is not a function of the telescope, epoch, filter, or image
quality (see Sect.~\ref{expl}). \looseness=-2

%%%%%%%%%%%
\subsection{An external check}
%%%%%%%%%%%
\label{extcheck}

\begin{figure}[t!]
  \centering
  \includegraphics[width=0.95\columnwidth]{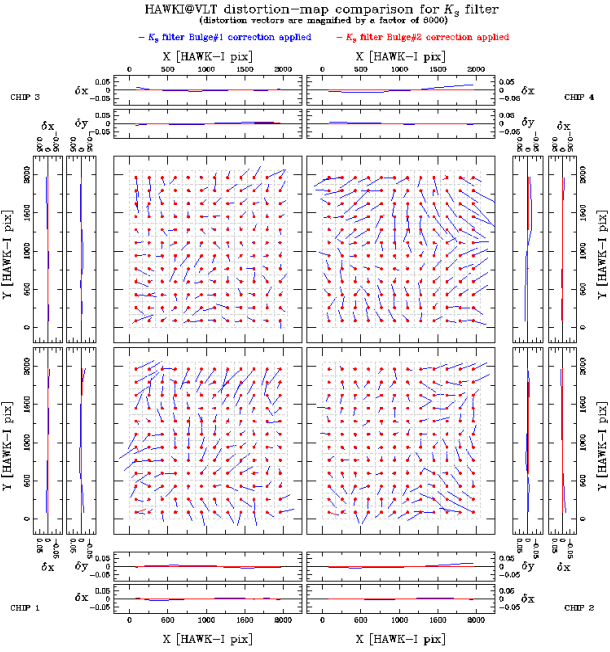}
  \caption{As in Fig.~\ref{fig:filtcomp} but for the Bulge\#2 case.}
  \label{fig:b1tob2comp}
\end{figure}

As done in the previous section for the case of NGC~6656, we apply the
distortion correction obtained by self-calibration of Bulge images
(hereafter, Bulge\#1) to a different data set, which is collected for
the same field, but with the de-rotator at a different position angle
at $\sim$135$^{\circ}$ (hereafter, Bulge\#2). In this case, we
obtained a $\sigma_{\rm perc}$ of $\sim$5.8 mas, which is
significantly larger than that obtained in Sect.~\ref{acc} for
Bulge\#1 ($\sigma_{\rm perc}$$\sim$2.8 mas). This may give the
impression that the distortion solution obtained for Bulge\#1 is not
suitable for the rotated images of Bulge\#2. However, self-calibration
of these rotated images gives us a $\sigma_{\rm perc}$ of $\sim$5.6
mas, indicating only a marginal improvement. The lower accuracy of the
distortion solution of Bulge\#2 must be ascribed to the intrinsic
lower quality of this data set (worse average seeing, higher airmass,
worse weather conditions, guiding, and instrument$+$telescope
conditions). The distortion maps obtained applying the two solutions
to the same data set (Fig.~\ref{fig:b1tob2comp}) highlight different
trends (even if the residuals are lower than 0.05 pixel), in the
upper-right corner of chip[4], which recommends again the
auto-calibration for the distortion solution of each data set for
high-accuracy astrometry.  \looseness=-2

\begin{figure}
  \centering
  \includegraphics[width=\columnwidth]{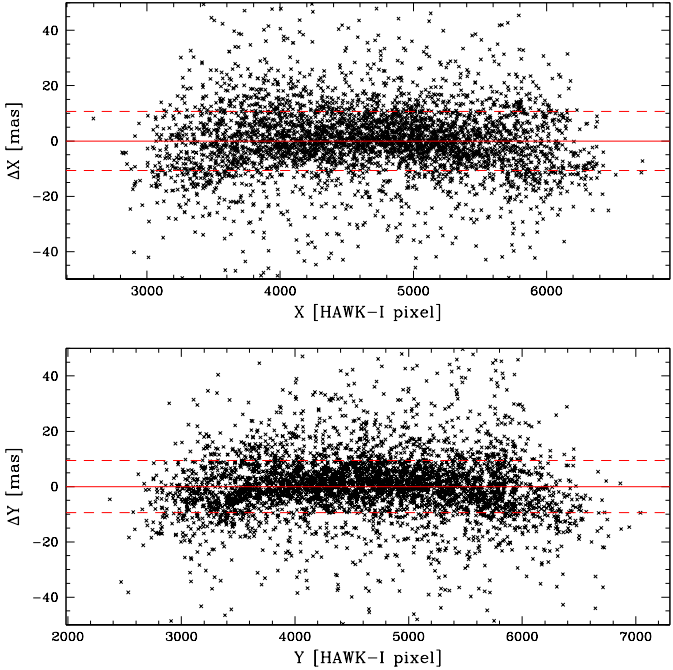}
  \caption{(\textit{Top}): $\Delta$x vs. X between the two Bulge
    fields. We plotted only bright unsaturated
    stars. (\textit{Bottom}): as on \textit{Top} but with $\Delta$y
    vs. Y. The red solid line is set at 0 mas, while the dashed lines
    are set at $\pm$10.7 mas on the top and $\pm$9.4 mas on the
    bottom.}
  \label{fig:dxdy}
\end{figure}

Nevertheless, even if rotated Bulge\#2 images were taken under worse
conditions, the value of their $\sigma_{\rm perc}$$\sim$5.6 mas allows
us to make an important external check of our solution down to this
level. The astrometric quantity $\sigma_{\rm perc}$ tells us how
accurately we can expect to register the \textit{relative} position of
a star among different dithered images. However, these are internal
estimates of the error, and do not account for all of the sources of
systematic errors. For a better estimate of the uncertainty on the
relative position of stars, we compared the two calibrated master
frames of Bulge\#1 and Bulge\#2 and measured how much the two frames
deviate from each other. We know the linear terms could be different
due to change in the thermal- or flexure-induced focal lengths,
differential atmospheric refractions, etc. For this reason, we
transformed the two master frames of Bulge\#1 and Bulge\#2 using
general linear transformations and measured the amount of residuals in
the non-linear part of the distortion. For this test, we only used
those regions of both master frames where stars were measured in at
least 10 images (out of 25). In Fig.~\ref{fig:dxdy}, we show the
residual trend of bright, unsaturated stars between the two
frames. The 68.27$^{\rm th}$ percentile of the $\Delta$X distribution
is about 10.7 mas, while that of $\Delta$Y is about 9.4 mas. Thus, the
non-linear terms of our distortion solution can be transferred between
observing runs at the 10 mas level.

%%%%%%%%%%%
\subsection{Description of the geometric-distortion-correction subroutines released}
%%%%%%%%%%%                                    
\label{gdrel}

We release FORTRAN routines to correct the geometric distortion, using
the solution computed for the Bulge\#1
field\footnote{{\url{http://vizier.u-strasbg.fr/viz-bin/VizieR}}.}
(Sect.~\ref{poly}, \ref{step}, \ref{fine} and \ref{interp}). There are
three different routines with one for each filter ($J$, $H$, $K_{\rm
  S}$). They require $x^{\rm raw}$ and $y^{\rm raw}$ coordinates and
the chip number. In output, the codes produce $x^{\rm corr}$ and
$x^{\rm corr}$ corrected coordinates. Both raw and corrected
coordinates are in the single-chip reference frame ($1 \le x^{\rm
  raw/corr},y^{\rm raw/corr} \le 2048$). In addition to these codes,
we release our distortion solution as FITS images (one per each
coordinate/chip/filter) to make the distortion solutions also
available for other program languages. Bi-linear interpolation must be
used to compute the amount of the distortion correction in inter-pixel
locations. We refer to Appendix~\ref{appendixB} for a brief
description of these corrections. Furthermore, we release FITS images
(one per chip/filter) that could be used to correct the variation in
the pixel area across the field of view. These images are useful for
improving the HAWK-I photometry. \looseness=-2

\begin{figure}
  \centering
  \includegraphics[width=\columnwidth]{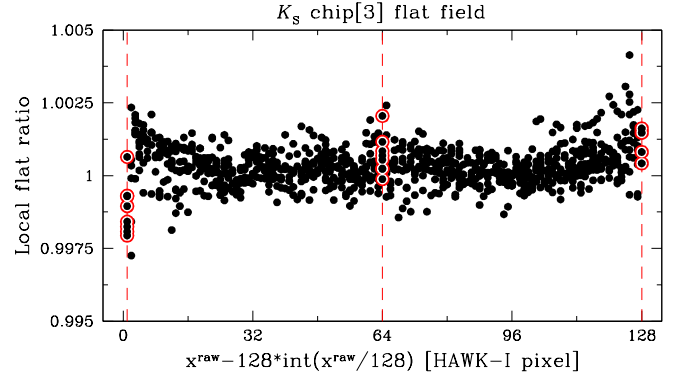}
  \caption{Local flat ratio of $K_{\rm S}$ chip[3] flat field. Dashed
    red lines mark the boundaries of the possible
    discontinuities. There are not significant local flat ratio
    variations at the $1^{\rm st}$, $64^{\rm th}$, and $128^{\rm th}$
    columns (highlighted by red open circles).}
  \label{fig:flat}
\end{figure}

%%%%%%%%
\section{A possible explanation of the periodicity}
%%%%%%%%
\label{expl}

In Sect.~\ref{step}, we have corrected for the periodic trend observed
in the $\delta x$ positions. At first, this component might suggest
the presence of some irregularities in the pixel grid, due to
manufacturing defects, such as an imperfect alignment in the placement
of the lithographic stencils that established the pixel boundaries on
the detector.  Examples include, the well-known $34^{\rm th}$-row
error found by Anderson \& King (1999) in the case of the CCDs of the
WFPC2 of the \textit{Hubble Space Telescope} (\textit{HST}), or, for a
more recent example, the pattern observed on the detectors of the
\textit{HST}'s WFC3/UVIS channel (see Bellini, Anderson \& Bedin 2011
for details).  If the square wave that we see here in the HAWK-I
$\delta x$ residuals (from a-1 to a-4 panels of
Fig.~\ref{fig:stepcor}) is due to a geometric effect, then the
variation in pixel spacing would cause periodic features in the flat
fields, since wider pixels collect more light when the detector is
illuminated by a flat surface brightness. In this case, the observed
$\delta x$-residual trend would imply that the $64^{\rm th}$ and the
$65^{\rm th}$ pixels in each row would be physically smaller than the
$128^{\rm th}$ and the $129^{\rm th}$ pixels. \looseness=-2

To verify this hypothesis, we computed the local flat ratio as
described in Bellini, Anderson \& Bedin (2011).  We took the ratio of
the pixel values over the median of the 32-pixel values on either side
along X direction (independently for each of the amplifiers). We
computed the median value of this ratio for all pixels within
$400<y^{\rm raw}<1900$ in each column. We did this for each column
between $100<x^{\rm raw}<1900$ pixels. We chose this particular area
to avoid some artifacts in the flat field near the edge of each
chip. We then plotted the local flat ratio as a function of the
128-column pattern. The plot for chip[3] of the $K_{\rm S}$-filter
flat field is shown in Fig.~\ref{fig:flat} as example. The variation
of the flat ratio of all chips in the vicinity of the columns 1, 64,
and 128 is lower than 0.25\%, thus suggesting an uniform pixel
grid. \looseness=-2

\begin{figure}[!t]
  \centering 
  \includegraphics[width=\columnwidth]{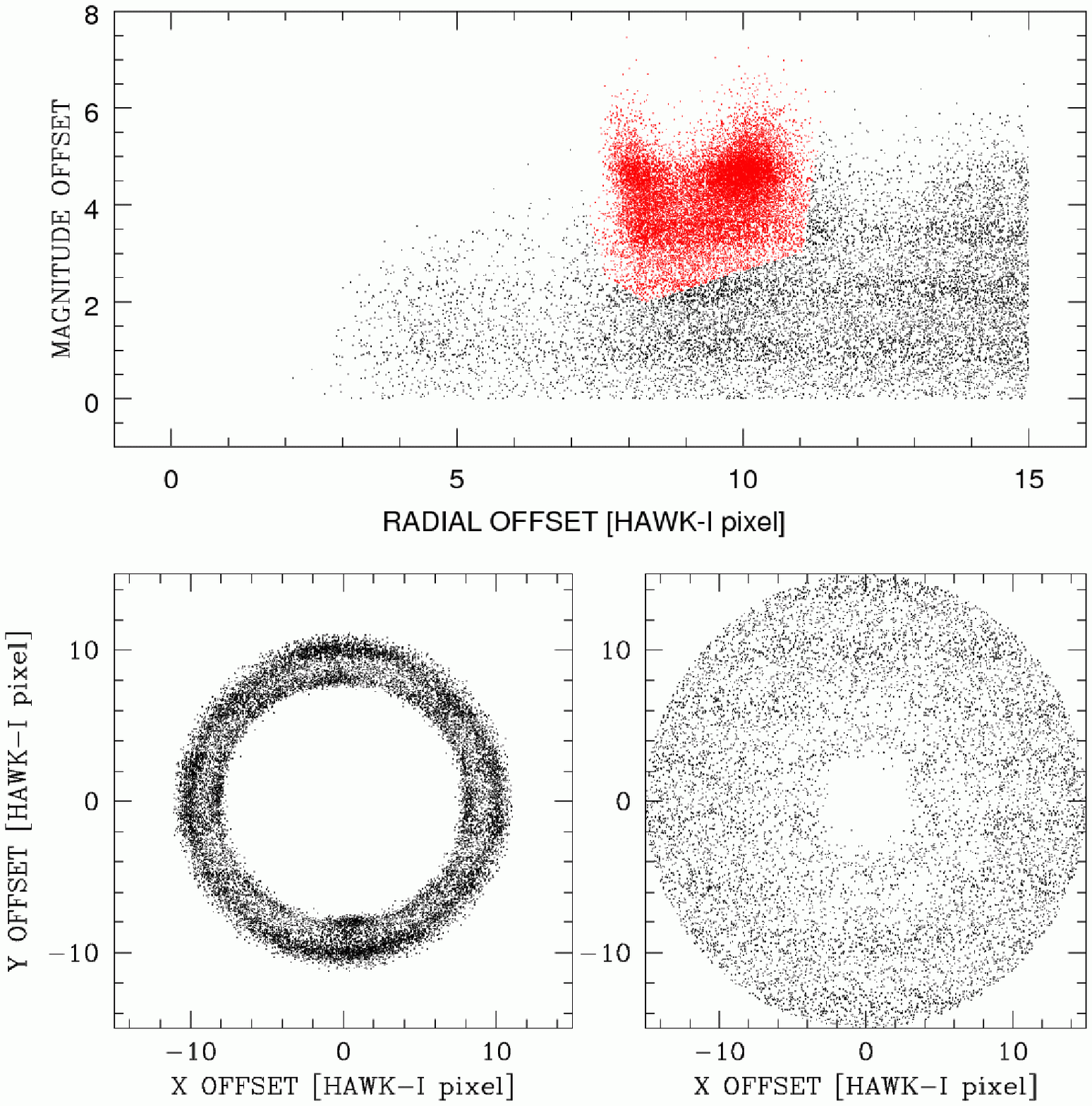}
  \includegraphics[width=\columnwidth]{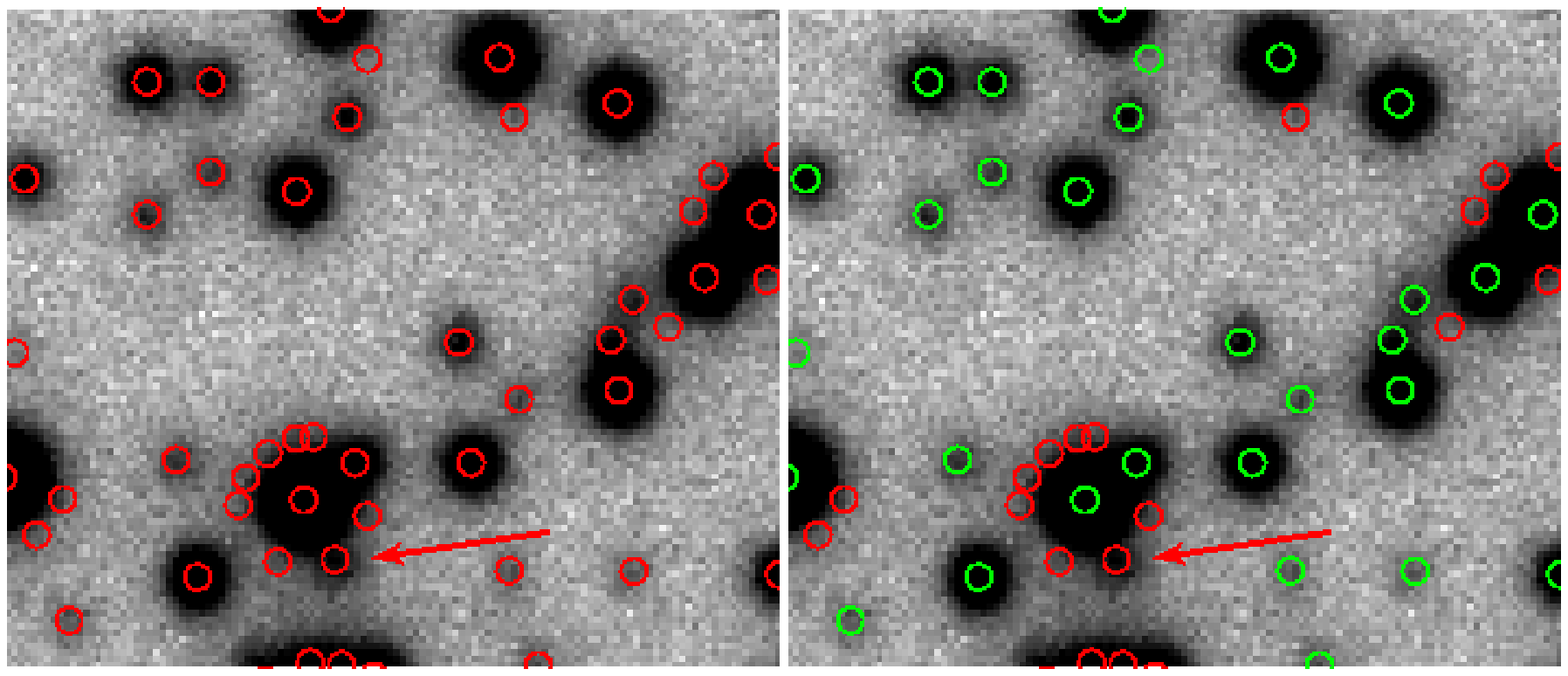}
  \caption{Example of the weeding process for the 47\,Tuc
    catalog. (\textit{Top}): Magnitude offset versus radial offset for
    sources found around bright stars. We plotted the spurious objects
    that have been rejected in red. (\textit{Middle}): Y vs. X
    separation of the rejected (left) and accepted (right) detections
    from the corresponding bright stars. (\textit{Bottom}): On the
    left, we show all the objects (red circles) in a
    $\sim$100$\times$100 pixels region in the 47\,Tuc catalog, and on
    the right, we highlight the detections that have been accepted
    (green circles). As described in the text, a faint star near the
    bright one pointed by the red arrow on the bottom has been flagged
    even if it was not an artifact.}
  \label{fig:unweed}
\end{figure}

Another possible explanation of this effect can be ascribed to the
readout process of the Rockwell detectors. The Rockwell HgCdTe
detectors are designed to have three output modes. It can use 1, 4 or
32 amplifiers. The HAWK-I detector is set up to use all 32 amplifiers,
and it takes 1.3 s to read out the entire chip. In the 32-amp mode,
the chip is divided into 32 64-pixel-wide strips, each fed into a
different amplifier. The adopted operating mode performs the read-out
from left-to-right in even amplifiers and from right-to-left in odd
amplifiers. \looseness=-2

An apparent shift of the stars' position along the x axis may happen
if there is a ``periodic lag'' during the read out, since the
amplifier reads the pixels in sequence. This effect is very similar to
the ``bias shift'' observed in ACS/WFC of the \textit{HST} after part
of the electronics, in particular the new amplifiers, has been
replaced during service mission 4 (Golimowski et al. 2012). As for
ACS, the readout electronics of HAWK-I's detectors take a while to
settle to a new value when the charge of another pixel is loaded.
Without waiting an infinite amount of time to settle down, there is
some imprint left from the previous pixel. Although HAWK-I's
NIR-detectors are very different from the ACS/WFC CCDs, a similar
effect, or, an inertia of discharging the capacitors to reset to a new
value, could cause the observed periodicity. Furthermore, each
amplifier in the 32-amp mode reads 64 pixels, and 64 pixels is the
observed periodicity of the effect that we found. This suggests that
the cause of the periodic trend we see in the distortion could be
related to the amplifiers' setup. \looseness=-2

%%%%%%%%%%%
\section{Weeding out spurious objects}
%%%%%%%%%%%
\label{weed}

We applied our Bulge-based distortion correction, derived as in
Sect.~\ref{GDC}, to the entire data set with the exceptions of the
Bulge\#2 and NGC~6656 fields, where we applied the distortion
correction computed from their own exposures. We also produced stacked
images to have a visible reference for our catalogs. \looseness=-2

While analyzing the catalogs and inspecting the stacked images, we
noted the presence of some faint, spurious objects identified as stars
and close to brighter stars. Many of these objects were found to be in
fact PSF artifacts and are not real stars. These artifacts are called
``PSF bumps''. These bumps can easily mimic faint, real stars in the
proximity of much brighter stars. Normally, PSF bumps are located at
the same radial distance from the PSF center and have about the same
brightness level. Because of these two characteristics, PSF bumps can
be removed. Therefore, we introduced a flag to purge the catalogs from
these detections. This flag may necessary exclude a few real stars,
but faint stars close to a bright one could not be well measured
anyway. \looseness=-2

To flag these spurious objects, we followed the same approach as
described in Sect.~6.1 of Anderson et al. (2008). We first selected
all stars fainter than a specific instrumental magnitude (e.g. for
47\,Tuc, shown in Fig.~\ref{fig:unweed}, we chose $J \ge -10$) and
with a $\sf QFIT \ge$ 0.4. For each of such objects, we computed the
magnitude difference and the distance from the closest bright star
(e.g. $J \le -11$ for the case of 47\,Tuc) out to 15 pixels. We then
plotted those magnitude differences as a function of the radial
distance (top panel of Fig.~\ref{fig:unweed}). Different clumps show
up on the plot. We drew-by-hand a region around them that encloses
most of these spurious objects (in red). In this way, we built a mask
(one for each filter/field) used to purge PSF artifacts. \looseness=-2

The selection we made is a compromise between missing faint objects
near bright stars and including artifacts in the catalog. In the
bottom panel of Fig.~\ref{fig:unweed}, the red circles show all the
detected objects in the 47\,Tuc catalog. In the bottom right panel of
Fig.~\ref{fig:unweed}, the green circles highlights the objects that
have been finally accepted as real stars. The bright star close to the
bottom of the figure has a faint neighbor (pointed by the arrow) that
clearly is not an artifact but has been unfortunately flagged-out by
our mask. These flagged stars represent only a very small fraction
with respect to the total number of PSF artifacts removed by our
procedure. \looseness=-2

In each final catalog, we added a column for each filter called $\rm
F_{\rm weed}$. The flag $\rm F_{\rm weed}$ is equal to 0 for those
objects rejected by our mask. The only exception is the $K_{\rm
  S}$-filter catalog of Bulge\#1, for which the purging was not
possible because of the too-high number of objects in the catalog that
did not allow us to build a reliable mask for the PSF
artifacts. \looseness=-2

%%%%%%%%
\section{Photometric calibration}
%%%%%%%%
\label{calib}

\begin{table*}[!t]
  \caption{List of the HAWK-I filter zero-points, r.m.s., number of
    stars used, and zero-point formal uncertains ($\sigma/\sqrt{N-1}$)
    to which 2MASS error of the stars used to calibrate should be
    added in quadrature. The values listed in columns from (3) to (6)
    are those obtained in the 2MASS system, and from (7) to (10) in
    the MKO system.} 
  \label{tab:zp}     
  \centering
  \begin{tabular}{cc|cccc|cccc}          
    \hline\hline    
    \textbf{Field} & \textbf{Filter} & \textbf{Zero-point} & \textbf{$\sigma$} & \textbf{$N$} & \textbf{$\sigma/\sqrt{N-1}$} & \textbf{Zero-point} & \textbf{$\sigma$} & \textbf{$N$} & \textbf{$\sigma/\sqrt{N-1}$} \\
    & & \multicolumn{4}{c}{\textbf{2MASS}} & \multicolumn{4}{|c}{\textbf{MKO}} \\
    \hline          
    & & & & & & & \\
    Bulge --- Baade's Window (\#1) & $J$ & $-28.31$ & 0.07 & 995 & 0.01 & $-28.25$ & 0.07 & 966 & 0.01 \\
    & $H$ & $-28.60$ & 0.05 & 57 & 0.01 & $-28.57$ & 0.05 & 37 & 0.01 \\
    & $K_{\rm S} $ & $-28.01$ & 0.09 & 543 & 0.01 & $-27.98$ & 0.09 & 543 & 0.01 \\ 
    & & & & & & & \\
    Bulge --- Baade's Window (\#2) & $K_{\rm S}$ & $-27.77$ & 0.09 & 543 & 0.01 & $-27.75$ & 0.09 & 542 & 0.01 \\
    & & & & & & & \\
    NGC~6121 (M~4) & $J$ & $-28.76$ & 0.06 & 298 & 0.01 & $-28.71$ & 0.06 & 298 & 0.01 \\
    & $K_{\rm S}$ & $-28.69$ & 0.09 & 86 & 0.01 & $-28.67$ & 0.09 & 86 & 0.01 \\
    & & & & & & & \\
    NGC~6822 & $J$ & $-28.40$ & 0.04 & 28 & 0.01 & $-28.36$ & 0.04 & 28 & 0.01 \\
    & $K_{\rm S}$ & $-27.59$ & 0.07 & 26 & 0.01 & $-27.57$ & 0.07 & 26 & 0.01 \\
    & & & & & & & \\
    NGC~6656 (M~22) & $K_{\rm S}$ & $-27.95$ & 0.15 & 83 & 0.02 & $-27.93$ & 0.15 & 81 & 0.02 \\
    & & & & & & & \\
    NGC~6388 & $J$ & $-28.46$ & 0.06 & 289 & 0.01 & $-28.41$ & 0.06 & 286 & 0.01 \\
    & $K_{\rm S}$ & $-27.73$ & 0.05 & 229 & 0.01 & $-27.72$ & 0.05 & 229 & 0.01 \\
    & & & & & & & \\
    JWST calibration field (LMC) & $J$ & $-28.78$ & 0.07 & 233 & 0.01 & $-28.74$ & 0.07 & 226 & 0.01 \\
    & $K_{\rm S}$ & $-27.84$ & 0.09 & 122 & 0.01 & $-27.82$ & 0.09 & 123 & 0.01 \\
    & & & & & & & \\
    NGC~104 (47\,Tuc) & $J$ & $-28.26$ & 0.05 & 80 & 0.01 & $-28.22$ & 0.05 & 80 & 0.01 \\
    & $K_{\rm S}$ & $-27.71$ & 0.05 & 128 & 0.01 & $-27.69$ & 0.05 & 128 & 0.01 \\
    & & & & & & & \\
    \hline
  \end{tabular} 
\end{table*} 

In this section, we provide two calibrations of the zero-points. The
first calibration was performed using the 2MASS photometric system
(Skrutskie et al. 2006) and the second calibration using the native
system of the HAWK-I filters.

\begin{figure}[!t]
  \centering
  \includegraphics[width=\columnwidth]{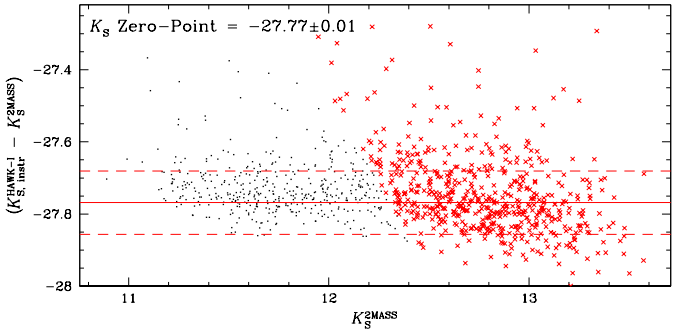}
  \caption{Magnitude difference between HAWK-I and 2MASS as function
    of the 2MASS magnitude. The black dots represent all the stars
    matched between HAWK-I and 2MASS with good photometry. Red crosses
    show stars from the saturation limit ($K_{\rm S} =$ $-15.46$) to
    two magnitudes fainter. The red solid line is the zero-point
    (median of the magnitude difference of the red crosses); the
    dashed line are set at a zero-point $\rm\pm 1\sigma$ (defined as
    the 68.27$^{\rm th}$ percentile of the distribution around the
    median). The label on the top left corner gives the zero-point
    $\rm \pm \sigma/\sqrt{N-1}$, where $N$ is the number of stars used
    to compute the zero-point.}
  \label{fig:zp}
\end{figure}

%%%%%%%%%%%
\subsection{2MASS system}
%%%%%%%%%%%
\label{cal2mass}

The first photometric calibration was performed using the 2MASS
catalog.  Since 2MASS is a shallow survey, we only got a small overlap
between unsaturated stars in HAWK-I images and 2MASS data covering a
very narrow magnitude range near the faint limit of 2MASS. Therefore,
we can apply a single zero-point calibration only. We selected
well-measured bright unsaturated stars within two magnitudes from
saturation in our catalogs to calculate these photometric
zero-points. In Fig.~\ref{fig:zp}, we show the case of the Bulge\#2
catalog, as an example. For the Bulge\#1 $K_{\rm S}$-filter, we first
registered the zero-point to that of the Bulge\#2, and then used the
Bulge\#2 zero-point due the low number of unsaturated stars in common
with 2MASS. \looseness=-2

\begin{figure*}[!t]
  \centering
  \includegraphics[width=\textwidth]{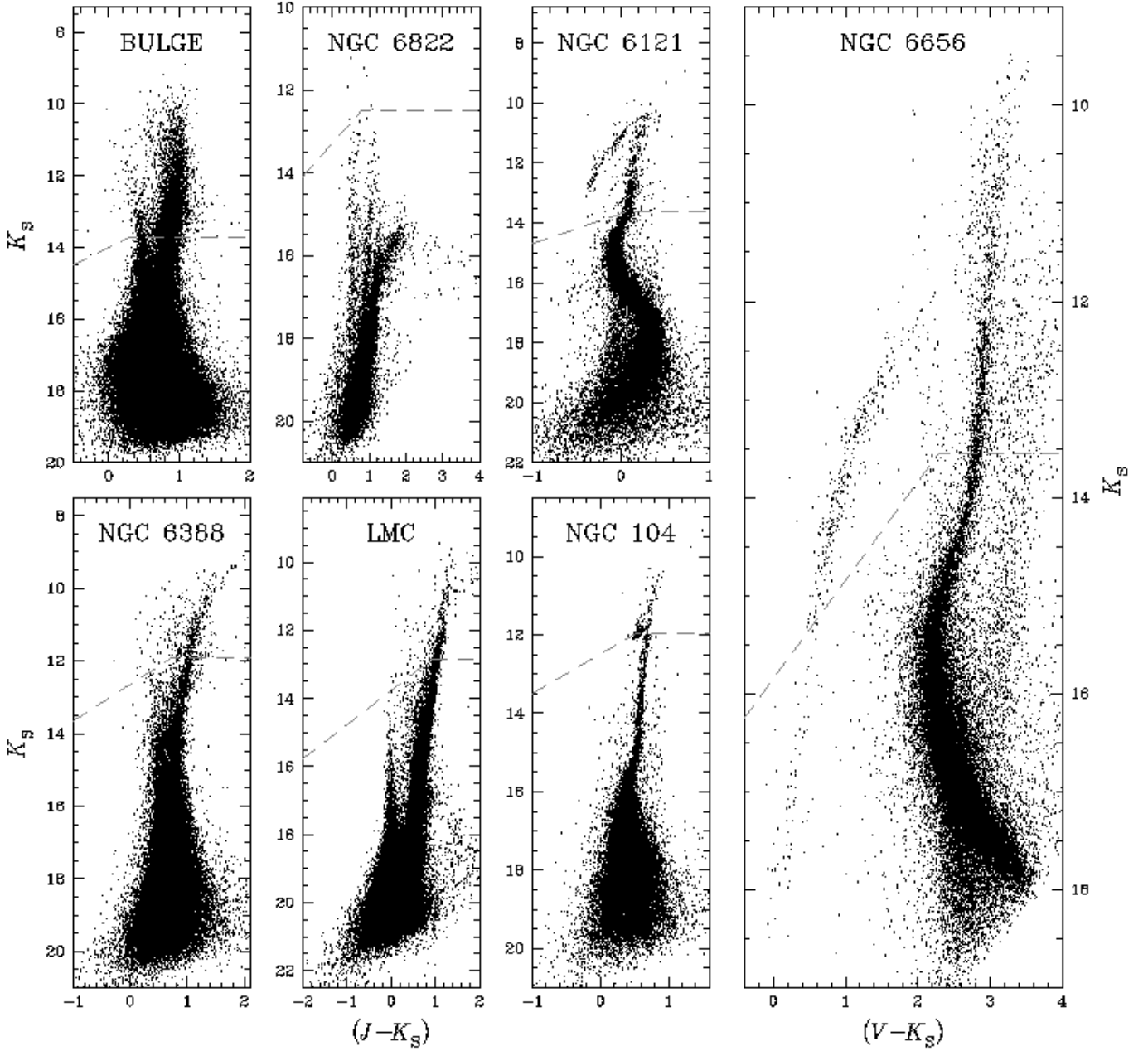}
  \caption{Full set of CMDs of the fields used in this paper. The
    dotted gray lines set the saturation threshold in $K_{\rm S}$
    filter.}
  \label{fig:review}
\end{figure*}

In Table~\ref{tab:zp}, we list the zero-points, their
r.m.s. ($\sigma$), the number of stars used to compute the zero-points
($N$) and $\sigma/\sqrt{N-1}$ (formal error values) for all
fields. These are estimates of the zero-point uncertainties and not
the errors because the 2MASS errors are not added in quadrature. Note
that we only provide one zero-point for each filter/field and not one
for each chip/filter/field. We registered all chips in the
flat-fielding phase to the common reference system of chip[1], and
while building the master frame, we iteratively registered the
zero-point of all chips to that of the chip[1]. \looseness=-2

As clearly visible in Fig.~\ref{fig:zp}, this calibration is not
perfect, but there is a more conceptual limitation of this
calibration. The filters of HAWK-I are in the Mauna Kea Observatory
(hereafter, MKO) photometric system, which have pass-bands slightly
different from the 2MASS pass-bands and are likely to contain a color
term. \looseness=-2

%%%%%%%%%%%
\subsection{MKO system}
%%%%%%%%%%%

As suggested by the Referee, determining the zero-points in the native
MKO photometric system would be a more rigorous zero-point
calibration.

Therefore, we transformed the 2MASS magnitudes into the MKO system
using the transformations described in the “2MASS Second Incremental
Release”
website\footnote{{\url{http://www.ipac.caltech.edu/2mass/releases/allsky/doc/sec6\_4b.html}}.\looseness=-2}
for 2MASS stars in common with our catalogs:

\begin{displaymath}
 (K_{\rm S})_{\rm 2MASS}\!=\! (K)_{\rm MKO} + (0.002 \pm 0.004) + (0.026
  \pm 0.006)(J-K)_{\rm MKO} ,
\end{displaymath}
\begin{displaymath}
  (J-H)_{\rm 2MASS} = (1.156 \pm 0.015)(J-H)_{\rm MKO} + (-0.038 \pm
  0.006) ,
\end{displaymath}
\begin{displaymath}
  (J-K_{\rm S})_{\rm 2MASS} = (1.037 \pm 0.009)(J-K)_{\rm MKO} + (-0.001 \pm
  0.006) ,
\end{displaymath}
\begin{displaymath}
  (H-K_{\rm S})_{\rm 2MASS} = (0.869 \pm 0.021)(H-K)_{\rm MKO} + (0.016 \pm
  0.005) .
\end{displaymath}

We used only 2MASS stars that, once transformed in the MKO system,
were in the color range $-0.2$$<$($J$$-$$K$)$_{\rm MKO}$$<$$1.2$. We
then registered the zero-points as described in the previous
section. We found an average difference between the MKO and 2MASS
zero-points of 0.05, 0.03, and 0.02 mag in $J$-, $H$-, and $K_{\rm
  S}$-filter, respectively. For the $K_{\rm S}$ filter in NGC~6121
catalog, we did not have enough stars to compute the zero-point in the
color range in which the transformations are valid, so we added the
average $K_{\rm S}$-filter offset (0.02 mag) between the two systems
to the 2MASS-based zero-point. In Table~\ref{tab:zp}, we also list the
MKO zero-points (with their $\sigma$, $N$, and
$\sigma/\sqrt{N-1}$). \looseness=-2

\begin{figure*}[!t]
  \centering
  \includegraphics[width=\columnwidth]{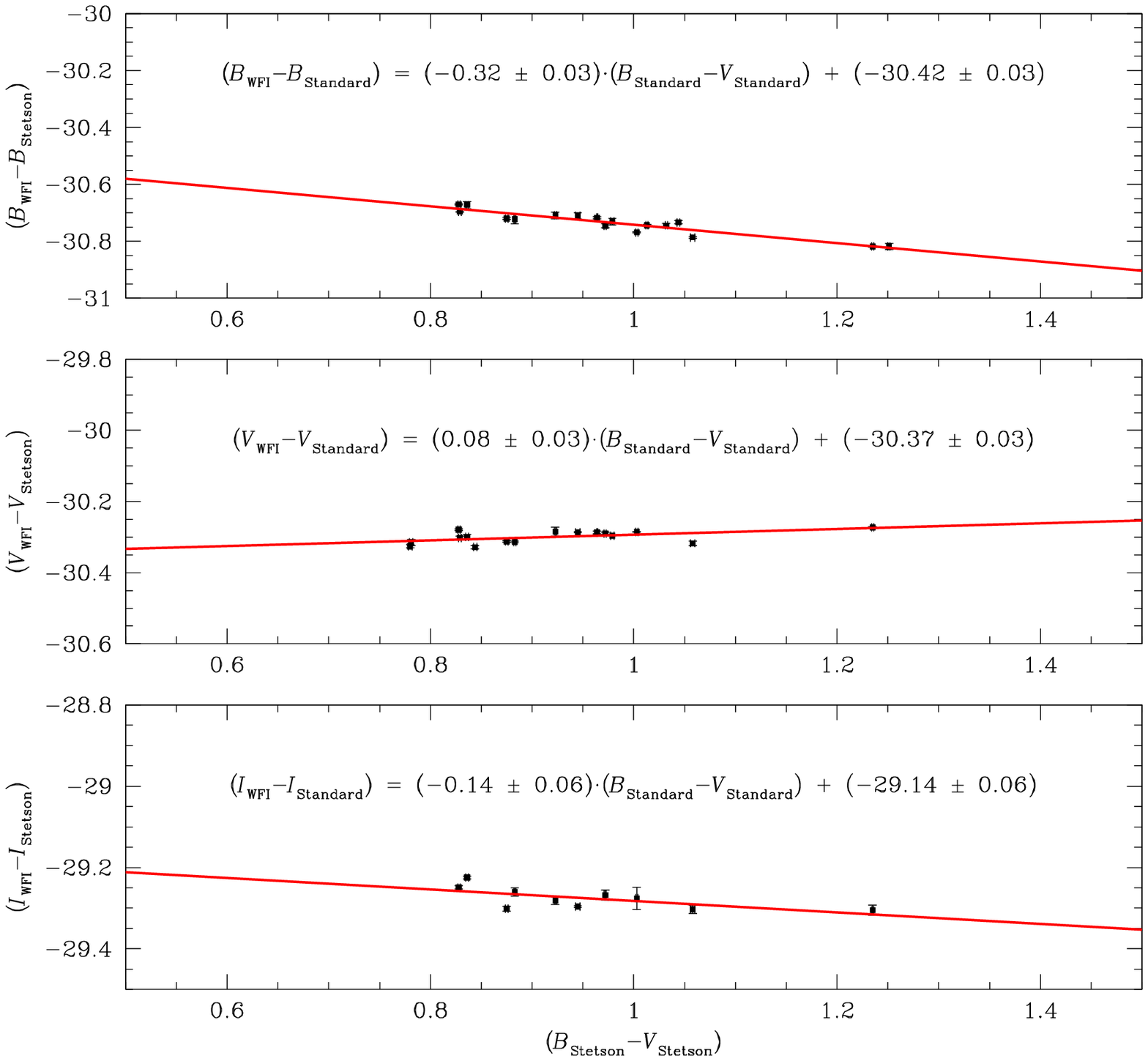}
  \includegraphics[width=\columnwidth]{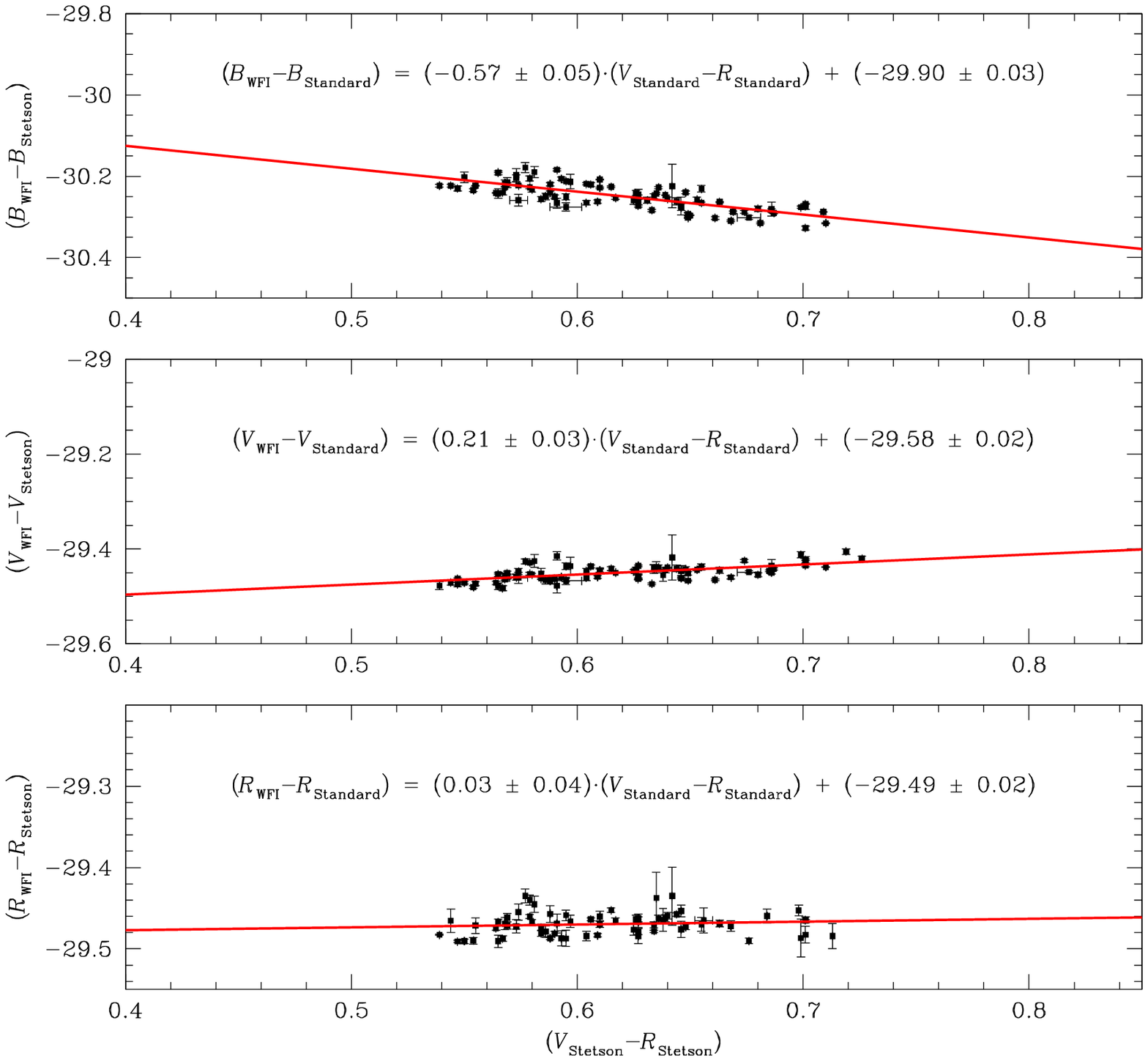}
  \caption{(\textit{Left}): Calibration fits and equations for $B$,
    $V$, and $I$ filters for NGC~6656. (\textit{Right}): Same as on
    the left but for $B$, $V$, and $Rc$ filters for NGC~6121.}
  \label{fig:calib}
\end{figure*}

%%%%%%%%
\section{Applications: NGC~6656 and NGC~6121}
%%%%%%%%
\label{app}

Figure~\ref{fig:review} shows a full set of CMDs with one for each
field. For the HAWK-I data used in this section, the photometric
zero-points are those obtained in the 2MASS system
(Sect.~\ref{cal2mass}). We chose two possible targets among the
observed fields to illustrate what can be done with HAWK-I. The two
close globular clusters NGC~6656 and NGC~6121 have high proper motions
relative to the Galactic field. In the ESO archive, we found
WFI@2.2\,m MPI/ESO exposures of the same fields taken $\sim$8 years
before our HAWK-I observations. These exposures allow us to obtain a
nearly-perfect separation between cluster and field stars, as we will
show in the following subsections. \looseness=-2

%%%%%%%%%%%
\subsection{WFI data set: photometric calibration and differential reddening correction}
%%%%%%%%%%%

We downloaded multi-epoch images of NGC~6656 from the ESO archive
(data set 163.O-0741(C), PI: Renzini), taken between May 13 and 15
1999 in $B$, $V$, and $I$ filters with the WFI@2.2\,m MPI/ESO. These
images were not taken for astrometric purposes and only have small
dithers, thus preventing us from randomizing the distortion-error
residuals. Photometry and astrometry were extracted with the
procedures and codes described in Paper~I. Photometric measurements
also were corrected for “sky concentration” effects (light
contamination caused by internal reflections of light in the optics,
causing a redistribution of light in the focal plane) using recipes in
Paper~III. The WFI photometry was calibrated matching our catalogs
with the online secondary-standard stars catalog of Stetson (Stetson
2000, 2005) using well-measured, bright stars, and least-square
fitting. We found that a linear relation between our instrumental
magnitudes and Stetson standard magnitudes was adequate to register
our photometry.  The calibration equations are shown in
Fig.~\ref{fig:calib} in the left panels. \looseness=-2

As for NGC~6656, we downloaded the NGC~6121 images from the ESO
archive taken with the WFI@2.2\,m MPI/ESO between August 16 and 18
1999, in $B$, $V$, and $Rc$ filters. We performed the photometric
calibration as described above. The corresponding calibration fit and
equations are shown in Fig.~\ref{fig:calib} in the right
panels. \looseness=-2

\begin{figure}[!t]
  \centering
  \includegraphics[width=\columnwidth]{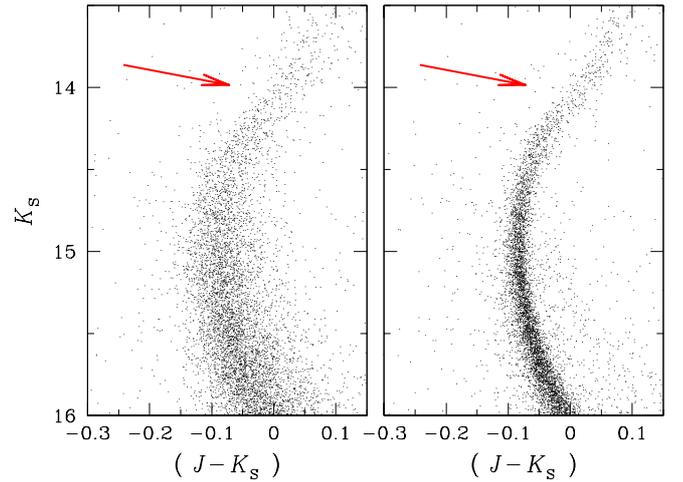}
  \caption{Zoom-in of the $K_{\rm S}$ vs. ($J-K_{\rm S}$) CMD of
    NGC~6121. We show the CMDs before and after the differential
    reddening correction is applied (left and right panels
    respectively). The red arrow indicates the reddening direction.}
  \label{fig:m4field}
\end{figure}

In the following subsections, we explore some applications in which
the photometry has been corrected for differential reddening. We
performed a differential reddening correction following the iterative
procedure described in Milone et al. (2012). As described in detail by
Milone et al., the correction to apply to a given star is measured
from the differential reddening of the selected reference stars that
are spatially close to the target. The number of stars to use should
be a compromise between the need to have an adequate number of
reference stars to compute the correction and the need for spatial
resolution. We chose the nearest 45 reference stars from the faint
part of the red giant branch (RGB) to the brighter part of the main
sequence (MS) to compute the correction. In Fig.~\ref{fig:m4field}, we
present an example to demonstrate how the CMDs change by taking the
differential reddening into account and correcting for it. We show a
zoom-in of the NGC~6121 $K_{\rm S}$ vs. ($J-K_{\rm S}$) CMD before
(left panel) and after (right panel) the correction. Around the MS
turn-off ($K_{\rm S}$$\sim$15), the sequence narrowed from $\Delta
(J-K_{\rm S})$$\sim$0.08 to $\sim$0.03 mag. Hereafter, all CMDs for
both NGC~6656 and NGC~6121 are corrected for differential
reddening. \looseness=-2

%%%%%%%%%%%
\subsection{NGC~6656 proper motion}
%%%%%%%%%%%
\label{m22pm}

\begin{figure*}[!t]
  \centering
  \includegraphics[width=\textwidth]{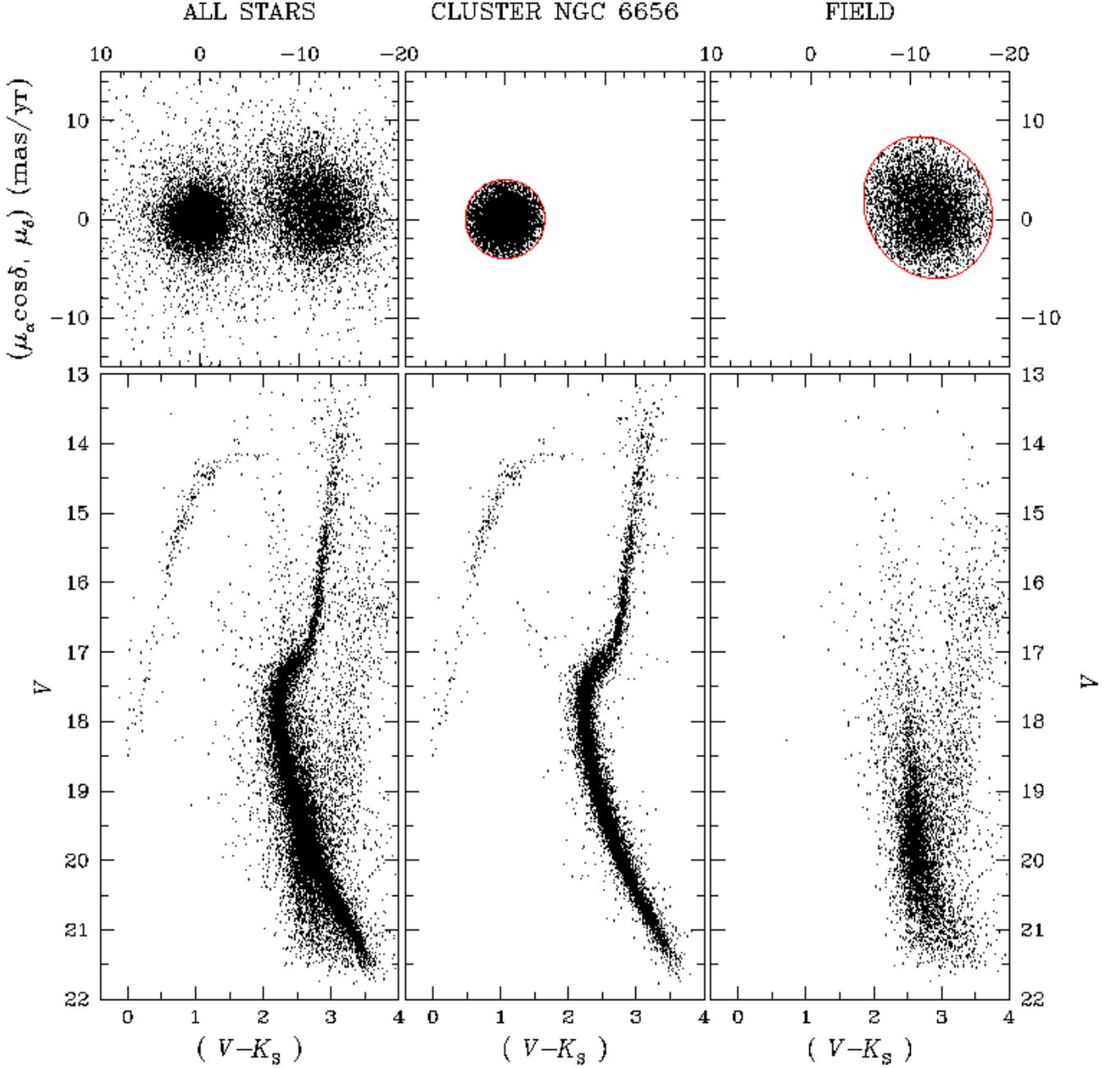}
  \caption{(\textit{Top panels}): Proper motion vector-point diagram
    with a $\sim$8-yr time baseline between HAWK-I and WFI data. The
    (0,0) location in VPD is the mean motion of cluster stars
    candidates. (\textit{Bottom panels}): $V$ vs. ($V-K_{\rm S}$)
    color-magnitude diagram. (\textit{Left}): The entire
    sample. (\textit{Center}): Stars in VPD with proper motion within
    4 mas yr$^{-1}$ around the cluster mean. (\textit{Right}):
    Probable background/foreground field stars in the area of NGC~6656
    studied in this paper. The ellipse that encloses most of the field
    stars is centered at ($-11.8$,1.2) mas yr$^{-1}$ with major and
    minor axes of 12.5 and 14.8 mas yr$^{-1}$, respectively.}
  \label{fig:cmd1}
\end{figure*}

To compute proper motions, we followed the method described in
Paper~I, to which we refer for the detailed description of the
procedure. For the WFI images, we only used those chips that overlap
with the HAWK-I field and with an exposure time of $\sim$239 s for a
total of 19 first-epoch catalogs that include $B$, $V$, and $I$
filters. With the 100 catalogs for the second epoch (HAWK-I) in
$K_{\rm S}$ band, we computed the displacements for each star. As
described in Paper~I, the local transformations used to transform the
star's position in the 1$^{\rm st}$ epoch system into that of the
2$^{\rm nd}$ epoch minimize the effects of the residual geometric
distortion. \looseness=-2

In Fig.~\ref{fig:cmd1}, we show our derived proper motions for
NGC~6656. We show only stars with well-measured proper motions. In the
left panels of Fig.~\ref{fig:cmd1}, we show the entire sample of
stars; the middle panels display likely cluster members. The right
panels show predominantly field stars. In the middle vector-point
diagram (VPD), we drew a circle around the cluster's motion centroid
of radius 4 mas yr$^{-1}$ to select proper-motion-based cluster
members. The chosen radius is a compromise between missing cluster
members with larger proper motions and including field stars that have
velocities equal to the cluster's mean proper motion. This example
demonstrates the ability of our astrometric techniques to separate
field and cluster stars. To enclose most of the field stars, we drew
an ellipse centered at ($-11.8$,1.2) mas yr$^{-1}$ in the right VPD
with major and minor axes of 12.5 and 14.8 mas yr$^{-1}$,
respectively. \looseness=-2

We analyzed the impact of the differential chromatic effects in our
astrometry for this cluster as done in Anderson et al. (2006). Using
unsaturated stars and with a color baseline of about 3 mag, the
effects seem to be negligible (less than 1 mas/yr in each direction)
within the airmass range of our data set. Thus we assumed to be
negligible and did not correct it.

%%%%%%%%%%%%%%
\subsubsection{The radial distribution of NGC~6656 SGB stars}
%%%%%%%%%%%%%%
\label{m22sgb}

\begin{figure}
  \centering
  \includegraphics[width=\columnwidth]{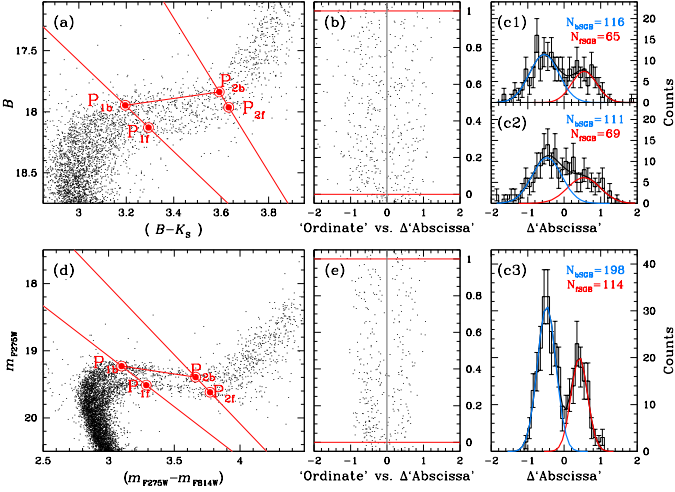}
  \caption{(\textit{a}): SGB zoom-in on $B$ vs. ($B-K_{\rm S}$) CMD of
    NGC~6656. The four points (and the two straight lines) used to
    perform the linear transformation are plotted in
    red. (\textit{b}): Rectified SGB. The red horizontal lines are set
    at `Ordinate' 0 and 1; the gray solid line is set at
    $\Delta$`Abscissa'$=$0. (\textit{c}1): Dual-Gaussian fit in black;
    individual Gaussian in blue and red are used for bright and faint
    SGB respectively, for the SGB-star sample between 1.5 and 3.0
    arcmin from the cluster center. (\textit{c}2): As for
    (\textit{c}1) but in the range 3.0-9.0
    arcmin. (\textit{c}3,\textit{e},\textit{d}): Same as in panels
    (\textit{a},\textit{b},\textit{c}1,\textit{c}2) but for the {\it
      HST} data in the $m_{\rm F275W}$ vs. ($m_{\rm F275W}-m_{\rm
      F814W}$) plane.}
  \label{fig:tuttisgb}
\end{figure}

\begin{figure}
  \centering
  \includegraphics[width=\columnwidth]{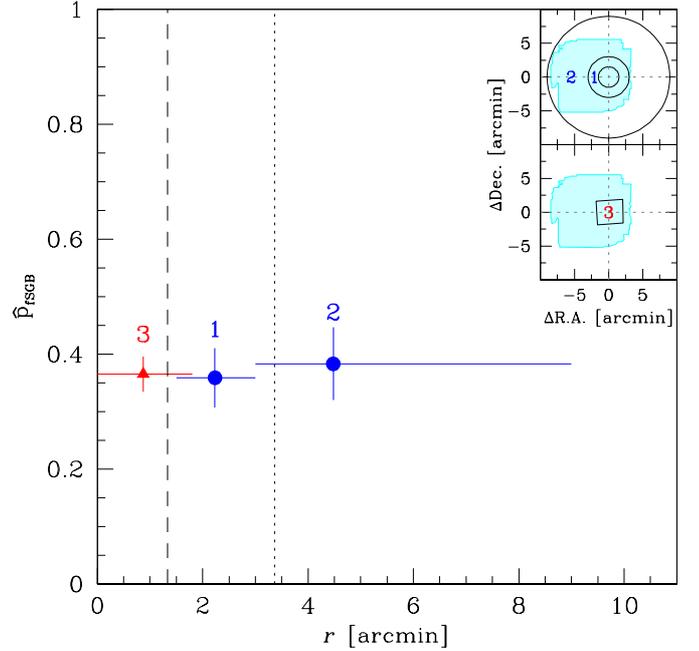}
  \caption{Radial trend of $\rm \hat{p}_{\rm fSGB}$. The numbers 1, 2,
    and 3 correspond to panels (c) in Fig.~\ref{fig:tuttisgb}. The
    points are placed at the average distance of the SGB stars used to
    compute the ratio in each radial bin. In blue, we plotted the
    ratios obtained with the HAWK-I data set; in red, we show the
    ratio obtained with the {\it HST} data set. The vertical error
    bars are computed as described in the text. The horizontal error
    bars cover the radial intervals. The two vertical lines indicate
    the core radius and half-light radius ($1\farcm33$ and $3\farcm36$
    respectively; from Harris 1996, 2010 edition). In the top-right
    panels, the cyan region highlights the HAWK-I field. The cluster
    center is set at (0,0). The three circles have radius 1.5, 3.0,
    and 9.0 arcmin. The black parallelogram represents the field
    covered by the {\it HST} data. The regions used to compute the
    ratios are labeled with the numbers 1, 2, and 3, respectively.}
  \label{fig:radial}
\end{figure}

The sub-giant branch (SGB) based on HAWK-I data remains broadened even
after the differential reddening correction.  This is not surprising
since NGC~6656 is known to have a split SGB (Piotto et al. 2012).  The
large FoV of our data set allowed us to study the behavior of the
radial trend of the ratio $\rm \hat{p}_{\rm fSGB} = \rm N_{\rm fSGB} /
(N_{\rm fSGB}+N_{\rm bSGB})$, where $\rm N_{\rm bSGB}$ and $\rm N_{\rm
  fSGB}$ are the number of stars belonging to the bright (bSGB) and
the faint SGB (fSGB), respectively. First of all, we computed this
ratio for SGB stars between 1.5 and 3.0 arcmin from the center of the
cluster (we adopted the center given by Harris 1996, 2010 edition),
and between 3.0 and 9.0 arcmin (close to the edge of the FoV). We
chose these two radial bins to have about the same number of SGB stars
in both samples. Since the innermost region (within $1\farcm5$ from
the center) of the cluster is too crowded to be analyzed with the
HAWK-I data, we also used the data set of Piotto et al. (2012) to have
a third, inner point. \looseness=-2

\begin{figure*}[!t]
  \centering \includegraphics[width=\textwidth]{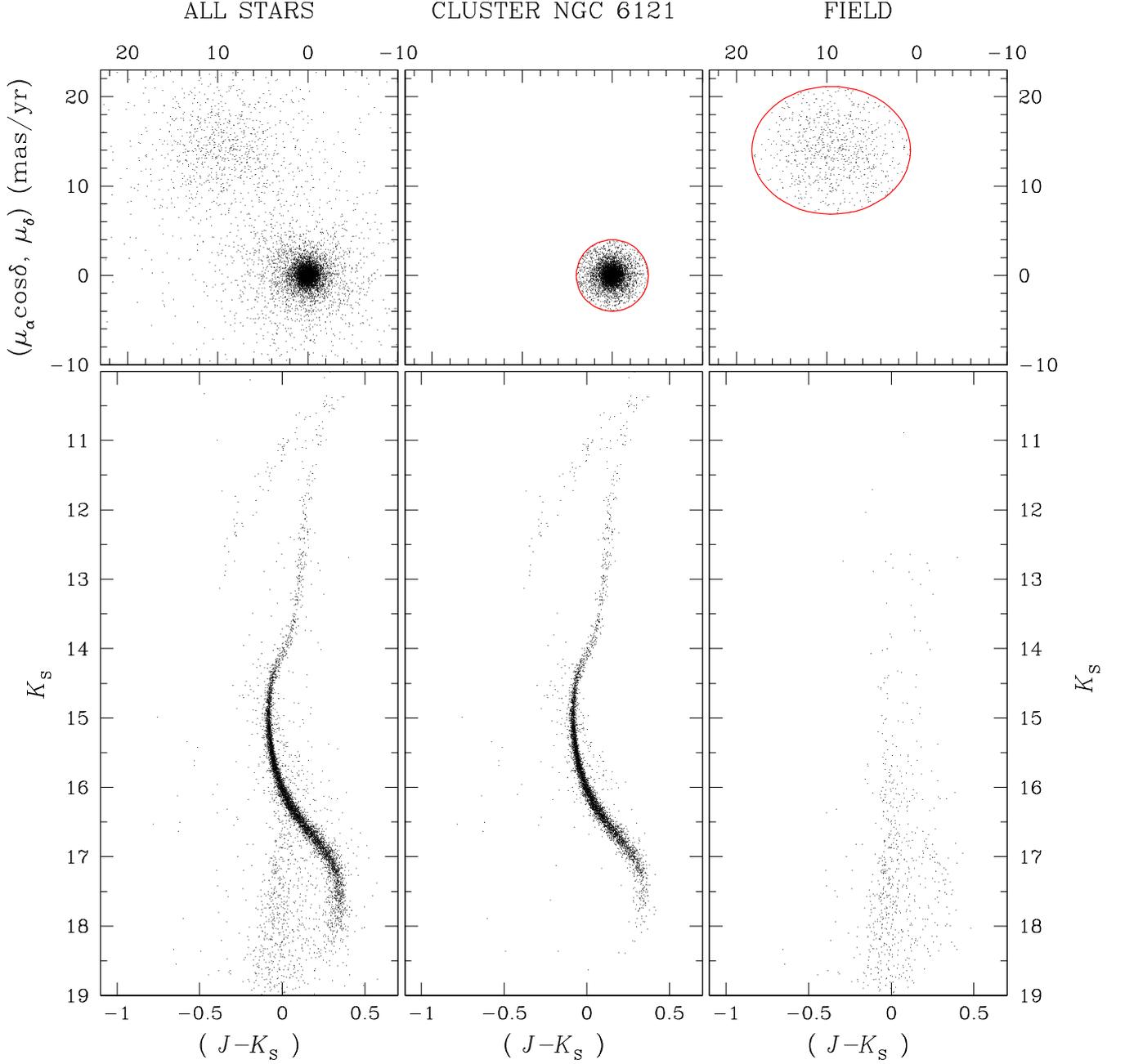}
  \caption{Same as Fig.~\ref{fig:cmd1} but for NGC~6121. The radius of
    the circle centered on the origin of the VPD is 4 mas yr$^{-1}$,
    while the ellipse in the right VPD defining probable field stars
    is centered at (9.5,14.0) mas yr$^{-1}$ with major and minor axes
    of 17.6 and 14.3 mas yr$^{-1}$, respectively. The ellipse mainly
    encloses stars in the outer part of the Bulge.}
  \label{fig:cmd2}
\end{figure*}

To compute the $\rm N_{fSGB} / (N_{\rm fSGB}+N_{\rm bSGB})$ ratio, we
rectified the SGBs using an approach similar to that described in
Milone et al. (2009). In this analysis, we used only cluster members
with good photometry. The results are shown in
Fig.~\ref{fig:tuttisgb}. We started by using HAWK-I data only. As
described by Milone et al., we need four points (P1b, P1f, P2b, and
P2f in Fig.~\ref{fig:tuttisgb}) to rectify the SGBs. Once selected
these points, we transformed the CMD into a new reference system in
which the points P1b, P1f, P2b, and P2f have coordinates (0,0), (1,0),
(0,1), and (1,1), respectively. We drew by hand a line to separate the
two sequences.  We derived a fiducial line for each SGB by dividing it
into bins of 0.12 `Ordinate' value and fitting the 3.5$\sigma$-clipped
median `Abscissa' and `Ordinate' for each of them with a spline. We
rectified the two sequences using the average of the two
fiducials. The rectification was performed by subtracting, from the
`Abscissa' of each star, the `Abscissa' of the fiducial line at the
same `Ordinate' level (panel (b)). In panels (c1) and (c2), we show
the resulting final $\Delta$`Abscissa' histogram (between
$\Delta$`Abscissa' $-0.19$ and $+0.19$ and `Ordinate' 0 and 1) for
stars between 1.5 and 3.0 arcmin (c1) and between 3.0 ad 9.0 arcmin
(c2) from the cluster center. The individual Gaussians for the bright
and the faint SGB are shown in blue and red, where the sum of the two
in black. After this complicated procedure, we were finally able to
estimate the fraction of stars belonging to the fSGB and bSGB. We used
binomial statistics to estimate the error $\sigma$ associated with the
fraction of stars. We defined $\sigma_{\hat{p}_{\rm fSGB}} =
\sqrt{\hat{p}_{\rm fSGB} (1-\hat{p}_{\rm fSGB}) / (N_{\rm fSGB}+N_{\rm
    bSGB})}$. \looseness=-2

As noted by Piotto et al. (2012), points P1b-P2b and P1f-P2f define a
mass interval for stars in the two SGB segments. If we want to
calculate the absolute value of the ratio $\rm \hat{p}_{\rm fSGB}$, we
need to make sure that the same mass interval is selected in the two
SGBs and at all radial distances.  Due to the lack of appropriate
isochrones for the HAWK-I data, this was not feasible.  Still, we can
estimate the radial trend of $\rm \hat{p}_{\rm fSGB}$ by taking
advantage of both \textit{HST} (for the inner region) and HAWK-I (for
the outer region) data by making sure that we use the same mass
interval for the bSGB (and the same mass interval for the fSGB) in
both data sets. \looseness=-2

For this reason, we cross-correlated our HAWK-I catalog with that of
Piotto et al. (2012). First, we selected the sample of SGBs stars in
the $B$ vs. ($B-K_{\rm S}$) CMD between the P1b, P1f, P2b, and P2f
points of panel (a) in Fig.~\ref{fig:tuttisgb}.  In the $m_{\rm
  F275W}$ vs. ($m_{\rm F275W}-m_{\rm F814W}$) CMD, we selected the
same stars.  In this CMD, we fixed four points that enclose these
stars, used them to rectify the SGBs, and then calculated the ratio by
following the same procedure as described for HAWK-I data (panels
(c3), (d), and (e) in Fig.~\ref{fig:tuttisgb}). We emphasize that
these intervals are not the same as the ones used in Piotto et
al. (2012), but we used approximately (within the uncertainties due to
difficulty to select the limiting points) the same mass intervals in
calculating the three SGB population ratios. \looseness=-2

The trend of the $\rm \hat{p}_{\rm fSGB}$ ratio is shown in
Fig.~\ref{fig:radial}. To give a more reliable estimate of the error
bars, we also included the histogram binning uncertainty (We computed
the ratio varying the starting point/bin width in the histogram and
estimate the $\sigma$ of these values.) and quadratically added them
to $\sigma_{\hat{p}_{\rm fSGB}}$. In any case, these error bars still
represent an underestimate of the total error because other sources of
uncertainty should to be taken in account (e.g., the uncertainty in
the location of the limiting points of the selected SGB segments).
The error bars are larger for the two HAWK-I points because of the
smaller number of objects in the sample. \looseness=-2

Our conclusion is that the radial trend of the two SGB populations
within the error bars is flat. \looseness=-2

%%%%%%%%%%%
\subsection{NGC~6121 proper motion}
%%%%%%%%%%%
\label{m4pm}

As before, we chose all WFI images with an exposure time of about 180
s and only used the chips overlapping HAWK-I data. In this way, we had
36 catalogs for the first epoch. For the second HAWK-I epoch, we had
400 catalogs (100 images $\times$ 4 chips). Using local
transformations, we iteratively computed the star's
displacements. \looseness=-2

The resulting CMD and VPD are shown in Fig.~\ref{fig:cmd2}. We only
show here the best-measured stars. Unfortunately, the stars fainter
than $J \sim$ 18.5 in the NIR catalogs were not detectable in the
optical-band WFI images. Even though the lower part of the CMD is
poorly sampled (near the MS kink at $K_{\rm S}$$\sim$17.5), the
separation between cluster and field objects is still good. Cluster
members are those within 4 mas yr$^{-1}$ of the cluster mean motion,
while we drew an ellipse centered at (9.5,14.0) mas yr$^{-1}$ with
major and minor axes of 17.6 and 14.3 mas yr$^{-1}$ for field stars
(right VPD), respectively. \looseness=-2

Unlike NGC~6656, we did not estimate the differential chromatic
refraction effects, since the color baseline is not large enough to
study the effect using only unsaturated stars. Saturated stars' proper
motions are less precise, and we could confuse differential chromatic
refraction with systematic trends in saturated stars' proper
motions. \looseness=-2

%%%%%%%%%%%
\subsection{Cluster membership probability}
%%%%%%%%%%%
\label{cmp_sec}

For the two globular clusters with new proper motions, NGC~6656 and
NGC~6121, we calculated cluster membership probability, $P_\mu$, for
each star. Recently, these two clusters have been analyzed by
Zloczewsky et al. (2012) but, instead of giving membership
probabilities, these authors simply divided all stars with measured
proper motions into field stars, possible cluster members, and likely
cluster members. This approach can be justified on the grounds of a
clear separation between field and cluster in the VPD
(Fig.~\ref{fig:cmd1},~\ref{fig:cmd2}). However, a more rigorous
cluster membership calculation technique would help to better
characterize each star's membership probability. We selected a
well-tested local sample method (e.g., van Altena 2013, Chapter
25). In this method, a limited subset of stars is selected for each
target star with properties close to those of a target. Then, a
cluster membership probability, $P_\mu$ of a star is calculated using
the density functions defined by the local sample. This approach
delivers more accurate membership probabilities over the entire range
of magnitudes.  In the case of globular clusters, the potential bias
in $P_\mu$ at various magnitudes is less significant because the
cluster stars dominate a relatively small number of field stars. In
the presence of a highly varying precision of calculated proper
motions (ranging from 0.2 to 5.5 mas yr$^{-1}$ for NGC~6656 and
NGC~6121), however, using an aggregate density function for a cluster
and field can produce unreliable membership probabilities for
low-precision proper motions. This is due to a significant widening of
cluster's density function at the low-precision end of proper
motions. Therefore, we adopted the mean error $\sigma_\mu$ of proper
motions as a single parameter allowing us to find a local sample,
which is similar to what was applied to the catalog of proper motions
in $\omega$~Cen (Paper~III). There are a few differences from the
study of $\omega$~Cen. First, we used a fixed window in the error
distributions with the total width not exceeding 0.75 mas yr$^{-1}$ so
that a target star is located in the middle of this window. The total
number of stars in local sample never exceeds 3000, hence the window
size for well-measured proper motions, which dominate the catalog, can
be as small as 0.1 mas yr$^{-1}$.  At the extreme values of
proper-motion errors, the window size is fixed and the placement of a
target may no longer be in the middle of this window. Second, we used
a modified mean $\sigma_\mu$= $\sqrt{(\mu_\alpha \cos
  \delta)^{2}+\mu_\delta^{2}} / \sqrt{2}$. Third, the Gaussian width
of a cluster density function was interpolated by using an empirical
relationship: $\sigma_{c}$=$(0.04\times(K_s - 12)+1)\times\sigma_\mu$,
where $K_{\rm S}$ is the measured near-infrared magnitude of a target
star.  In addition, $\sigma_{c}$ was never let to be lower than 0.7
mas yr$^{-1}$. \looseness=-2

While the cluster density function is always a 2-D Gaussian, it is
often convenient to use a flat sloping density function for the
field-star distribution in the VPD. This is related to the binning of
VPD. The adopted size of a binning area, centered on the cluster, is
5$\sigma_c\times$5$\sigma_c$ which formally should contain all cluster
members. If a star's $\sigma_\mu$$<$2 mas yr$^{-1}$, then it also
means that the binning area never reaches the center of a field-star
centroid in the VPD for both globular clusters. In the regime of high
proper motion errors ($>3$ mas yr$^{-1}$), the distribution of field
stars is so diffuse that a significant portion of its wings falls
outside the VPD area covered by proper motions. The few free
parameters of both cluster and field density distributions, $\Phi_c$
and $\Phi_f$, are calculated according to Kozhurina-Platais et
al. (1995) but the resulting cluster membership probability $P_\mu$ is
defined by Eq.~25.8 from van Altena (2013). \looseness=-2

We note that likely clusters stars have $P_\mu >$75\%, but almost
certain field stars have $P_\mu <$1\%. The stars with intermediate
membership probabilities are more likely to be cluster members than
field stars. This follows from the fact that the respective centroids
in the VPD have a significant separation and that there is a
relatively small fraction of field stars among all stars with measured
proper motions, namely $\sim$24\% for NGC~6656 and $\sim$9\% for
NGC~6121. This should be considered when examining the astrometric
cluster membership of rare stars, such as variables, blue stragglers,
and horizontal branch stars. \looseness=-2

%%%%%%%%%%%%%%
\subsubsection{Membership of variables in NGC~6656}
%%%%%%%%%%%%%%
\label{m22memb}

\begin{figure}[!t]
  \centering
  \includegraphics[width=\columnwidth]{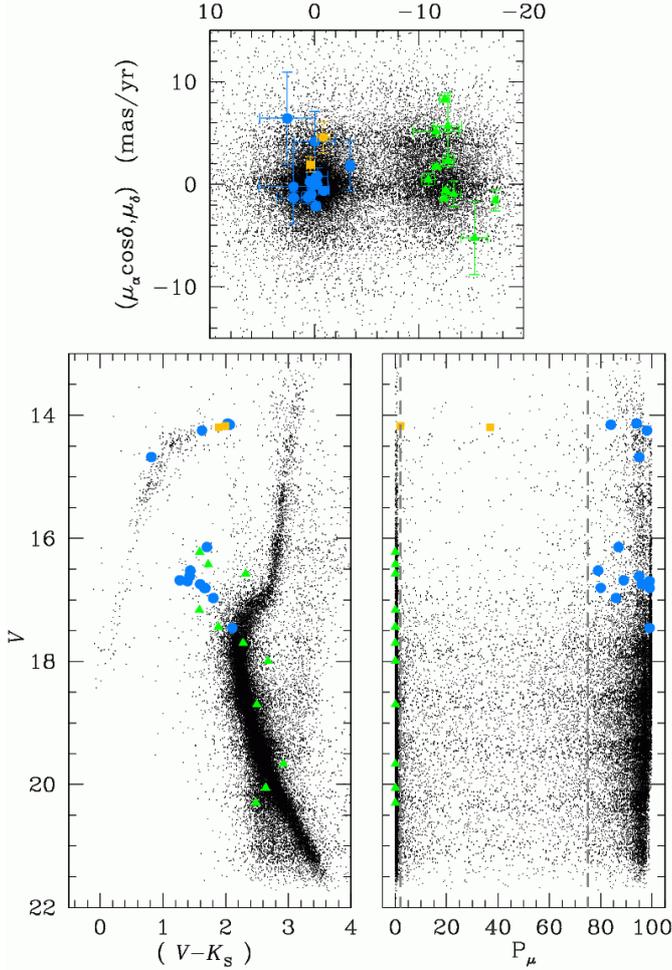}
  \caption{(\textit{Bottom-left panel}): NGC~6656 $V$ vs. ($V-K_{\rm
      S}$) CMD for all stars in our catalog (black dots) that have a
    membership probability measure. We plotted variable stars from
    Kaluzny \& Thompson (2001) with green triangles that are
    cross-identified in our catalog with P$_\mu < 2$\%; yellow squares
    represent the star with 2\% $\rm \le P_\mu < 75$\%. The azure
    circles are those stars with P$_\mu \ge
    75$\%. (\textit{Bottom-right panel}): $V$
    vs. P$_\mu$. (\textit{Top}): VPD. We also drew proper motion error
    bars for matched variable stars.}
  \label{fig:m22memb}
\end{figure}

Kaluzny \& Thompson (2001) published a catalog of 36 variable stars in
the central field of NGC~6656. We cross-identified these sources in
our catalog and found 27 stars in our proper motion catalog. In
Table~\ref{tab:m22memb}, we report the membership probability for
these stars (ID$_{\rm KT}$ are Kaluzny \& Thompson labels; ID$_{\rm
  L13}$ are the identification labels in our catalog.). In
Fig.~\ref{fig:m22memb}, we show $V$ vs. ($V-K_{\rm S}$) CMD
(bottom-left panel), $V$ vs. P$_\mu$ (bottom-right panel) and the VPD
(top panel) for all stars in our sample with a membership probability
measure. We set two thresholds in P$_\mu$ (P$_\mu = 2$\% and P$_\mu =
75$\%) and divided our catalog in three samples: likely-field stars
with P$_\mu < 2$\%, dubious membership stars with 2\% $\rm \le P_\mu <
75$\% and likely cluster members with P$_\mu \ge 75$\%. Eleven stars
have P$_\mu < 2$\% (green triangles); two stars have 2\% $\rm \le
P_\mu < 75$\% (yellow squares), and the remaining 14 stars have P$_\mu
\ge 75$\% (azure circles). The two variable stars with 2\% $\rm \le
P_\mu < 75$\%, namely M22\_V37 and M22\_V14, are saturated
horizontal-branch stars for which the motion is generally consistent
with the cluster's mean motion. As we stated in Sect.~\ref{cmp_sec},
they should be considered likely cluster members although their formal
membership probabilities are below 75\%. Note that not all the
saturated stars could have well-measured proper motion due to the
less-accurate PSF-fitting process and less-constrained positions (QFIT
could be higher than 0.1). Saturated stars (while they will have less
precise astrometry) are of intrinsic interest, since they are the best
candidates for follow-up spectroscopy if identified as cluster
members. Nonetheless, caution must be taken in interpreting their
astrometry. \looseness=-2

\begin{table}
  \caption{Membership probability for the NGC~6656 variable star
    catalog of Kaluzny \& Thompson (2001). ID$_{\rm KT}$ is the ID
    used in Kaluzny \& Thompson (2001), and ID$_{\rm L13}$ is in our
    catalog.}
  \label{tab:m22memb}     
  \centering
  \begin{tabular}{ccccccc}
    \hline
    \hline
    ID$_{\rm KT}$ & ID$_{\rm L13}$ & P$_{\mu}$ & & ID$_{\rm KT}$ & ID$_{\rm L13}$ & P$_{\mu}$ \\
    \hline
    & \\
    \multicolumn{7}{c}{\textbf{Members}} \\
    & \\
    M22\_V02 & 45566  & 99 & & M22\_V29 & 161746 & 96 \\ 
    M22\_V04 & 56333  & 99 & & M22\_V33 & 181768 & 86 \\ 
    M22\_V10 & 93320  & 89 & & M22\_V34 & 181631 & 80 \\ 
    M22\_V16 & 104996 & 98 & & M22\_V36 & 185551 & 84 \\
    M22\_V20 & 131248 & 99 & & M22\_V45 & 155768 & 79 \\
    M22\_V23 & 144995 & 95 & & M22\_V51 & 77720  & 95 \\
    M22\_V28 & 155692 & 87 & & M22\_V55 & 148685 & 94 \\
    & \\
    \multicolumn{7}{c}{\textbf{Probably Members}} \\
    & \\
    M22\_V14 & 102946 & 2 & & M22\_V37 & 193535 & 37 \\
    & \\
    \multicolumn{7}{c}{\textbf{Non Members}} \\
    & \\
    M22\_V03 & 47877  & 0 & & M22\_V18 & 113038 & 0 \\ 
    M22\_V05 & 57746  & 0 & & M22\_V42 & 79935  & 0 \\
    M22\_V07 & 68695  & 0 & & M22\_V46 & 157407 & 0 \\
    M22\_V08 & 74698  & 0 & & M22\_V48 & 183116 & 0 \\
    M22\_V12 & 101344 & 0 & & M22\_V54 & 138212 & 0 \\
    M22\_V15 & 106319 & 0 \\
    & \\
    \hline
  \end{tabular} 
\end{table}

%%%%%%%%%%%%%%
\subsubsection{Membership of variables in NGC~6121}
%%%%%%%%%%%%%%
\label{m4memb}

\begin{figure}[!t]
  \centering
  \includegraphics[width=\columnwidth]{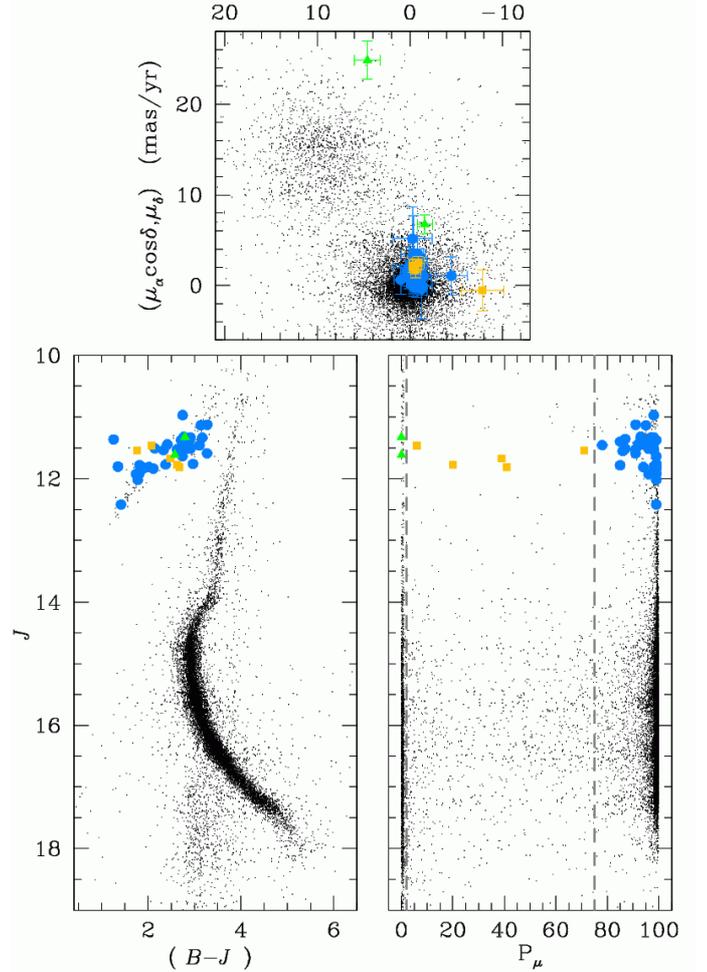}
  \caption{(\textit{Bottom-left panel}): NGC~6121 $J$ vs. ($B-J_{\rm
      S}$) CMD for all stars with a membership probability measure. As
    in Fig.~\ref{fig:m22memb}, all stars are shown with black dots. We
    used azure circles, yellow squares, and green triangles to
    highlight cross-identified stars in the Shokin \& Samus (1996)
    with P$_\mu \ge 75$\%, 2\% $\rm \le P_\mu < 75$\%, and P$_\mu <
    2$\%, respectively. (\textit{Bottom-right panel}): $J$
    vs. P$_\mu$. (\textit{Top}): VPD. We show proper motion error bars
    for Shokin \& Samus variable stars. Star V17 (P$_{\mu} = 0$\%) is
    not shown because it lies outside the VPD.}
  \label{fig:m4memb}
\end{figure}

We can similarly use our proper motion data to assign membership
probabilities to candidate variable star members of NGC~6121. Shokin
\& Samus (1996) cataloged 53 NGC~6121 variable stars from the
literature and provided equatorial coordinates. We cross-checked our
proper motion catalog with that provided by the authors, and we found
42 sources in common. Figure~\ref{fig:m4memb} shows these variable
stars in $J$ vs. ($B-J$) CMD and VPD. As for NGC~6656, we set two
thresholds at P$_\mu = 2$\% and P$_\mu = 75$\%, dividing our catalog
in three samples. The membership probabilities are listed in
Table~\ref{tab:m4memb} (ID$_{\rm S}$ and ID$_{\rm L13}$ are the labels
in Shokin \& Samus and in this paper, respectively.). All
cross-identified variable stars are saturated in our proper motion
catalog. Five stars (V6, V7, V11, V19, and V31) have 2\% $\rm \le
P_\mu < 75$\%.  For these stars, the same considerations we made for
M22\_V37 and M22\_V14 should be applied. All the known variable stars
in NGC~6121 seem to be horizontal-branch stars. As field contamination
is small in that portion of the CMD, their membership is strengthened
considering that these are poorly-measured saturated
stars. \looseness=-2

\begin{table}
  \caption{Membership probability for the NGC~6121 variable star
    catalog of Shokin \& Samus (1996). ID$_{\rm S}$ is the ID used in
    Shokin \& Samus (1996); ID$_{\rm L13}$ is in our catalog.}
  \label{tab:m4memb}   
  \centering
  \begin{tabular}{ccccccc}
    \hline
    \hline
    ID$_{\rm S}$ & ID$_{\rm L13}$ & P$_{\mu}$ & & ID$_{\rm S}$ & ID$_{\rm L13}$ & P$_{\mu}$ \\
    \hline
    & \\
    \multicolumn{7}{c}{\textbf{Members}} \\
    & \\  
    V1 & 176589 & 99  & & V28 &  30831 & 99 \\
    V2 & 173396 & 98  & & V30 &  16737 & 99 \\ 
    V5 & 164050 & 97  & & V36 (NE) & 167082 & 91 \\
    V8 & 148455 & 87  & & V36 (SW) & 167583 & 96 \\
    V9 & 146061 & 92  & & V37 & 125921 & 94 \\ 
    V10 & 136435 & 93 & & V38 & 119757 & 98 \\
    V12 & 129429 & 98 & & V39 & 111350 & 99 \\
    V14 & 127336 & 91 & & V41 &  87105 & 99 \\
    V15 & 123997 & 99 & & A381 &  98514 & 95 \\
    V16 & 121622 & 78 & & A382 &  96072 & 99 \\
    V18 & 111157 & 95 & & A505 &  98523 & 85 \\
    V20 & 107627 & 98 & & A519 & 112683 & 99 \\
    V22 & 100193 & 93 & & L1610 & 142443 & 87 \\    
    V24 &  94766 & 91 & & L1717 & 115979 & 96 \\
    V25 &  87432 & 97 & & L2630 & 155330 & 98 \\
    V26 &  76423 & 86 & & L3602 &  60119 & 99 \\
    V27 &  68948 & 97 & & L3732 &  97216 & 85 \\
    & \\
    \multicolumn{7}{c}{\textbf{Probably Members}} \\
    & \\
    V6 & 149664 & 39  & & V19 & 109197 & 20 \\
    V7 & 149106 &  6  & & V31 &  14721 & 71 \\
    V11 & 133287 & 41 & & \\
    & \\
    \multicolumn{7}{c}{\textbf{Non Members}} \\
    & \\
    V17 & 114412 &  0 & & A246 & 174902 &  0 \\
    V23 &  98450 &  0 & & \\      
    & \\
    \hline
  \end{tabular} 
\end{table}

%%%%%%%%
\section{Catalogs}
%%%%%%%%
\label{catalog}

We constructed eight different catalogs for our seven fields. We split
the Baade's Window field into two different catalogs (Bulge\#1 and the
rotated Bulge\#2). These catalogs are electronically available in the
VizieR on-line
database\footnote{{\url{http://vizier.u-strasbg.fr/viz-bin/VizieR}}.}. \looseness=-2

We also converted pixel-based coordinates into equatorial coordinates
using the UCAC\,4 catalog as reference. We only used bright,
unsaturated stars to compute the coefficients of the 6-parameter
linear transformation between HAWK-I and UCAC\,4 frames. The choice of
using 6-parameter linear transformations (see Sect.~\ref{acc} for a
description of these transformations) to transform star positions in
our catalogs into the UCAC~4 reference system allows us to solve not
only for shift, orientation, and scale, but it also minimizes most of
the telescope+optics-system residuals, as well as most of the
atmospheric refraction effects. Furthermore, linking our catalogs to
the UCAC~4 catalog, we did not only provide the equatorial coordinates
for all stars but we determined the linear terms of our
distortion. The equatorial coordinates are truly our best calibrated
coordinates. \looseness=-2

The first eight columns are the same in all catalogs. Column (1)
contains the ID of the star; columns (2) and (3) give J2000.0
equatorial coordinates in decimal degrees. Note that positions are
given at the epoch of HAWK-I observations because of proper
motion. Columns (4) and (5) contain the pixel coordinates x and y of
the distortion-corrected reference frame.  Columns (6) and (7) contain
the corresponding positional r.m.s.; column (8) gives the number of
images where the star was found. \looseness=-2

Columns (9) and the following columns contain the photometric data and
proper motions when present. Values are flagged to $-1$ when their
measures are not available with the only exception of the proper
motions and their errors, for which we set to 99. Table~\ref{tab:cat1}
lists all columns contained in the Bulge\#1
catalog. Tables~\ref{tab:cat2} to \ref{tab:cat5} only contain those
columns that are not in common. In the following subsections, we
describe more in detail the remaining columns (9+) of each
catalog. \looseness=-2

%%%%%%%%%%%
\paragraph{\bf Baade's Window (Bulge\#1).}
%%%%%%%%%%%

Columns (9) to (23) contain the photometric data: i.e., $J$, $H$,
$K_{\rm S}$ (both with 2MASS- and MKO-based zero-points added) , their
errors, the number of images used to compute the magnitude of the star
in the master frame, and the {\sf QFIT} in this order. Columns (24)
and (25) contain a flag to weed out PSF artifacts from the $J$ and $H$
filter (see Sect.~\ref{weed}). All stars in this catalog have a
measure of the magnitude in $K_{\rm S}$ filter. \looseness=-2

%%%%%%%%%%%
\paragraph{\bf NGC~6822, NGC~6388, LMC, and 47\,Tuc.}
%%%%%%%%%%%

Columns (9) to (18) contain the $J$ and $K_{\rm S}$ photometric data;
columns (19) and (20) contain the weed-out flags (see
Table~\ref{tab:cat2}). In the NGC~6822, LMC, and 47\,Tuc catalogs, all
stars have a $J$-magnitude measurement, and NGC~6388 has a $K_{\rm
  S}$-magnitude measurement. \looseness=-2

%%%%%%%%%%%
\paragraph{\bf Baade's Windows rotated by 135$^{\circ}$ (Bulge\#2).}
%%%%%%%%%%%

Columns (9) to (14) contain $K_{\rm S}$ magnitudes, errors, the number
of stars used to compute the average magnitudes, {\sf QFIT} and the
weed-out flag values, respectively
(Table~\ref{tab:cat3}). \looseness=-2

%%%%%%%%%%%
\paragraph{\bf NGC~6656.}
%%%%%%%%%%%

Columns (9) to (26) contain the photometric data in $K_{\rm S}$, $B$,
$V$, and $I$ band. Finally, Columns (27) to (31) contain the proper
motion data: $\rm \mu_{\alpha} cos \delta$ (27), $\rm
\sigma_{\mu_{\alpha} cos \delta}$ (28), $\rm \mu_\delta$ (29), $\rm
\sigma_{\mu_\delta}$ (30), and the number of pairs of images in which
a given star's proper motion was measured (31). Finally, column (32)
contains the membership probability (Table~\ref{tab:cat4}). Stars
measured in only one exposure in either $B$, $V$, or $I$ filters have
photometric r.m.s. values of 9.9. As for Bulge\#1, all stars have a
measure of the magnitude in $K_{\rm S}$ filter. \looseness=-2

%%%%%%%%%%%
\paragraph{\bf NGC~6121.}
%%%%%%%%%%%

Columns (9) to (32) contain the photometric data in $J$, $K_{\rm S}$,
$B$, $V$, and $Rc$ bands. Columns (33) to (37) contain the proper
motion data and column (38) is the membership probability
(Table~\ref{tab:cat5}). As in the NGC~6656 catalog, a r.m.s. equal to
9.9 is used for those stars measured in only 1 exposure in $K_{\rm
  S}$, $B$, $V$, or $Rc$ filter. In this catalog, all stars have a
$J$-magnitude measurement. \looseness=-2

%%%%%%%%
\section{Conclusions}
%%%%%%%%

We derived an accurate distortion solution in three broad band filters
for the HAWK-I detector and release the tools to correct the geometric
distortion with our solution. We also produced astro-photometric
catalogs of seven stellar fields. We release catalogs with astrometric
positions, photometry, proper motions, and membership probabilities of
NGC~6121 (M~4) and NGC~6656 (M22), while the remaining fields (the
Baade's Window, NGC~6822, NGC~6388, NGC~104, and the James Webb Space
Telescope calibration field) studied in the present paper only
contains astrometry and photometry. These catalogs are useful for
selecting spectroscopic targets, and can serve as distortion-free
frames with respect to which one can solve for the geometric
distortion of present/future imagers. The astronomical community has
started to focus its attention on wide-field cameras equipped with NIR
detectors, and the quantity and quality of NIR devices have improved
considerably. This is a first effort to develop the expertise with
these detectors to fully exploit the data coming from large-field NIR
surveys, such as the VVV survey taken with VIRCAM@VISTA. Finally, an
additional goal of this work is to get ready for the upcoming James
Webb Space Telescope, whose imagers define the state-of-the-art in
astrometry, in particular in crowded environments not reachable by
GAIA. \looseness=-2

We analyzed both photometric and astrometric performance of the NIR
mosaic HAWK-I@VLT using images of seven different fields observed
during commissioning in 2007. We computed a geometric-distortion
solution for each chip of HAWK-I in three different broad band filters
($J$, $H$, $K_{\rm S}$). Our dithered-observation strategy using the
self-calibration technique allowed us to randomize the systematic
errors and to compute the average stars' positions that provide an
approximation of the true positions in the distortion-free master
frame. A fifth-order polynomial solution highlighted a periodic
pattern in the distortion residuals. We have demonstrated that this
pattern is not a geometric effect (as it is the case for the WFPC2 or
the WFC3/UVIS@\textit{HST}) but it is a periodic lag introduced by
alternating readout amplifiers. To remove it, we used a square-wave
function and a 64-pixel step table of residuals. Finally we used four
additional look-up tables (one per chip) to perform a bi-linear
interpolation to take all uncorrected residuals into account and to
further improve our solutions. Thanks to our 5-step distortion
correction, we are able to reach a positional r.m.s. of $\sim$3.5 mas
in each coordinate. Using a general 6-parameter linear transformations
to match-up different images, the effects due to
telescope$+$instrument and atmosphere are absorbed, and the
$\sigma$(Radial residual) further decreases, reaching $\sim$2.8 mas
under good seeing conditions. We emphasize that this is a relative
positioning precision -- i.e., it indicates how accurately we can
measure the differential position of a star in multiple images of the
same field. \looseness=-2

We have also shown that the non-linear terms of our distortion
solution can be transferred between observing runs at the 10 mas
level. The astrometric accuracy contained in the pixel-coordinate
system degrades moving toward the edges of the FoV because the stars'
positions were obtained as the average of fewer images than in the
center of the field. Therefore, the average positions are more
vulnerable to poorly-constrained transformations of the individual
exposures into the master frame. The accuracy can decrease from
$\sim$10 mas to $\sim$100 mas ($\sim$1 pixel) going from the center to
the edges of the FoV. For this reason, we advise to use the inner part
of the detector for high-precision astrometry. To achieve a higher
astrometric accuracy, we also advise to link the pixel-coordinate
catalog to a reference frame such as UCAC~4 to determine the linear
terms of the distortion. Finally local transformations (as those used
to compute the proper motion in Sect.~\ref{app}) should be used to
minimize the effects of residuals in the geometric distortion
corrections as described in Paper~I. \looseness=-2

In the second part of the paper, we showed the potential applications
of our astrometric techniques and computed the \textit{relative}
proper motion of stars in the field of the globular clusters NGC~6656
and NGC~6121. With a time baseline of about 8 years, we have clearly
separated cluster members from field stars. Accuracy of proper-motion
measurements is limited by the depth and the precision of first-epoch
data set. We note that the stellar positions in our catalogs have been
derived from only a single epoch of HAWK-I data. A second-epoch HAWK-I
(or another wide-field infrared camera) data set is needed to provide
proper-motion solutions that allow these data to be extended with
confidence to arbitrary future epochs. We exploit photometry and
proper motions of stars in NGC\,6656 to study its stellar
populations. We find that the bimodal SGB, previously discovered from
visual and ultraviolet \textit{HST} photometry (Piotto et al. 2012),
is also visible in the $K_{\rm S}$ versus ($B-K_{\rm S}$) CMD. We
combined information from HAWK-I observation of the outer part of
NGC\,6656 and from \textit{HST} images of the innermost cluster region
(Piotto et al. 2012) to study the radial distribution of the two SGBs.
To do this, we calculated the number ratio of the faint SGB $\rm
\hat{p}_{\rm fSGB}$ for stars at different radial distances from the
cluster center to 9$\arcmin$ ($\sim$6.8 core radii). We found that the
two SGBs have the same radial distribution within our
uncertainty. \looseness=-2

\begin{acknowledgements}

ML and GP acknowledge partial support by the Universit$\rm\grave{a}$
degli Studi di Padova CPDA101477 grant. ML acknowledges support by the
STScI under the 2013 DDRF program. APM acknowledges the financial
support from the Australian Research Council through Discovery Project
grant DP120100475. We thank Dr. Jay Anderson for careful reading of
the manuscript and for thoughtful comments. We thank the anonymous
referee for the useful comments and suggestions that considerably
improved the quality of our paper. \looseness=-2

\end{acknowledgements}

\begin{table*}
  \caption{Bulge\#1 catalog.}
  \label{tab:cat1}    
  \centering
  \begin{tabular}{ccc}
    \hline
    \hline
    Column \# & Name & Description \\
    \hline
    & & \\
    (1) & ID & ID number of the star \\
    (2) & $\alpha$ & Right Ascension $[^\circ]$ \\
    (3) & $\delta$ & Declination $[^\circ]$ \\
    (4) & x & x-master frame position [pixel] \\
    (5) & y & y-master frame position [pixel] \\
    (6) & $\rm \sigma_x$ & r.m.s. error in the x-position [pixel] \\
    (7) & $\rm \sigma_y$ & r.m.s. error in the y-position [pixel] \\
    (8) & n$\rm _{pos}$ & Number of images where the star was found in used to compute the master-frame position \\
    (9) & $J$ & Calibrated $J$ magnitude in 2MASS system \\
    (10) & $H$ & Calibrated $H$ magnitude in 2MASS system \\
    (11) & $K_{\rm S}$ & Calibrated $K_{\rm S}$ magnitude in 2MASS system \\
    (12) & $J$ & Calibrated $J$ magnitude in MKO system \\
    (13) & $H$ & Calibrated $H$ magnitude in MKO system \\
    (14) & $K_{\rm S}$ & Calibrated $K_{\rm S}$ magnitude in MKO system \\
    (15) & $\sigma_J$ & r.m.s. error in $J$ photometry \\
    (16) & $\sigma_H$ & r.m.s. error in $H$ photometry \\
    (17) & $\sigma_{K_{\rm S}}$ & r.m.s. error in $K_{\rm S}$ photometry \\
    (18) & n$_J$ & Number of images where the star was found in used to compute the $J$ magnitude \\
    (19) & n$_H$ & Number of images where the star was found in used to compute the $H$ magnitude \\
    (20) & n$_{K_{\rm S}}$ & Number of images where the star was found in used to compute the $K_{\rm S}$ magnitude \\
    (21) & {\sf QFIT}$_J$ & Quality of $J$ PSF-fit \\
    (22) & {\sf QFIT}$_H$ & Quality of $H$ PSF-fit \\
    (23) & {\sf QFIT}$_{K_{\rm S}}$ & Quality of $K_{\rm S}$ PSF-fit \\
    (24) & weed$_J$ & $J$ weed-out flag ($1 =$ star, $0 =$ PSF-artifact, $-1 =$ star not found in $J$ exposures) \\
    (25) & weed$_H$ & $H$ weed-out flag \\
    & & \\
    \hline
  \end{tabular} 
\end{table*} 

~\\ 

\begin{table*}
  \caption{NGC~6822, NGC~6388, LMC and 47\,Tuc catalogs.}
  \label{tab:cat2}     
  \centering
  \begin{tabular}{ccc}
    \hline
    \hline
    Column \# & Name & Description \\
    \hline
    & & \\
    (...) & (...) & (...) \\
    (9) & $J$ & Calibrated $J$ magnitude in 2MASS system \\
    (10) & $K_{\rm S}$ & Calibrated $K_{\rm S}$ magnitude in 2MASS system  \\
    (11) & $J$ & Calibrated $J$ magnitude in MKO system \\
    (12) & $K_{\rm S}$ & Calibrated $K_{\rm S}$ magnitude in MKO system \\
    (13) & $\sigma_J$ & r.m.s. error in $J$ photometry \\
    (14) & $\sigma_{K_{\rm S}}$ & r.m.s. error in $K_{\rm S}$ photometry \\
    (15) & n$_J$ & Number of images where the star was found in used to compute the $J$ magnitude \\
    (16) & n$_{K_{\rm S}}$ & Number of images where the star was found in used to compute the $K_{\rm S}$ magnitude \\
    (17) & {\sf QFIT}$_J$ & Quality of $J$ PSF-fit \\
    (18) & {\sf QFIT}$_{K_{\rm S}}$ & Quality of $K_{\rm S}$ PSF-fit \\
    (19) & weed$_J$ & $J$ weed-out flag ($1 =$ star, $0 =$ PSF-artifact, $-1 =$ star not found in $J$ exposures) \\
    (20) & weed$_{K_{\rm S}}$ & $K_{\rm S}$ weed-out flag \\
    & & \\
    \hline
  \end{tabular} 
\end{table*}

\begin{table*}
  \caption{Bulge\#2 catalog.}
  \label{tab:cat3}     
  \centering
  \begin{tabular}{ccc}
    \hline
    \hline
    Column \# & Name & Description \\
    \hline
    & & \\
    (...) & (...) & (...) \\
    (9) & $K_{\rm S}$ & Calibrated $K_{\rm S}$ magnitude in 2MASS system \\
    (10) & $K_{\rm S}$ & Calibrated $K_{\rm S}$ magnitude in MKO system \\
    (11) & $\sigma_{K_{\rm S}}$ & r.m.s. error in $K_{\rm S}$ photometry \\
    (12) & n$_{K_{\rm S}}$ & Number of images where the star was found in used to compute the $K_{\rm S}$ magnitude \\
    (13) & {\sf QFIT}$_{K_{\rm S}}$ & Quality of $K_{\rm S}$ PSF-fit \\
    (14) & weed$_{K_{\rm S}}$ & $K_{\rm S}$ weed-out flag  ($1 =$ star, $0 =$ PSF-artifact, $-1 =$ star not found in $K_{\rm S}$ exposures) \\
    & & \\
    \hline
  \end{tabular} 
\end{table*}

\begin{table*}
  \caption{NGC~6656 catalog.}
  \label{tab:cat4}     
  \centering
  \begin{tabular}{ccc}
    \hline
    \hline
    Column \# & Name & Description \\
    \hline
    & & \\
    (...) & (...) & (...) \\
    (9) & $K_{\rm S}$ & Calibrated $K_{\rm S}$ magnitude in 2MASS system \\
    (10) & $K_{\rm S}$ & Calibrated $K_{\rm S}$ magnitude in MKO system \\
    (11) & $B$ & Calibrated $B$ magnitude \\
    (12) & $V$ & Calibrated $V$ magnitude \\
    (13) & $I$ & Calibrated $I$ magnitude \\
    (14) & $\sigma_{K_{\rm S}}$ & r.m.s. error in $K_{\rm S}$ photometry \\
    (15) & $\sigma_B$ & r.m.s. error in $B$ photometry \\
    (16) & $\sigma_V$ & r.m.s. error in $V$ photometry \\
    (17) & $\sigma_I$ & r.m.s. error in $I$ photometry \\
    (18) & n$_{K_{\rm S}}$ & Number of images where the star was found in used to compute the $K_{\rm S}$ magnitude \\
    (19) & n$_B$ & Number of images where the star was found in used to compute the $B$ magnitude \\
    (20) & n$_V$ & Number of images where the star was found in used to compute the $V$ magnitude \\
    (21) & n$_I$ & Number of images where the star was found in used to compute the $I$ magnitude \\
    (22) & {\sf QFIT}$_{K_{\rm S}}$ & Quality of $K_{\rm S}$ PSF-fit \\
    (23) & {\sf QFIT}$_B$ & Quality of $B$ PSF-fit \\
    (24) & {\sf QFIT}$_V$ & Quality of $V$ PSF-fit \\
    (25) & {\sf QFIT}$_I$ & Quality of $I$ PSF-fit \\
    (26) & weed$_{K_{\rm S}}$ & $K_{\rm S}$ weed-out flag ($1 =$ star, $0 =$ PSF-artifact, $-1 =$ star not found in $K_{\rm S}$ exposures) \\
    (27) & $\mu_\alpha \cos \delta$ & Proper-motion value along $\mu_\alpha \cos \delta$ [mas yr$^{-1}$] \\
    (28) & $\sigma_{\mu_\alpha \cos \delta}$ & r.m.s. of $\mu_\alpha \cos \delta$ [mas yr$^{-1}$] \\
    (29) & $\mu_\delta$ & Proper-motion value along $\mu_\alpha \cos \delta$ [mas yr$^{-1}$] \\
    (30) & $\sigma_{\mu_\delta}$ & r.m.s. of $\mu_\delta$ [mas yr$^{-1}$] \\
    (31) & n$_{\rm pairs}$ & Number of pairs of first-second epoch images used to compute the proper motion of the star \\
    (32) & P$_\mu$ & Membership probability \\
    & & \\
    \hline
  \end{tabular} 
\end{table*}

~\\

\begin{table*}
  \caption{NGC~6121 catalog.}
  \label{tab:cat5}     
  \centering
  \begin{tabular}{ccc}
    \hline
    \hline
    Column \# & Name & Description \\
    \hline
    & & \\
    (...) & (...) & (...) \\
    (9) & $J$ & Calibrated $J$ magnitude in 2MASS system \\
    (10) & $K_{\rm S}$ & Calibrated $K_{\rm S}$ magnitude in 2MASS system \\
    (11) & $J$ & Calibrated $J$ magnitude in MKO system \\
    (12) & $K_{\rm S}$ & Calibrated $K_{\rm S}$ magnitude in MKO system \\
    (13) & $B$ & Calibrated $B$ magnitude \\
    (14) & $V$ & Calibrated $V$ magnitude \\
    (15) & $Rc$ & Calibrated $Rc$ magnitude \\
    (16) & $\sigma_J$ & r.m.s. error in $J$ photometry \\
    (17) & $\sigma_{K_{\rm S}}$ & r.m.s. error in $K_{\rm S}$ photometry \\
    (18) & $\sigma_B$ & r.m.s. error in $B$ photometry \\
    (19) & $\sigma_V$ & r.m.s. error in $V$ photometry \\
    (20) & $\sigma_{Rc}$ & r.m.s. error in $Rc$ photometry \\
    (21) & n$_J$ & Number of images where the star was found in used to compute the $J$ magnitude \\
    (22) & n$_{K_{\rm S}}$ & Number of images where the star was found in used to compute the $K_{\rm S}$ magnitude \\
    (23) & n$_B$ & Number of images where the star was found in used to compute the $B$ magnitude \\
    (24) & n$_V$ & Number of images where the star was found in used to compute the $V$ magnitude \\
    (25) & n$_{Rc}$ & Number of images where the star was found in used to compute the $Rc$ magnitude \\
    (26) & {\sf QFIT}$_J$ & Quality of $J$ PSF-fit \\
    (27) & {\sf QFIT}$_{K_{\rm S}}$ & Quality of $K_{\rm S}$ PSF-fit \\
    (28) & {\sf QFIT}$_B$ & Quality of $B$ PSF-fit \\
    (29) & {\sf QFIT}$_V$ & Quality of $V$ PSF-fit \\
    (30) & {\sf QFIT}$_{Rc}$ & Quality of $Rc$ PSF-fit \\
    (31) & weed$_J$ & $J$ weed-out flag ($1 =$ star, $0 =$ PSF-artifact, $-1 =$ star not found in $J$ exposures) \\
    (32) & weed$_{K_{\rm S}}$ & $K_{\rm S}$ weed-out flag \\
    (33) & $\mu_\alpha \cos \delta$ & Proper-motion value along $\mu_\alpha \cos \delta$ [mas yr$^{-1}$] \\
    (34) & $\sigma_{\mu_\alpha \cos \delta}$ & r.m.s. of $\mu_\alpha \cos \delta$ [mas yr$^{-1}$] \\
    (35) & $\mu_\delta$ & Proper-motion value along $\mu_\alpha \cos \delta$ [mas yr$^{-1}$] \\
    (36) & $\sigma_{\mu_\delta}$ & r.m.s. of $\mu_\delta$ [mas yr$^{-1}$] \\
    (37) & n$_{\rm pairs}$ & Number of pairs of first-second epoch images used to compute the proper motion of the star \\
    (38) & P$_\mu$ & Membership probability \\
    & & \\
    \hline
  \end{tabular} 
\end{table*}

\clearpage
\clearpage

\begin{appendix}

%%%%%%%%
\section{Study of the HAWK-I PSF}
%%%%%%%%
\label{appendixA}

%%%%%%%%%%%
\subsection{PSF spatial variability}
%%%%%%%%%%% 

\begin{figure}[t!]
  \centering
  \includegraphics[height=4.cm]{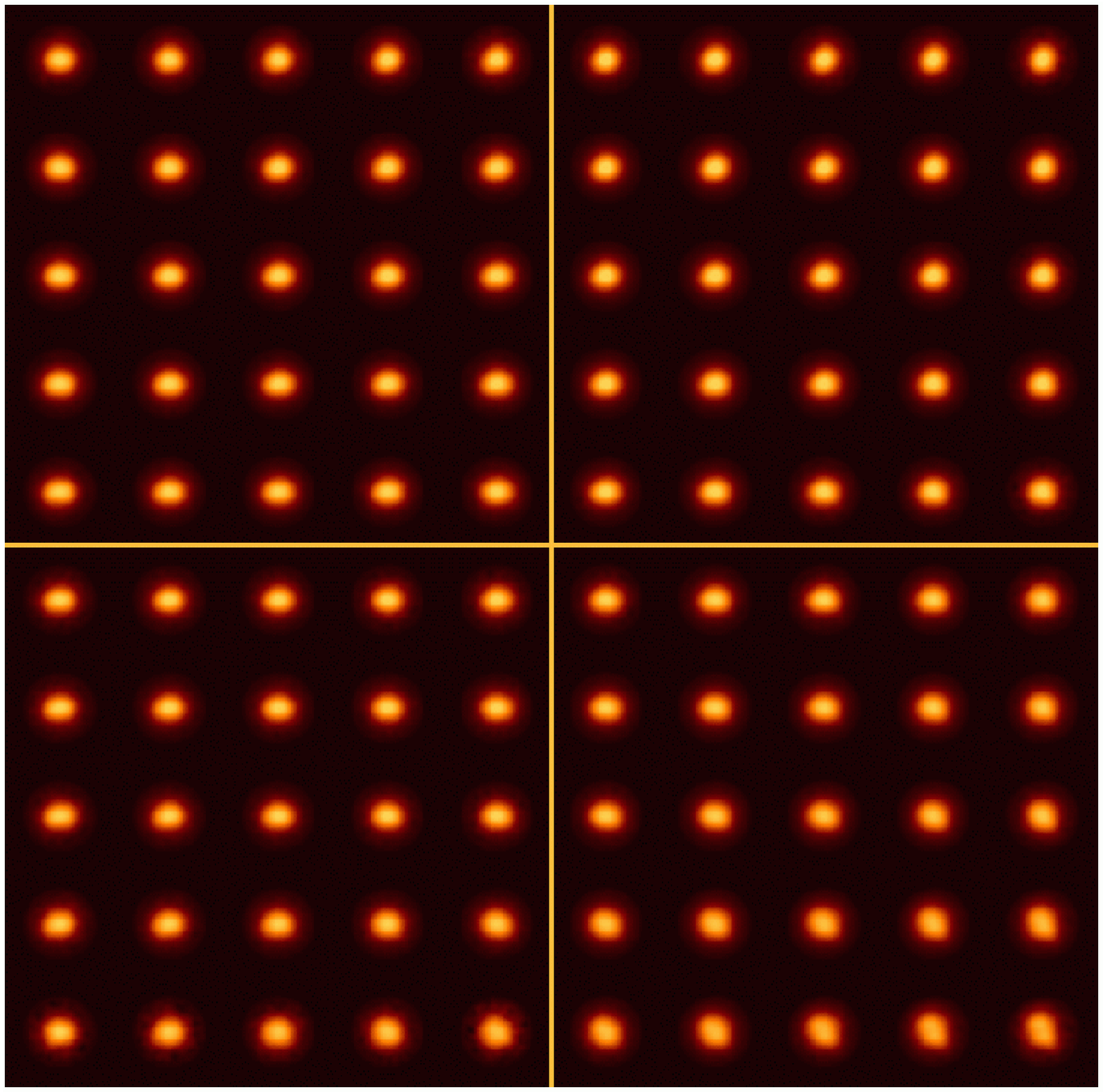}
  $\phantom{a}$\includegraphics[height=4.cm]{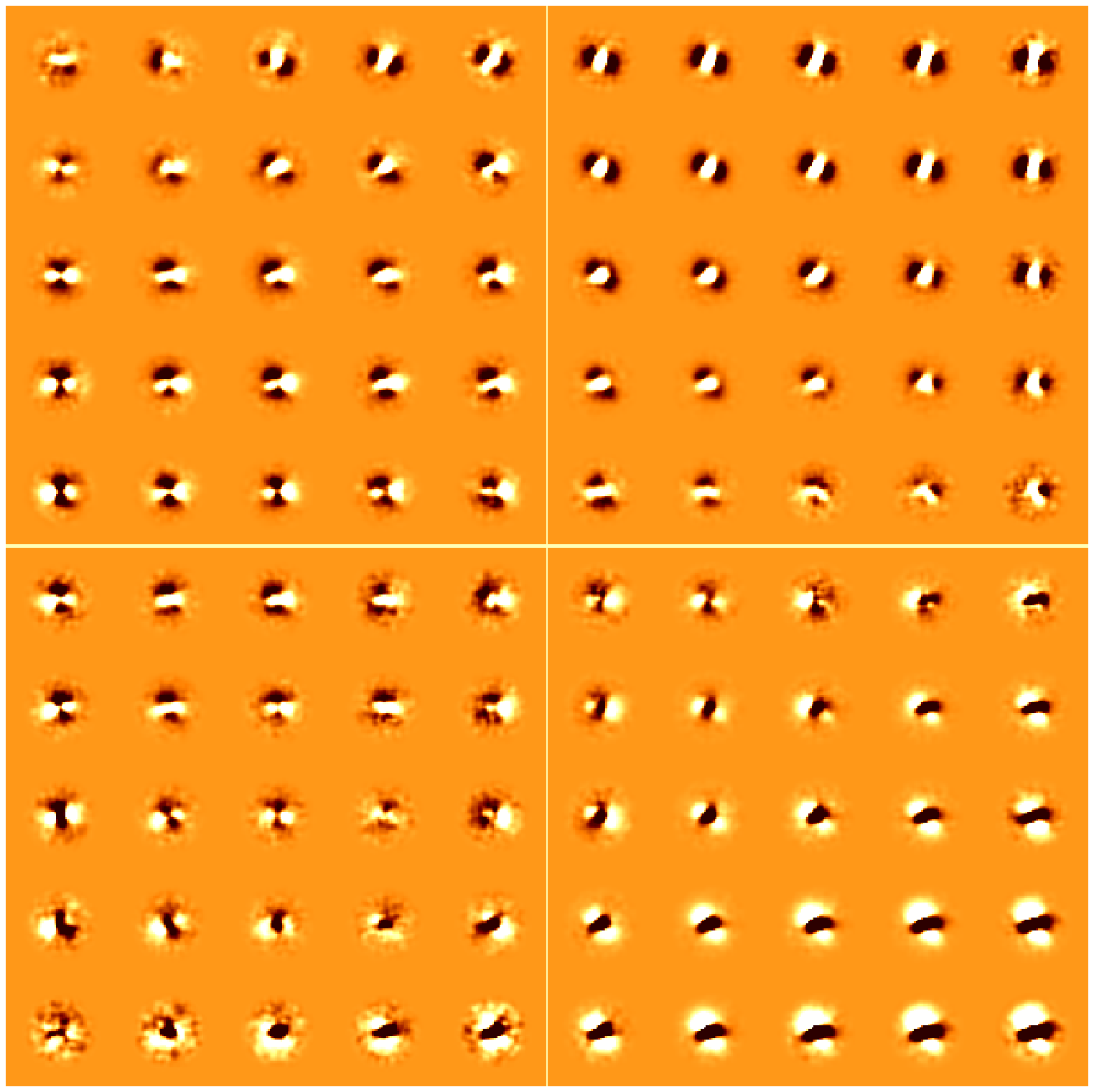}
  \caption{\textit{Left}: 10$\times$10 PSFs for the whole HAWK-I
    detector (5$\times$5 PSFs per chip). \textit{Right}: Spatial
    variation of the PSFs. To each local PSF, we subtracted a single
    average PSF for the whole detector. A darker color means less flux
    than the average PSF, and a lighter color means more flux.}
  \label{fig:psf1}
\end{figure}

As in the case of the WFI@2.2\,m (Paper~I), the PSF shape for the
HAWK-I@VLT detector is different from one chip to the other and from
side to side within the same chip. To fully take this spatial
variation into account, we decided to solve for an array of 25 PSFs
per chip (5 across and 5 high). A bi-linear interpolation is used to
derive the proper PSF model in each location of the detector (see
Paper~I). The left panel of Fig.~\ref{fig:psf1} shows these $5\times
5$ PSFs for the whole HAWK-I detector. At a first glance, we can
already see that the PSFs' shape and orientation vary across the
detector (especially close to the edge). To better quantify the size
of these variations, we built a single, average PSF for the whole
detector, and we subtracted it to each of the local PSFs. The result
is shown on the right panel of the Fig.~\ref{fig:psf1}. The actual
spatial variation of the PSF is indeed large, even between two
adjacent PSFs. The darker color of Fig.~\ref{fig:psf1} means less flux
than the average PSF, while a lighter one means more flux. The maximum
variation of the local PSF with respect to the average one is
0.02\%. \looseness=-2

In sparse fields (or short exposures), there might not be enough
bright and isolated stars to build the full $5\times 5 \times 4 = 100$
set of PSFs. We customized the software to allow for different
PSF-array solutions:\ (1) one PSF per chip; (2) four PSFs per chip (at
the corners); (3) six PSFs per chip, (4) nine PSFs per chip; and (5)
the full 5$\times$5 array. The user can choose to include fainter and
more crowded stars to model the PSF to increase the statistics of each
PSF model. According to the crowding of the field of interest and the
image quality, the user must determine the best compromise between how
finely to model the PSF's spatial variability and the need to have an
adequate number of stars to model each PSF.  Choosing the best
solution is a delicate matter. To obtain the best results, we
investigated for every single exposure whether it was better to have
more or less PSFs by analyzing the trend of the {\sf QFIT} across the
image: better PSF models provide smaller {\sf QFIT} values.
\looseness=-2

\begin{figure}[t!]
  \centering
  \includegraphics[width=0.98\columnwidth]{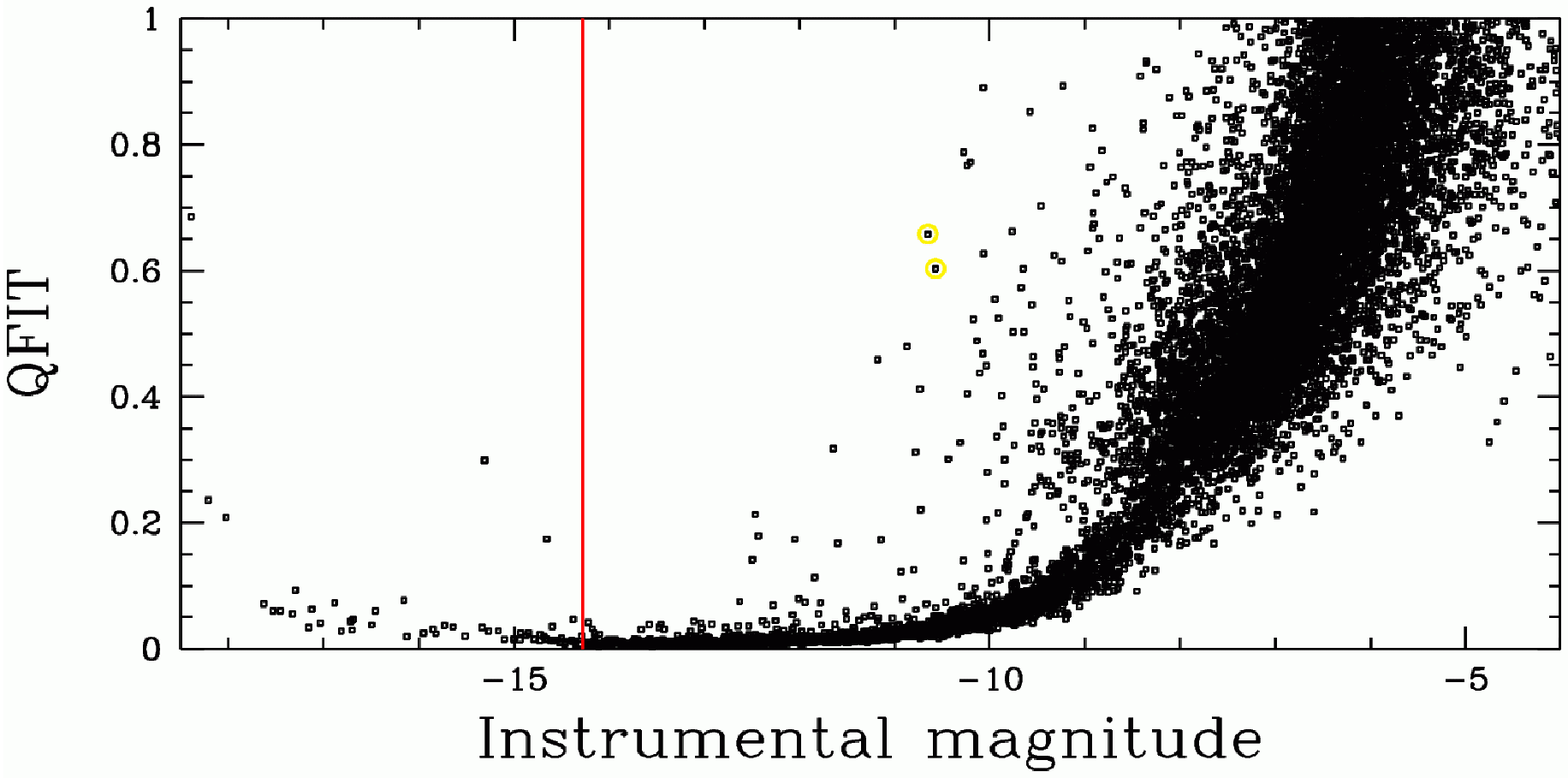}\\
  \includegraphics[width=\columnwidth]{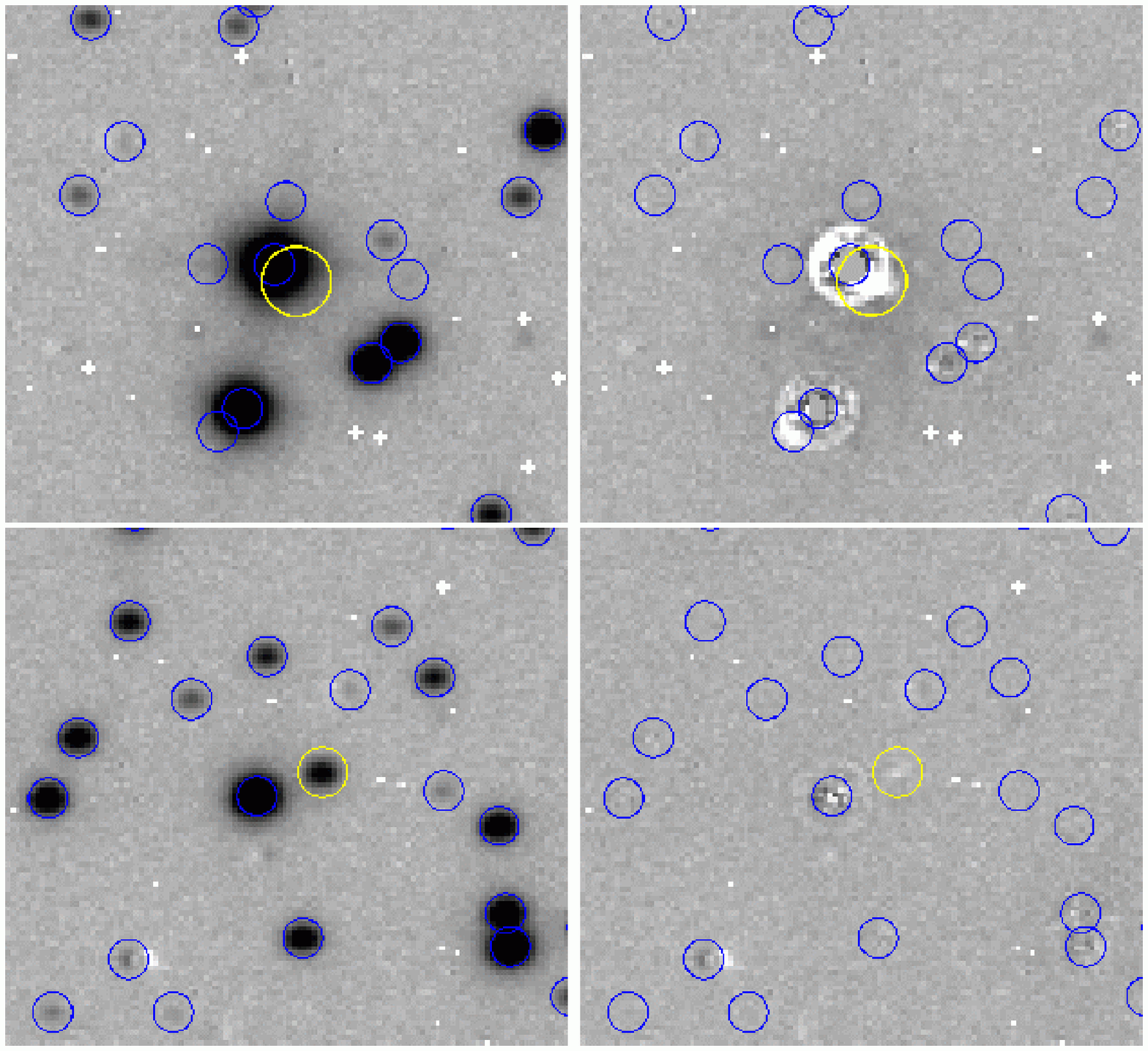}\\
  \caption{\textit{Top}: {\sf QFIT} parameter as a function of the
    instrumental magnitude $J$ for chip[1] of exposure
    HAWKI.2007-08-03T01:41:29.785.fits. In yellow, we highlighted two
    sources with anomalously high {\sf QFIT}. The red line shows the
    saturation limit. \textit{Middle/Bottom-left}: Location of the two
    sources (in yellow) on the image. Blue circles mark all stars for
    which we were able to measure a position and a flux. White pixels
    are those flagged according to the bad-pixel mask. We do not find
    stars too close to these bad pixels. \textit{Middle/Bottom-right}:
    The corresponding subtracted images. The first of the two
    high-{\sf QFIT} stars (middle-right panel) we selected is in close
    proximity to a saturated star; it has been poorly measured and
    therefore has a large {\sf QFIT} value. The second star
    (bottom-right panel) has a cosmic ray event close to its center,
    increasing its total apparent flux and shifting its center on the
    image. This star has been over-subtracted, resulting again in a
    larger {\sf QFIT} value.}
  \label{fig:psf2}
\end{figure}

The top panel of Fig.~\ref{fig:psf2} shows {\sf QFIT} values as a
function of the instrumental magnitude for chip[1] of exposure
HAWKI.2007-08-03T01:41:29.785.fits. This is an image in the field of
M~4 taken through $J$ filter. The instrumental magnitude is defined as
$-2.5\times \log (\sum \rm{counts})$, where $\sum\rm{counts}$ is the
sum of the total counts under the fitted PSF. The red line in the
figure indicates when stars start to be saturated\footnote{The maximum
  central pixel value of the PSFs for this exposure is 0.058 (i.e.,
  5.8\% of the star's flux falls within its central pixel). We set
  saturation to take place at 30$\,$000 counts, which means at
  instrumental magnitude $-2.5\times
  \log(30000/0.058)\simeq-14.28$.}. \looseness=-2

For well-exposed stars (e.g., with instrumental magnitude $J$ between
$-14$ and $-10$), {\sf QFIT} values are typically below 0.05 and
increase for fainter or saturated stars. However, there are a few
sources with anomalously high {\sf QFIT} in this interval. To find out
what kind of outliers these sources are, we selected two of them
(highlighted in yellow in Fig.~\ref{fig:psf2}). Their location on the
image is shown in the bottom-left panels of Fig.~\ref{fig:psf2} in
yellow. Blue circles mark all stars for which we were able to measure
a position and a flux. White pixels are those flagged using the
bad-pixel mask. We did not find stars too close to these bad pixels.
Bottom-right panels in the figure show the corresponding subtracted
images. We can see that our PSF-fitting procedure is able to leave
very small residuals in the subtracted image, except for saturated
stars.  The first of the two high-{\sf QFIT} stars we selected (see
top-right panel) is in close proximity of a saturated star; it has
been poorly measured because of the light contamination from the
neighboring star, and therefore has a large {\sf QFIT} value. The
second star has a cosmic ray event close to its center, increasing its
total apparent flux and shifting its center on the image. This star
has been over-subtracted (see bottom-right panel), resulting again in
a larger {\sf QFIT} value. \looseness=-2

To better quantify how our PSF models adequately represent star
profiles across the detector, we perform the following test. A total
of 100 exposures in the field of M~4 were taken between August 3 and
5, 2007 in four runs of 25 images each. We used here only the first 25
images taken consecutively on August 3, 2007 during a time span of
about 34 minutes.  We derived an array of $5\times5$ PSFs for each
chip of these images. Then, we selected all the $k$ bright,
unsaturated stars (instrumental magnitude $J<-10$), which have no
brighter neighbors within 10 pixels, and we measured positions and
fluxes for them using our $5\times 5$ PSF arrays. Usually, there are
over 1000 such stars per chip in our exposures, uniformly distributed
over the detector. \looseness=-2

If our PSFs are well characterized, we should be able to obtain
subtracted images where removed stars leave nearly no flux
residuals. Therefore, the size of these residuals tell us how much our
PSF models differ from real star profiles.  \looseness=-2

We extracted $11\times 11$ pixel rasters around each star $k$ (i.e.,
$\pm$ 5 pixels from its center) in each exposure for a total of 121
${\rm P}_{i,j}$ pixel values per star. We subtracted the local sky
value to all of them, which is computed as the 2$\sigma$-clipped
median value of the counts in an annulus between 8 and 12 from the
star's center. This is the net star's flux at any given location on
the raster. The fractional star's flux is obtained by dividing these
values by the total star's flux $z$. Besides Poisson errors, these
values should reflect what our PSF models predict for those pixels
($\psi_{i,j}$), so that we should always have in principle:
\looseness=-2

\begin{displaymath}
\frac{{\rm P}^k_{i,j}-{\rm sky}^k}{z^k}-\psi^k_{i,j}=0.
\end{displaymath}

\begin{figure}[t!]
  \centering
  \includegraphics[width=\columnwidth]{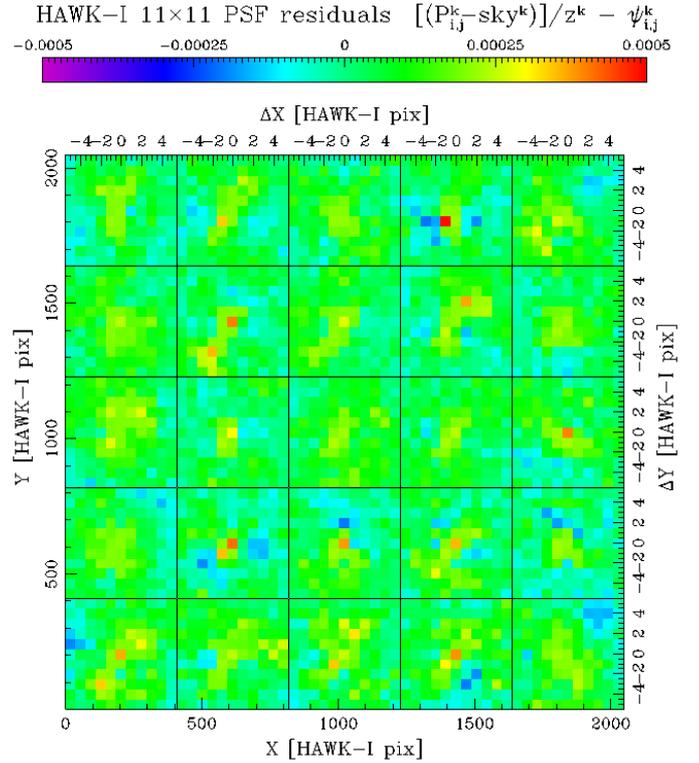}
  \caption{PSF spatial variability for chip[1]. Pixel values are
    color-coded as shown on top.}
  \label{fig:psf3}
\end{figure}

\noindent Deviations of these values from zero tell us how much our
PSFs over- or underestimate the true star's profile.  Results of this
test are reported in Fig.~\ref{fig:psf3} for chip[1].  We divided the
$2048\times 2048$ pixels of the chip into $5\times5$ sub-regions with
one for each PSF we built. Within each area, we computed the
$3\sigma$-median values of the residuals for each pixel of the
raster. Pixel values are color-coded as shown on top of
Fig.~\ref{fig:psf3}. From the Fig.~\ref{fig:psf3}, we can easily see
that PSF residuals are in general smaller than 0.05\% even in the
central pixel (where Poisson noise is most effective). This proves
that our spatial-dependent PSF models are able to adequately represent
a star profile at any given location of the chip. \looseness=-2

%%%%%%%%%%%
\subsection{PSF time variability}
%%%%%%%%%%% 

\begin{figure*}[t!]
  \centering
  \includegraphics[width=0.98\textwidth]{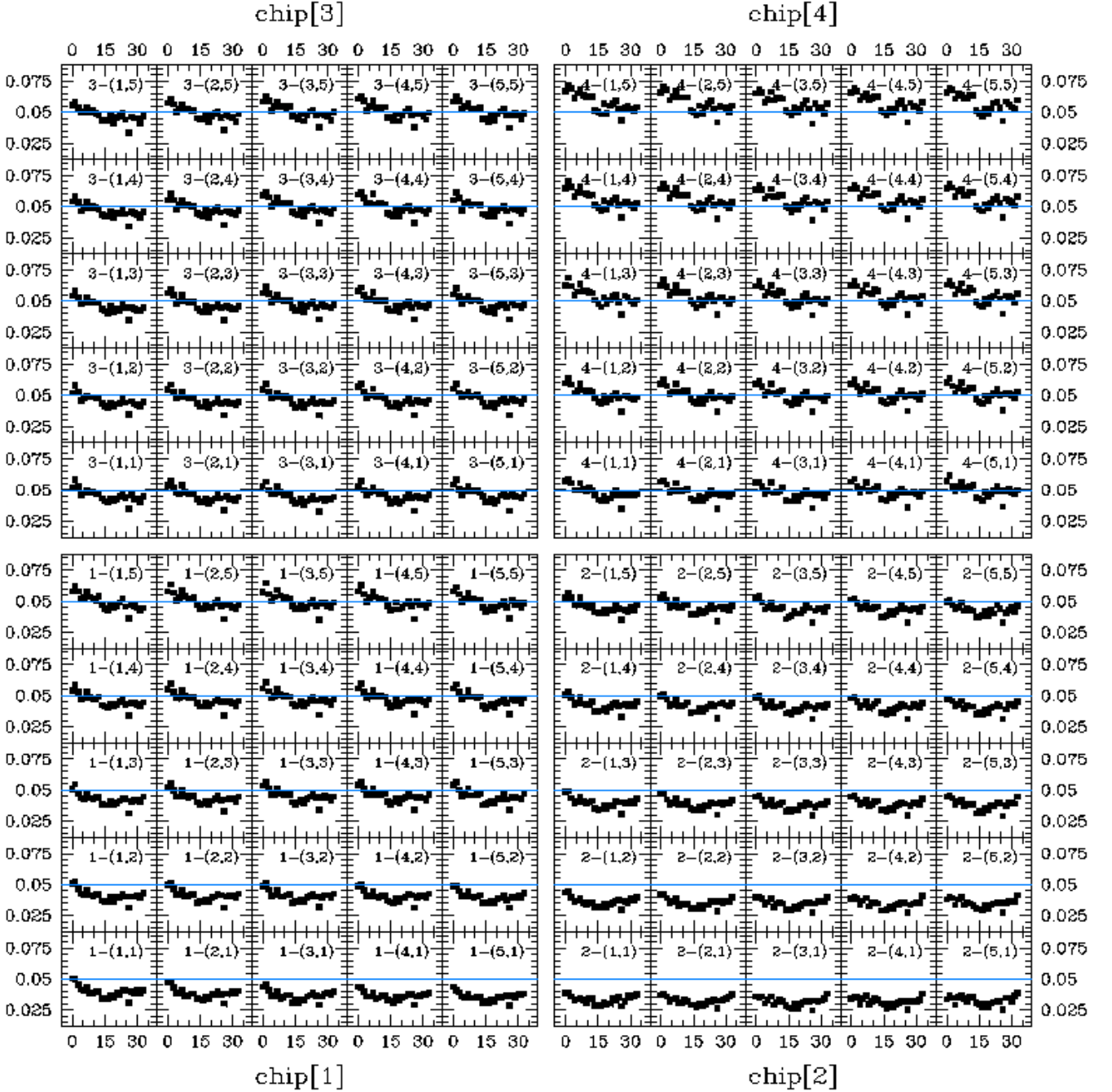}
  \caption{Central PSF values as function of the exposure sequence. We
    highlighted the central PSF value of 0.05 in blue.}
  \label{fig:psf4}
\end{figure*}

Ground-based telescopes suffer from varying seeing and airmass
conditions, telescope flexures, and changes in focus. These are all
effects that may severely alter the shape of the PSF.
Figure~\ref{fig:psf4} illustrates how much the seeing can actually
affect the PSFs. In the figure, we show the first 25 exposures of M\,4
that were consecutively taken on August 3, 2007. The total time
baseline is 34 minutes. For each of our PSF models, we considered the
value of its central pixel as a function of the exposure sequence,
starting from the first exposure. On average, chip[4] PSFs are
sharper, while chip[2] stars have the least amount of flux in their
central pixels. As a reference, we highlight a central PSF value of
0.05 (i.e., 5\% of the total star's flux in its center pixel) in
blue. Within the same exposures, central PSF values can range from
0.03 to 0.07 (see also Fig.~\ref{fig:psf1} for the PSF to PSF
variation). \looseness=-2

In a time span as short as half an hour, we can already see some
interesting PSF time-variability effects. First of all, central PSF
values vary in an inhomogeneous way across the detector. For instance,
there are specific locations on the detector (e.g., the top PSFs of
chip[4]) where central PSF values can change by up to 40\%. On the
other hand, central PSF values are more stable in different locations
(e.g., the bottom PSFs of chip[2]). \looseness=-2

Moreover, while for some PSFs (e.g., the one labeled as 4-(1,5) on
chip[4]), we have a general decrease of the central values. For other
PSFs (e.g., 1-(2,1) on chip[1]), we have a decrease of the central
values during the first 15 minutes, followed by an increase
afterwards. [Note that no focus adjustments have been made during this
  34 minutes.] \looseness=-2

Figure~\ref{fig:psf4} clearly shows that there are large variations in
the PSF shape even from one exposure to the other, and this variation
is not constant across the field. \textit{HST}'s PSFs are very stable
over time with variations on the order of at most a few percent,
mostly due to the so-called telescope breathing\footnote{\textit{HST}
  focus is known to experience variations on the orbital time scale,
  which are attributed to thermal contraction/expansion of the
  \textit{HST} optical telescope assembly as the the telescope warms
  up during its orbital day and cools down during orbital night.}. For
\textit{HST}, one spatially-constant perturbation PSF for is generally
enough to take into account this effect (e.g., ACS/WFC PSFs, Anderson
\& King 2006). To achieve high-precision astrometry and photometry
with the HAWK-I camera, we have to derive a specific set of PSFs for
each individual exposure.  \looseness=-2

\begin{figure}[t!]
  \centering
  \includegraphics[width=\columnwidth]{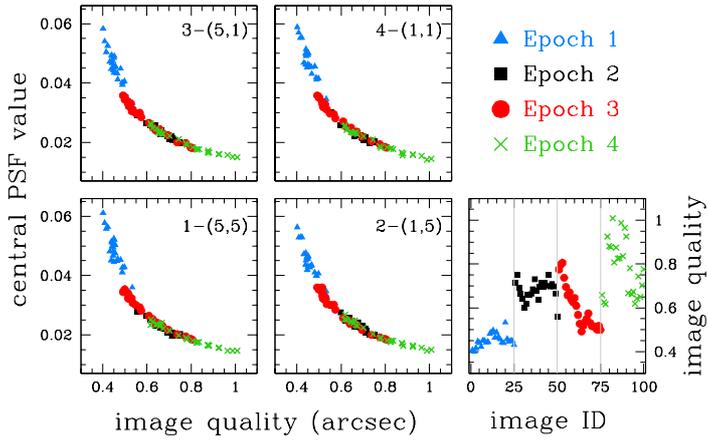}
  \caption{The central PSF value of these PSFs as a function of the
    image quality.}
  \label{fig:psf5}
\end{figure}

\begin{figure}[t!]
  \centering
  \includegraphics[width=\columnwidth]{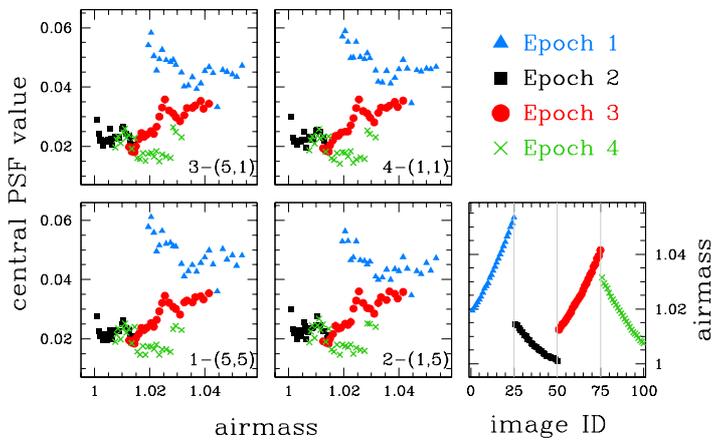}
  \caption{Same as Fig.~\ref{fig:psf5} but for airmass variations.}
  \label{fig:psf6}
\end{figure}

\noindent To further infer the effects of time variation on our PSF
models, we performed the following additional analysis using the 100
images in the field of M~4. As already mentioned, these images were
taken in blocks of 25 consecutive exposures in four different runs,
spanning three nights. Because each observing run lasted about 30
minutes, we can safely assume that focus variations have played a
little role in changing the shape of PSFs, if compared to airmass and
seeing variations. Seeing should actually be the most important factor
in changing the PSF shape from one exposure to the next one. We
focused on the centermost four PSFs, namely:\ 1-(5,5); 2-(1,5);
3-(5,1); and 4-(1,1), following labels of Fig.~\ref{fig:psf4}. In
Fig.~\ref{fig:psf5}, we plot the central value of these PSFs as a
function of the image quality (i.e., the average stars' FWHM as
measured directly on the exposures). Different observing runs are
marked with different colors and symbols. On the bottom right panel of
the figure, we plot the variation in the image quality during the four
runs. Here, we want to emphasize that these variations occurred within
30 minutes within the same run. As we expected, there is a strong
correlation between our PSF shapes and image quality. The correlation
between PSF shapes and airmass is shown in
Fig.~\ref{fig:psf6}. Airmass variations seem to play a secondary role
in changing PSF shape with respect to image quality. \looseness=-2

\end{appendix}

\clearpage

\begin{appendix}
\section{Geometric distortion: 2-D maps and size of the corrections}
\label{appendixB}

\begin{table*}[!h]
  \caption{Size of the $J$ corrections. All the values are given in pixel.}
  \label{tab:jcor}
  \centering 
  \begin{tabular}{c|cccc|cc|cc|cccc}
    \hline\hline
    \textbf{Chip} & \multicolumn{2}{|c}{\textbf{X-axis}} & \multicolumn{2}{c|}{\textbf{Y-axis}} & \multicolumn{2}{c}{\textbf{X-axis}} & \multicolumn{2}{|c}{\textbf{X-axis}} & \multicolumn{2}{|c}{\textbf{X-axis}} & \multicolumn{2}{c}{\textbf{Y-axis}} \\
    & Min & Max & Min & Max & Min & Max & Min & Max & Min & Max & Min & Max \\
    \hline  
    & & & & & & & & & & & \\
    & \multicolumn{4}{c|}{\textbf{P Correction}} & \multicolumn{2}{c|}{\textbf{S Correction}} & \multicolumn{2}{c|}{\textbf{FS Correction}} & \multicolumn{4}{c}{\textbf{TP Correction}} \\
    & & & & & & & & & & & \\
    \hline 
    & & & & & & & & & & & \\
    1 & $-$2.4974 & 0.0001 & $-$3.4467 & 0.7815 & $-$0.0354 & 0.0294 & $-$0.0922 & 0.0828 & $-$0.3698 & 0.1842 & $-$0.0638 & 0.1439 \\
    2 & $-$0.2483 & 2.7147 & $-$3.7959 & 0.8621 & $-$0.0354 & 0.0294 & $-$0.0221 & 0.0391 & $-$0.0903 & 0.0602 & $-$0.0847 & 0.0694 \\
    3 & $-$2.2199 & 0.0001 & $-$0.9715 & 3.2343 & $-$0.0354 & 0.0294 & $-$0.0197 & 0.0288 & $-$0.0784 & 0.0979 & $-$0.0713 & 0.0974 \\
    4 & $-$0.2678 & 2.8462 & $-$0.7153 & 3.8896 & $-$0.0354 & 0.0294 & $-$0.0323 & 0.0266 & $-$0.0929 & 0.1281 & $-$0.0803 & 0.0751 \\
    & & & & & & & & & & & \\
    \hline
  \end{tabular} 
\end{table*} 

\begin{table*}[t]
  \caption{As in Table~\ref{tab:jcor} but for the $H$ corrections (in pixel).}
  \label{tab:hcor}
  \centering 
  \begin{tabular}{c|cccc|cc|cc|cccc}
    \hline\hline
    \textbf{Chip} & \multicolumn{2}{|c}{\textbf{X-axis}} & \multicolumn{2}{c|}{\textbf{Y-axis}} & \multicolumn{2}{c}{\textbf{X-axis}} & \multicolumn{2}{|c}{\textbf{X-axis}} & \multicolumn{2}{|c}{\textbf{X-axis}} & \multicolumn{2}{c}{\textbf{Y-axis}} \\
    & Min & Max & Min & Max & Min & Max & Min & Max & Min & Max & Min & Max \\
    \hline  
    & & & & & & & & & & & \\
    & \multicolumn{4}{c|}{\textbf{P Correction}} & \multicolumn{2}{c|}{\textbf{S Correction}} & \multicolumn{2}{c|}{\textbf{FS Correction}} & \multicolumn{4}{c}{\textbf{TP Correction}} \\
    & & & & & & & & & & & \\
    \hline 
    & & & & & & & & & & & \\
    1 & $-$2.2681 & 0.0001 & $-$2.7934 & 0.6161 & $-$0.0307 & 0.0308 & $-$0.0365 & 0.0547 & $-$0.0833 & 0.0643 & $-$0.0543 & 0.0505 \\
    2 & $-$0.0449 & 2.6406 & $-$4.0058 & 0.9549 & $-$0.0307 & 0.0308 & $-$0.0188 & 0.0245 & $-$0.0575 & 0.0497 & $-$0.0473 & 0.0506 \\
    3 & $-$2.2937 & 0.0001 & $-$1.3461 & 3.4982 & $-$0.0307 & 0.0308 & $-$0.0328 & 0.0257 & $-$0.0768 & 0.0920 & $-$0.0439 & 0.0624 \\
    4 & $-$0.3316 & 2.8668 & $-$0.4570 & 3.5777 & $-$0.0307 & 0.0308 & $-$0.0298 & 0.0395 & $-$0.1149 & 0.1084 & $-$0.0507 & 0.0538 \\
   & & & & & & & & & & & \\
    \hline
  \end{tabular} 
\end{table*} 

\begin{table*}[t]
  \caption{As above but for the $K_{\rm S}$ corrections (in pixel).}
  \label{tab:kcor}
  \centering
  \begin{tabular}{c|cccc|cc|cc|cccc}
    \hline\hline
    \textbf{Chip} & \multicolumn{2}{|c}{\textbf{X-axis}} & \multicolumn{2}{c|}{\textbf{Y-axis}} & \multicolumn{2}{c}{\textbf{X-axis}} & \multicolumn{2}{|c}{\textbf{X-axis}} & \multicolumn{2}{|c}{\textbf{X-axis}} & \multicolumn{2}{c}{\textbf{Y-axis}} \\
    & Min & Max & Min & Max & Min & Max & Min & Max & Min & Max & Min & Max \\
    \hline  
    & & & & & & & & & & & \\
    & \multicolumn{4}{c|}{\textbf{P Correction}} & \multicolumn{2}{c|}{\textbf{S Correction}} & \multicolumn{2}{c|}{\textbf{FS Correction}} & \multicolumn{4}{c}{\textbf{TP Correction}} \\
    & & & & & & & & & & & \\
    \hline 
    & & & & & & & & & & & \\
    1 & $-$2.3696 & 0.1417 & $-$3.4477 & 0.7932 & $-$0.0245 & 0.0275 & $-$0.0661 & 0.0678 & $-$0.2406 & 0.1095 & $-$0.0763 & 0.0632 \\
    2 & $-$0.1348 & 2.7460 & $-$3.7555 & 0.8715 & $-$0.0245 & 0.0275 & $-$0.0146 & 0.0256 & $-$0.0636 & 0.0557 & $-$0.0340 & 0.0403 \\
    3 & $-$2.1681 & 0.0001 & $-$1.0793 & 3.3528 & $-$0.0245 & 0.0275 & $-$0.0272 & 0.0212 & $-$0.0797 & 0.0799 & $-$0.0419 & 0.0320 \\
    4 & $-$0.1457 & 2.8005 & $-$0.5338 & 3.4126 & $-$0.0245 & 0.0275 & $-$0.0152 & 0.0144 & $-$0.0836 & 0.0532 & $-$0.0573 & 0.0375 \\
    & & & & & & & & & & & \\
    \hline
  \end{tabular} 
\end{table*} 

\begin{figure*}[t!]
  \centering
  \includegraphics[width=0.8\textwidth]{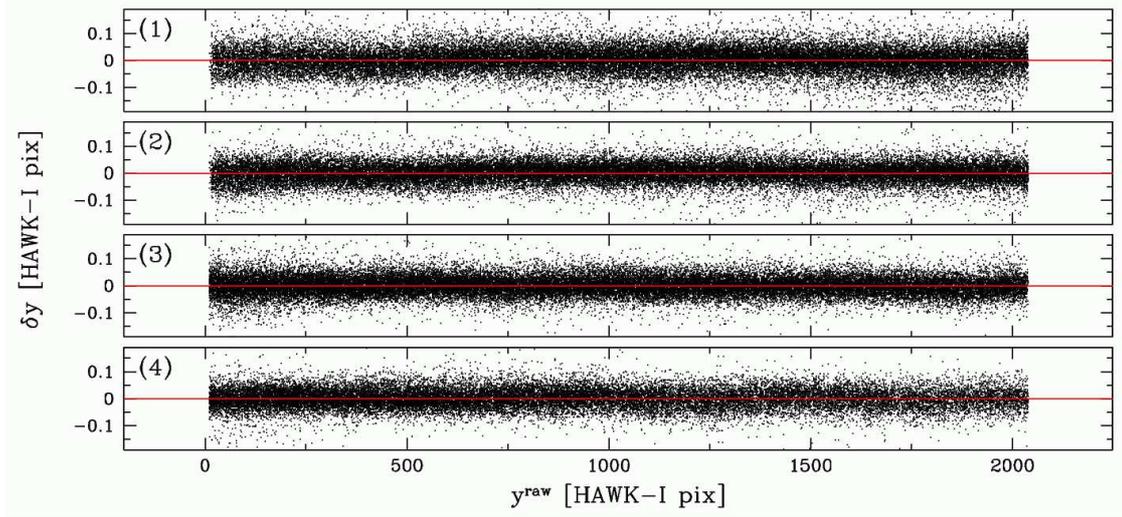}
  \caption{$\delta y$ as function of Y in units of HAWK-I pixels for all chips. The red lines is set at 0 HAWK-I pixel.}
  \label{fig:ytrend}
\end{figure*}

\begin{figure*}[t!]
  \centering
  \includegraphics[width=0.7\textwidth]{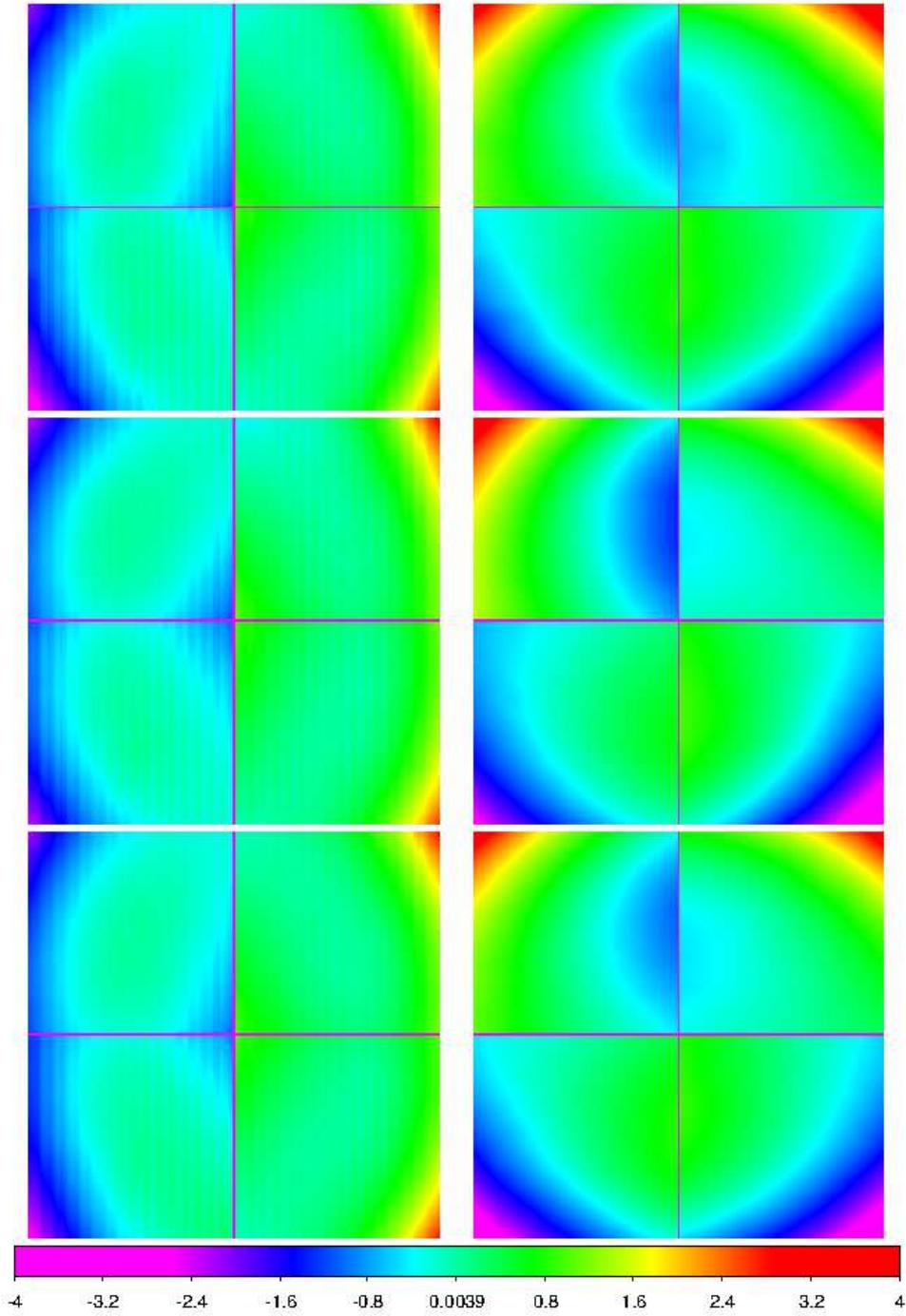}
  \caption{Maps of the corrections. From top to bottom, $J$-, $H$-,
    and $K_{\rm S}$-filter corrections. For each filter, the four
    chips on the left show the X-correction, while the four chips on
    the right show the Y-correction. A linear scale is used. Red means
    positive corrections; purple are negative corrections. The values
    in the color bar are expressed in pixels. For the $J$ filter, the
    x-corrections varies between $-2.95$ and 2.96 pixels across the
    whole detector, while the y-corrections between $-3.88$ and 3.95
    pixels. For the $H$ filter, the minimum and maximum corrections
    for the x-coordinate are $-2.40$ and 2.99 pixels, while the
    corrections for the y-coordinate are $-4.00$ and 3.59 pixels. The
    minimum and maximum x-corrections for the $K_{\rm S}$-filter
    solution are $-2.67$ and 2.85 pixels; for y-corrections, the
    minimum and maximum are $-3.77$ and 3.42 pixels.}
  \label{fig:jhkcorr}
\end{figure*}

\begin{figure*}[t!]
  \centering
  \includegraphics[width=0.4\textwidth]{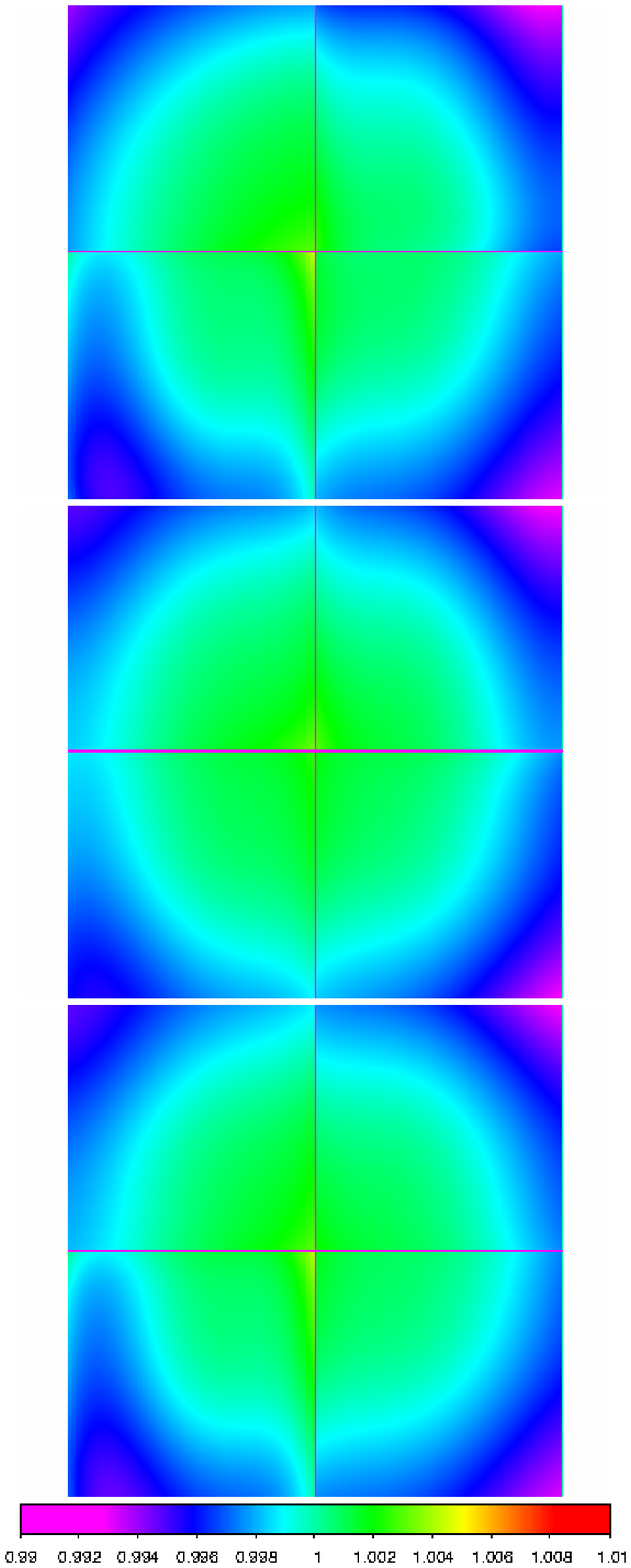}
  \caption{Maps of the pixel area corrections. From top to bottom,
    $J$-, $H$-, and $K_{\rm S}$-filter corrections. The values in the
    color bar represent the corrected area of the pixels. Before the
    correction, all pixels have an area of 1 pixel$^2$. The scale in
    the images is linear.}
  \label{fig:pixarea}
\end{figure*}

In this appendix, we report the size of the distortion corrections
released in this paper. In Table~\ref{tab:jcor}, \ref{tab:hcor}, and
\ref{tab:kcor}, we show the minimum and maximum values of each
correction in both coordinates for all chips. The largest correction
is applied with the P corrections, which decrease from the corner to
the center of the detector. The S and FS corrections are only applied
to the x-coordinates.  Figure~\ref{fig:ytrend} demonstrates there is
no $\delta y$ periodic pattern, so an FS correction along this axis is
not necessary. The residual distortion is corrected with the TP
correction. While the S correction is the same for all chips and has
only two values, the FS correction changes from chip to chip and
varies across the same chip. \looseness=-2

In Fig.~\ref{fig:jhkcorr}, we show a 2-D map of the correction for
each chip/filter. In the four left boxes of each row, we plot the
correction of each chip for the x-coordinates in the right boxes for
the y-coordinates. The polynomial correction creates the radial
pattern that changes from the corner to the center in all chips. The S
correction is visible only in the left boxes (x-coordinate
corrections) and creates a striped pattern. \looseness=-2

Another important correction that we are going to release (as FITS
images) is the correction for the pixel area variation across the
detector. This is a useful tool for improving HAWK-I photometry. On
average the size of the pixel area varies up to 0.7\% across the
detector. This value is reached at a point close to the edge and to
the center of the detector. We only applied the polynomial correction,
since it gives the maximum correction. Note that the corrections of
the periodic-lag effect should not be included in the pixel area
correction. The periodic lag is due to charges left in the amplifiers,
so the area of the pixel itself is not modified on sky. This is
different to what happens with the optics+filters distortion. In
Fig.~\ref{fig:pixarea}, we show three maps of the correction with one
for each HAWK-I filter. \looseness=-2

\end{appendix}


\begin{thebibliography}{}

\bibitem{AK99} Anderson, J., \& King, I.~R.\ 1999, \pasp, 111, 1095 

\bibitem{AK03} Anderson, J., \& King, I.~R.\ 2003, \pasp, 115, 113

\bibitem{AK04} Anderson, J., \& King, I.~R.\ 2004, Instrument Science Report ACS 2004-15, 51 pages, 3 

\bibitem{Ande06} Anderson, J., Bedin, L.~R., Piotto, G., Yadav, R.~S., \& Bellini, A.\ 2006, \aap, 454, 1029 , Paper~I

\bibitem{AK06} Anderson, J., \& King, I.~R.\ 2006, Instrument Science Report ACS 2006-01, 34 pages, 1 

\bibitem{Ande08} Anderson, J., King, I.~R., Richer, H.~B., et al.\ 2008, \aj, 135, 2114 

\bibitem{Arse06} Arsenault, R., Hubin, N., Stroebele, S., et al.\ 2006, The Messenger, 123, 6 

\bibitem{Bedi03} Bedin, L.~R., Piotto, G., King, I.~R., \& Anderson, J.\ 2003, \aj, 126, 247 

\bibitem{Bell09} Bellini, A., Piotto, G., Bedin, L.~R., et al.\ 2009, \aap, 493, 959 , Paper~III

\bibitem{BelBed09} Bellini, A., \& Bedin, L.~R.\ 2009, \pasp, 121, 1419

\bibitem{BelBed10} Bellini, A., \& Bedin, L.~R.\ 2010, \aap, 517, A34 , Paper~IV

\bibitem{Bell11} Bellini, A., Anderson, J., \& Bedin, L.~R.\ 2011, PASP, 123, 622 

\bibitem{Goli12} Golimowski, D., Suchkov, A., Loose, M., Anderson, J., \& Grogin, N.\ 2012, Instrument Science Report ACS 2012-02, 17 pages, 2 

\bibitem{Hadj12} Hadjiyska, E., Rabinowitz, D., Baltay, C., et al.\ 2012, IAU Symposium, 285, 324 

\bibitem{Han89} Han, I.\ 1989, \aj, 97, 607 

\bibitem{Har96} Harris, W.~E.\ 1996, VizieR Online Data Catalog, 7195, 0

\bibitem{Kais10} Kaiser, N., Burgett, W., Chambers, K., et al.\ 2010, \procspie, 7733,  

\bibitem{Kalu01} Kaluzny, J., \& Thompson, I.~B.\ 2001, \aap, 373, 899 

\bibitem{Kis08} Kissler-Patig, M., Pirard, J.-F., Casali, M., et al.\ 2008, \aap, 491, 941 

\bibitem{Kozhu95} Kozhurina-Platais, V., Girard, T.~M., Platais, I., et al.\ 1995, \aj, 109, 672 

\bibitem{Lin80} Lindegren, L.\ 1980, \aap, 89, 41 

\bibitem{Mad09} Madore, B.~F., Rigby, J., Freedman, W.~L., et al.\ 2009, \apj, 693, 936 

\bibitem{Milo09} Milone, A.~P., Stetson, P.~B., Piotto, G., et al.\ 2009, \aap, 503, 755 

\bibitem{Milo12} Milone, A.~P., Piotto, G., Bedin, L.~R., et al.\ 2012, \aap, 540, A16 

\bibitem{Min10} Minniti, D., Lucas, P.~W., Emerson, J.~P., et al.\ 2010, NA, 15, 433 

\bibitem{Pio12} Piotto, G., Milone, A.~P., Anderson, J., et al.\ 2012, \apj, 760, 39 

\bibitem{Pla02} Platais, I., Kozhurina-Platais, V., Girard, T.~M., et al.\ 2002, \aj, 124, 601

\bibitem{Pla06} Platais, I., Wyse, R.~F.~G., \& Zacharias, N.\ 2006, \pasp, 118, 107 

\bibitem{Saito12} Saito, R.~K., Hempel, M., Minniti, D., et al.\ 2012, \aap, 537, A107 

\bibitem{Shous96} Shokin, Y.~A., \& Samus, N.~N.\ 1996, Astronomy Letters, 22, 761 

\bibitem{Skru06} Skrutskie, M.~F., Cutri, R.~M., Stiening, R., et al.\ 2006, \aj, 131, 1163 

\bibitem{Stet87} Stetson, P.~B.\ 1987, \pasp, 99, 191 

\bibitem{Stet00} Stetson, P.~B.\ 2000, \pasp, 112, 925 

\bibitem{Stet05} Stetson, P.~B.\ 2005, \pasp, 117, 563

\bibitem{vanAlt13} van Altena, W.~F.\ 2013, Astrometry for Astrophysics, by William F.~van Altena, Cambridge, UK: Cambridge University Press, 2012

\bibitem{Yad08} Yadav, R.~K.~S., Bedin, L.~R., Piotto, G., et al.\ 2008, \aap, 484, 609 , Paper~II

\bibitem{Zach13} Zacharias, N., Finch, C.~T., Girard, T.~M., et al.\ 2013, \aj, 145, 44 

\bibitem{Zloc12} Zloczewski, K., Kaluzny, J., Rozyczka, M., Krzeminski, W., \& Mazur, B.\ 2012, \actaa, 62, 357

\end{thebibliography}
\end{document}